\documentclass[11pt,epsfig]{article} 
\usepackage[utf8]{inputenc}
\usepackage[top=1in, left=0.95in, bottom=1.1in, right=0.95in]{geometry}
\usepackage[colorlinks=true,
linkcolor=blue,
urlcolor=red,
citecolor=red]{hyperref}

\usepackage[T1]{fontenc} 
\usepackage{cite}

\usepackage[toc]{appendix}
\usepackage{amssymb, amsmath, mathrsfs}
\numberwithin{equation}{section}

\usepackage{graphicx}
\usepackage{tikz}
\usepackage{tikz-3dplot}

\usetikzlibrary{decorations.pathmorphing,positioning}

\usepackage{caption} 
\usepackage{float}      
\usepackage{placeins}   

\usepackage{multirow, multicol}
\usepackage{tabularx}
\usepackage[boxsize=1.1em,aligntableaux=top]{ytableau}
\usepackage{amsmath}
\usepackage{amssymb}
\usepackage{mathtools}
\usepackage{longtable, amsmath, array}
\usepackage{booktabs}

\begin{document}
\begin{center}


{\Large \textbf  {Covariant Spinor Formalism for Multipole Expanded Form Factor}}\\[10mm]

Hong Huang $^{a,b,c}$\footnote{huanghong23@mails.ucas.ac.cn}, Tuo Tan $^{a,b,c}$\footnote{tantuo24@mails.ucas.ac.cn}, Yi-Ning Wang $^{ b,c}$\footnote{wangyining@itp.ac.cn},  Jiang-Hao Yu $^{a, b, c, d}$\footnote{jhyu@itp.ac.cn}\\[10mm]

\noindent 
$^a${ \small School of Fundamental Physics and Mathematical Sciences, Hangzhou Institute for Advanced Study, UCAS, Hangzhou 310024, China}  \\
$^b${ \small Institute of Theoretical Physics, Chinese Academy of Sciences,   Beijing 100190, P. R. China}   \\
$^c${ \small School of Physical Sciences, University of Chinese Academy of Sciences,   Beijing 100049, P.R. China} \\
$^d${ \small International Centre for Theoretical Physics Asia-Pacific, Beijing/Hangzhou, China}\\
[10mm]

\end{center}

\begin{abstract}
We present a systematic technique for constructing Lorentz covariant orbital-spin ($LS$) bases for matrix elements of local operators and the associated form factors, thereby extending the traditional multipole expansion to a Lorentz covariant formalism. In the spinor-helicity formalism, matrix elements of local operators for spin-$j$ particles can be treated as several massive 3-point scattering amplitudes, each of which can be further decomposed into different $LS$ partial wave amplitudes. We obtain explicit complete and linearly independent $LS$ amplitude bases for scalar, vector, and rank 2 tensor form factor of particles with spin-$\frac{1}{2}$, $1$, and $\frac{3}{2}$. In the Breit frame, it recovers the traditional multipole expansion expression, and we show the explicit equivalence among the traditional multipole expansion, canonical $LS$ expansion, and the $\mathrm{SO}(3)$ Zemach tensor expansion. Finally noting covariant structures built from the relativistic external wave functions and momenta of the initial and final state particles, we give a universal construction formula for form factor of arbitrary Lorentz tensor operators for arbitrary external spin particles.

\end{abstract}

\newpage

\tableofcontents

\section{Introduction}
\label{sec:intro}

Form factors (FFs) are among the standard observables used to describe the internal structures and interactions of composite particles. They arise when matrix elements of local operators between one particle states are expanded in a basis of independent tensor or spinor structures, with Lorentz invariant scalar coefficients identified as FFs. Typical examples are provided by the electromagnetic current and the energy-momentum tensor (EMT), whose matrix elements define electromagnetic and gravitational FFs and thereby encode quantities such as charge, magnetization, momentum, angular momentum, and other mechanical characteristics of the system~\cite{Sachs:1962zzc,Arrington:2006zm,Perdrisat:2006hj,Pacetti:2014jai,Pagels:1966zza,Burkert:2023wzr,Lorce:2024ipy}. In the context of QCD, where quarks and gluons are not directly observed because of confinement, such FFs serve as especially important probes of nonperturbative hadron structure~\cite{Polyakov:2018zvc,Burkert:2018bqq,Perdrisat:2006hj,Lorce:2018egm,Leader:2013jra}.

Most of the early literature on matrix element decompositions concentrated on low spin targets, in particular spin-$\tfrac{1}{2}$ systems such as the proton~\cite{Sachs:1962zzc,Arrington:2006zm,Perdrisat:2006hj,Burkert:2023wzr,Pacetti:2014jai}, with substantial extensions later developed for spin-$1$ particles~\cite{Efremov:1981vs,Brodsky:1992px,Gross:2002ge,Holstein:2006ud,Taneja:2011sy,Boer:2016xqr,Polyakov:2019lbq,Cosyn:2019aio,Pefkou:2021fni}. In recent years, attention has increasingly turned to higher spin states, including spin-$\tfrac{3}{2}$~\cite{Nozawa:1990gt,Pascalutsa:2006up,Kim:2020lrs}, spin-$2$, and eventually arbitrary spin cases~\cite{Jaffe:1989xy,Lorce:2009br,Lorce:2019sbq,Cotogno:2019xcl,Pefkou:2021fni,JiLiu2022,Sun:2026wqa}. Correspondingly, a number of counting rules have been derived for scalar, vector, and rank two tensor operators acting on particles of various spin~\cite{Scadron:1968zz,Williams:1970ms,Cotogno:2019xcl,Cotogno:2019vjb,Lorce:2009bs,Sun:2026wqa}. Beyond counting, explicit covariant constructions of matrix element bases have also been developed. Besides case by case parameterizations for specific particles or specific processes, systematic methods have been proposed for constructing bases for particles of arbitrary spin and for operators in general Lorentz representations. For example, the spinor Young tableaux method provides an algorithmic construction of Lorentz covariant tensor structures and has been used to obtain complete bases for particles with spin up to $2$~\cite{Sun:2026wqa}. Taken together, these studies have significantly broadened the available toolkit for explicit covariant parameterizations of local operator matrix elements.

Within such covariant constructions, the same matrix element can be expressed using different choices of spin quantization axes for the external particles, which can highlight different properties of composite particles. For example, the intrinsic degrees of freedom of the external particles may be represented in the helicity basis or in the canonical spin basis~\cite{Jacob:1959at,Berman:1965gi,Wick:1962zz,Chung:1971ri,Chung:1998,Chung:1993da,Chung:2007nn}.  The canonical spin basis is particularly useful for the present purpose, because the traditional $LS$ partial wave expansion and the conventional low energy multipole expansion should be formulated in terms of canonical spin states. 
However, these traditional formulations are not manifestly Lorentz covariant.  Therefore, by themselves, they do not provide a sufficiently systematic framework for constructing complete covariant bases of general local operator matrix elements.  In this work, we extend the $LS$ coupling construction to a Lorentz covariant setting with canonical spin states. When the matrix element basis is organized by the $LS$ coupling, the canonical spin construction reduces in the Breit frame to the familiar $LS$ partial wave form and also matches naturally onto the conventional low energy multipole expansion.  In this sense, our construction can be regarded as a covariant extension of the $LS$ coupling description 

To make this $LS$ organization explicit, we use an organizing principle familiar from conventional angular momentum analyses, namely to regard the operator insertion as the third object in a three point structure. In this form, the basis can be classified by the coupling of the orbital angular momentum $L$ and the total spin $S$ of the external particles. The orbital part is carried by the relative momentum of external particles, while the spin part is carried by the little group degrees of freedom of the external particles. The two parts are coupled to a fixed total angular momentum by Clebsch-Gordan coefficients (CGCs).  Each allowed $(L,S)$ channel then gives one independent angular momentum structure, whose coefficient is the corresponding form factor.

This construction is closely connected with several traditional methods based on angular momentum.  On the one hand, the $LS$ coupling viewpoint provides a useful way to compare our construction with the conventional multipole expansion~\cite{DelNobile:2021wmp,DeForest:1966ycn,Donnelly:1975ze,Donnelly:1978tz,Fitzpatrick:2012ix,Walecka:2001,Edmonds}.  In the $LS$ basis, a matrix element is organized by first introducing the orbital part, which carries the momentum dependence, and the spin part, which is obtained by coupling the spin degrees of freedom of the initial and final states.  These two parts are then coupled to the total $\mathrm{SO}(3)$ angular momentum $J$ carried by the operator component.  
In tensor formulations, the same idea is usually implemented by first constructing orbital and spin tensors in the three dimensional rotation group, as in the Zemach tensor method~\cite{Zemach:1963bc,Zemach:1965ycj}.
By contrast, the multipole expansion follows a different ordering.  The current is first decomposed into the charge density and the spatial current, and the corresponding matrix elements are then expanded into Coulomb, longitudinal, electric, and magnetic multipoles $C_K,L_K,E_K,M_K$.  This comparison shows that the common point of these two constructions is that the multipole order $K$ is equivalent to the total spin $S$ of the spin part in the $LS$ basis, which is also built from the initial and final spin states.  In terms of the correspondence between the two bases, $C_K$ and $M_K$ each correspond to the single $LS$ channel $(L,S)=(K,K)$, whereas $L_K$ and $E_K$ for $K\geq1$ are two fixed orthogonal linear combinations of $(K-1,K)$ and $(K+1,K)$.  Thus, the conventional multipole basis classifies the current by its charge, longitudinal, electric, and magnetic components, while the $LS$ basis classifies the same structures by their orbital angular momentum and total spin.


However, the traditional $LS$ expansion is not covariant. It is usually defined in the center-of-mass (COM) frame. Once a boost is performed, different $LS$ components can mix with each other. Therefore, a covariant $LS$ method is needed, so that the $LS$ basis can be defined in an arbitrary reference frame. This motivates the canonical-spinor method~\cite{Huang:2026egv}.
This approach has several advantages. To begin with, it is simple and algorithmic, because the essential structural information has already been encoded in the amplitude basis, so one only needs to carry out the prescribed conversion. Moreover, no separate redundancy removal step is required, since completeness and linear independence are inherited from the three point amplitude basis itself. The resulting basis is also fully general and does not rely on any assumption of $P$ or $T$ conservation. Finally, the counting of independent structures becomes straightforward as well with the same $LS$ coupling analysis.

Following this strategy, we construct complete and linearly independent bases for matrix elements with arbitrary initial and final spins and with operator insertions in arbitrary Lorentz representations $(j_L,j_R)$. These bases can be written explicitly, and their counting follows directly from the allowed $LS$ couplings. For application, we present the construction results in a more concrete form for the cases in which the external particles carry spin-$\tfrac{1}{2}$, spin-$1$, and spin-$\tfrac{3}{2}$. As a supplement, the general construction for arbitrary spin and arbitrary representations is also presented in the appendix for completeness. Finally, we compare our results with the recent covariant construction of Ref.~\cite{Sun:2026wqa}, and find redundant structures in the proposed spin-$1$ vector and rank-2 tensor bases. Altogether, these results provide a systematic covariant multipole expansion for generalized FFs that is explicit, complete, and free of hidden redundancies.

The remainder of this paper is organized as follows. In section~\ref{sec:Non-relativistic multipole expansion}, we have reviewed the conventional electromagnetic multipole expansion and compared it with the $LS$ expansion and the Zemach $LS$ expansion. In section~\ref{sec:Canonical-spinor formulation of the multipole expansion} develops the canonical-spinor formulation, explains how the complete three point $LS$ amplitude basis is converted into a basis of matrix elements, and derives the kinematic constraints implied by the $LS$ coupling. The subsequent sections present explicit results for external particles of spin-$\tfrac{1}{2}$, spin-$1$, and spin-$\tfrac{3}{2}$.

\section{Multipole expansion and canonical $LS$ expansion}
\label{sec:Non-relativistic multipole expansion}

\subsection{Traditional multipole expansion}

The traditional electromagnetic multipole expansion starts from the separation of the electromagnetic current into the charge density and the spatial current,
\begin{equation}
    J^\mu(\mathbf{x})=(\rho(\mathbf{x}),\mathbf{J}(\mathbf{x})),
\end{equation}
where $\rho(\mathbf x)$ is the charge density and $\mathbf J(\mathbf x)$ is the spatial current density. Under spatial rotations, $\rho(\mathbf x)$ behaves as a scalar, whereas $\mathbf J(\mathbf x)$ behaves as a vector.

The momentum space current is obtained through the three-dimensional Fourier transformation,
\begin{equation}
    J^\mu(\mathbf q)=\int d^3x\, e^{i\mathbf q\cdot\mathbf x}J^\mu(\mathbf x).
\end{equation}
Equivalently, the charge density and the spatial current are transformed separately,
\begin{equation}\label{eq:current_charge_spatial}
    \rho(\mathbf q)=\int d^3x e^{i\mathbf{q}\cdot \mathbf{x}}\rho(\mathbf{x}),\qquad \mathbf J(\mathbf q)=\int d^3x\, e^{i\mathbf q\cdot\mathbf x}\mathbf J(\mathbf x).
\end{equation}
The multipole expansion follows from expanding the plane wave into spherical waves,
\begin{equation}\label{eq:planewaveexpend}
    e^{i\mathbf q\cdot\mathbf x}=4\pi\sum_{k,m}i^k j_k(qx)Y^{*k}_{m}(\hat q)Y^k_{m}(\hat x),
\end{equation}
where $j_k(qx)$ is the spherical Bessel function and $Y^k_m$ is the spherical harmonic. This formula separates the dependence on the direction of the momentum transfer, $\hat q$, from the dependence on the coordinate direction, $\hat x$. In the radial integrals below, $q=|\mathbf q|$ and $x=|\mathbf x|$.

First, consider the charge density part. Substituting the plane wave expansion in Eq.~\eqref{eq:planewaveexpend} into Eq.~\eqref{eq:current_charge_spatial}, one obtains
\begin{equation}
    \rho(\mathbf q) = 4\pi \sum_{k,m} i^k Y^{*k}_{m}(\hat q) \int d^3x j_k(qx) Y^k_{m}(\hat x) \rho(\mathbf x).
\end{equation}
Then, The charge density multipoles can be defined as
\begin{equation}
    C^k_{m}(q^2)=4\pi i^k\int d^3xj_k(qx)Y^k_{m}(\hat x)\rho(\mathbf x).
\end{equation}
Thus the charge density in momentum space can be written as
\begin{equation}
    \rho(\mathbf q)=\sum_{k,m}Y^{*k}_{m}(\hat q)C^k_m(q^2).
\end{equation}
This is the charge multipole expansion. The angular dependence is
carried by the spherical harmonic $Y^{*k}_{m}(\hat q)$, while the dynamical
information is contained in the charge density multipoles $C^k_m(q^2)$. 
In the comparison below, the common normalization factor $4\pi i^k$ will be
suppressed for notational simplicity and absorbed into the reduced multipole
coefficient $C^k_m$.

For the spatial current, the Fourier transformation gives
\begin{equation}
    \mathbf J(\mathbf q)=\int d^3x\,e^{i\mathbf q\cdot\mathbf x}\mathbf J(\mathbf x).
\end{equation}
Different from the charge density, $\mathbf J(\mathbf q)$ carries a Cartesian vector index. It can therefore be decomposed into components longitudinal and transverse to the momentum-transfer direction $\hat q$. The transverse part contains two independent angular structures, conventionally called electric and magnetic. This motivates the introduction of the vector spherical harmonics used below. 

A convenient vector harmonic basis is
\begin{equation}
Y^{(\mathrm L)i}_{km}(\hat q)=\hat q^iY^k_m(\hat q),
\label{eq:YL_def_current}
\end{equation}
\begin{equation}
Y^{(\mathrm E)i}_{km}(\hat q)=\frac{1}{\sqrt{k(k+1)}}q\nabla^iY^k_m(\hat q),
\label{eq:YE_def_current}
\end{equation}
and
\begin{equation}
Y^{(\mathrm M)i}_{km}(\hat q)=\epsilon^{ijr}\hat q^jY^{(\mathrm E)r}_{km}(\hat q)=-\frac{i}{\sqrt{k(k+1)}}(\mathbf q\times\boldsymbol{\nabla})^iY^k_m(\hat q).
\label{eq:YM_def_current}
\end{equation}
Here $\nabla^i_q=\partial/\partial q^i$. Since 
\begin{equation}
\hat q_iY^{(\mathrm E)i}_{km}(\hat q)=\frac{1}{\sqrt{k(k+1)}}\hat q_iq\nabla_q^iY^k_m(\hat q)=0 ,
\end{equation}
and
\begin{equation}
\hat q_iY^{(\mathrm M)i}_{km}(\hat q)=\epsilon^{ijr}\hat q_i\hat q_jY^{(\mathrm E)r}_{km}(\hat q)=0 .
\end{equation}
Thus the electric and magnetic vector harmonics are transverse to the
momentum transfer direction, while $Y^{(\mathrm L)i}_{km}$ is longitudinal. For $k=0$, the ordinary spherical harmonic is a constant, so only the longitudinal branch survives.

We now decompose the spatial current into longitudinal and transverse
parts,
\begin{equation}
    J^i(\mathbf q)=J^{(\mathrm L)i}(\mathbf q)+J^{(\mathrm E)i}(\mathbf q)+J^{(\mathrm M)i}(\mathbf q).
\end{equation}

The longitudinal component is obtained by projecting the current onto the
direction of the momentum transfer,
\begin{equation}
    J_L(\mathbf q)=\hat q^iJ^i(\mathbf q).
\end{equation}
Using
\begin{equation}
    \nabla_x^i e^{i\mathbf q\cdot\mathbf x}=i q^i e^{i\mathbf q\cdot\mathbf x},\qquad \nabla^i_x=\frac{\partial}{\partial x^i},
\end{equation}
one has
\begin{equation}
    \hat q^i e^{i\mathbf q\cdot\mathbf x}=-\frac{i}{q}
    \nabla^i_x e^{i\mathbf q\cdot\mathbf x}.
\end{equation}
Using Eq.~\eqref{eq:planewaveexpend}, we obtain
\begin{equation}
    \hat q^i e^{i\mathbf q\cdot\mathbf x}=4\pi\sum_{k,m}i^kY^{*k}_m(\hat q)\left[-\frac{i}{q}\nabla^i_x\left(j_k(qx)Y^k_m(\hat x)\right)\right].
\end{equation}
Therefore
\begin{equation}
    J_L(\mathbf q)=4\pi\sum_{k,m}i^kY^{*k}_m(\hat q)L^k_m(q^2),
\end{equation}
where the longitudinal multipole is defined by
\begin{equation}
    L^k_m(q^2)=-\frac{i}{q}\int d^3x\left\{\nabla^i\left[j_k(qx)Y^k_m(\hat x)\right]\right\}J^i(\mathbf x).
\label{eq:Lkm_def_current}
\end{equation}
This expression gives the scalar longitudinal component. The corresponding
longitudinal vector current is obtained by multiplying back $\hat q^i$,
\begin{equation}
    J^{(\mathrm L)i}(\mathbf q)
    =
    \hat q^iJ_L(\mathbf q)
    =
    4\pi
    \sum_{k,m}
    i^k
    \hat q^iY^{*k}_m(\hat q)
    L^k_m(q^2).
\end{equation}
Using Eq.~\eqref{eq:YL_def_current}, this becomes
\begin{equation}
    J^{(\mathrm L)i}(\mathbf q)
    =
    4\pi
    \sum_{k,m}
    i^k
    Y^{(\mathrm L)*i}_{km}(\hat q)
    L^k_m(q^2).
\label{eq:JL_vector_reconstruct}
\end{equation}

The transverse current contains two independent structures. In the
conventional electromagnetic multipole expansion, they are called the
electric and magnetic transverse multipoles. With the present notation, the
magnetic multipole is defined by projecting the current onto the magnetic
vector spherical wave,
\begin{equation}
    M^k_m(q^2)
    =
    \int d^3x
    j_k(qx)
    Y^{(\mathrm M)i}_{km}(\hat x)
    J^i(\mathbf x).
\label{eq:Mkm_def_current}
\end{equation}
The electric multipole is defined by the curl of the magnetic vector
spherical wave,
\begin{equation}
    E^k_m(q^2)
    =
    \frac{1}{q}
    \int d^3x
    \left\{
    \nabla_x\times
    \left[
    j_k(qx)\mathbf Y^{(\mathrm M)}_{km}(\hat x)
    \right]
    \right\}^i
    J^i(\mathbf x).
\label{eq:Ekm_def_current}
\end{equation}
Equivalently,
\begin{equation}
    E^k_m(q^2)=\frac{1}{q}\int d^3x\epsilon^{ijr}\nabla^j_x\left[j_k(qx)Y^{(\mathrm M)r}_{km}(\hat x)\right]J^i(\mathbf x).
\end{equation}

After the electric and magnetic multipole coefficients have been extracted,
the transverse vector current is reconstructed by multiplying them back by
the corresponding transverse vector harmonics. The electric part is
\begin{equation}
    J^{(\mathrm E)i}(\mathbf q)
    =
    4\pi
    \sum_{k,m}
    i^k
    Y^{(\mathrm E)*i}_{km}(\hat q)
    E^k_m(q^2),
\label{eq:JE_vector_reconstruct}
\end{equation}
and the magnetic part is
\begin{equation}
    J^{(\mathrm M)i}(\mathbf q)
    =
    4\pi
    \sum_{k,m}
    i^k
    Y^{(\mathrm M)*i}_{km}(\hat q)
    M^k_m(q^2).
\label{eq:JM_vector_reconstruct}
\end{equation}
Therefore the full transverse current is
\begin{equation}
    J^{(\mathrm T)i}(\mathbf q)
    =
    4\pi
    \sum_{k,m}
    i^k
    \left[
    Y^{(\mathrm E)*i}_{km}(\hat q)E^k_m(q^2)
    +
    Y^{(\mathrm M)*i}_{km}(\hat q)M^k_m(q^2)
    \right].
\end{equation}

Combining the longitudinal and transverse parts, the full spatial current is
\begin{equation}
    J^i(\mathbf q)=J^{(\mathrm L)i}(\mathbf q)+J^{(\mathrm E)i}(\mathbf q)+J^{(\mathrm M)i}(\mathbf q).
\end{equation}
Thus
\begin{equation}
    J^i(\mathbf q)=4\pi\sum_{k,m}i^k\left[Y^{(\mathrm L)*i}_{km}(\hat q)L^k_m(q^2)+Y^{(\mathrm E)*i}_{km}(\hat q)E^k_m(q^2)+Y^{(\mathrm M)*i}_{km}(\hat q)M^k_m(q^2)
    \right].
\label{eq:spatial_current_LEM_expansion}
\end{equation}
This is the conventional electromagnetic multipole expansion of the spatial current. The vector spherical harmonics carry the angular dependence in $\hat q$, while $L^k_m(q^2)$, $E^k_m(q^2)$, and $M^k_m(q^2)$ contain the longitudinal, transverse electric, and transverse magnetic multipole information.

The conventional expansion above is an operator level decomposition of the electromagnetic current. For form factor matrix elements, these operators are inserted between one-particle states with spin-$j$. Their matrix elements therefore carry the external spin labels of the final and initial states.

For example, the charge multipole operator gives
\begin{equation}
C^k_m(q^2;a_1,a_3)=\langle p_1;ja_1|C^k_m(q^2)|p_3;ja_3\rangle ,
\end{equation}
and similarly for the longitudinal, electric, and magnetic multipoles,
\begin{equation}
X^k_m(q^2;a_1,a_3)=\langle p_1;ja_1|X^k_m(q^2)|p_3;ja_3\rangle ,\qquad X=C,L,E,M .
\end{equation}
Thus, after taking matrix elements, the conventional multipole coefficients are no longer only functions of the momentum transfer. They also form matrices in the spin space of the external particles.

\subsection{Canonical $LS$ expansion}
We consider the elastic matrix element of a local operator $\mathcal{O}^{\mu\nu\cdots}$ between one-particle states with the same spin $j$,
\begin{equation}\label{eq:general_me}
    \mathcal{M}^{\mu\nu\cdots}(p_1,p_3;a_1,a_3)
    =
    \langle p_1;j a_1|\mathcal{O}^{\mu\nu\cdots}|p_3;j a_3\rangle.
\end{equation}
In this convention, $p_3$ labels the initial one-particle state and $p_1$ labels the final one-particle state. The indices $a_3$ and $a_1$ denote the corresponding canonical spin projections. Both external particles are taken on shell, so that $p_1^2=p_3^2=m^2$. We also introduce the average momentum and momentum transfer as
\begin{equation}
    P^\mu=\frac{p^\mu_1+p^\mu_3}{2},\qquad Q^\mu=p^\mu_3-p^\mu_1.
\end{equation}
To make the $\mathrm{SO}(3)$ decomposition explicit, we work in the Breit frame. The average momentum and momentum transfer can be written as
\begin{equation}
    P^\mu=\frac{p^\mu_1+p^\mu_3}{2}=(E,0,0,0),\qquad Q^\mu=p^\mu_3-p^\mu_1=(0,\mathbf Q).
\end{equation}
In the Breit frame, the momentum transfer has only spatial components,
\begin{equation}\label{eq:breit_kinematics}
    Q^\mu=p^\mu_3-p^\mu_1=(0,\mathbf Q),\qquad \mathbf p_3=-\mathbf p_1=\frac{\mathbf Q}{2}.
\end{equation}
We further denote
\begin{equation}
    \mathbf p_3=-\mathbf p_1=\frac{\mathbf Q}{2},\qquad
    q=|\mathbf Q|,
    \qquad
    \hat q=\frac{\mathbf Q}{|\mathbf Q|}.
\end{equation}
The matrix element evaluated in the Breit frame is denoted by
\begin{equation}\label{eq:breit_me}
    \mathcal{M}^{\mu\nu\cdots}(\mathbf q;a_1,a_3)
    =
    \mathcal{M}^{\mu\nu\cdots}(p_1,p_3;a_1,a_3).
\end{equation}

Once the Breit frame has been chosen, the angular dependence of the elastic matrix element is fully encoded in the direction $\hat q$. The Lorentz indices of the operator can then be separated into time and spatial components and classified according to $\mathrm{SO}(3)$. In the following discussion, we focus mainly on a spatial component with indices $i_1i_2\cdots i_n$. The corresponding matrix element can be expanded in a finite set of independent $\mathrm{SO}(3)$ tensor structures,
\begin{equation}
    \mathcal{M}^{i_1 i_2 \cdots i_n}(\mathbf q;a_1,a_3)
    =
    \sum_r F_r(q^2)\,
    T^{i_1 i_2 \cdots i_n}_r(\hat q;a_1,a_3).
\end{equation}
The tensor structure $T^{i_1 i_2 \cdots i_n}_r(\hat q;a_1,a_3)$ contains the dependence on the direction $\hat q$ and on the external spin projections, while the scalar coefficient $F_r(q^2)$ is the corresponding form factor. 
The scalar coefficient $F_r(q^2)$ depends on the chosen tensor basis rather than on the choice of reference frame. For an elastic form factor, the Breit frame is kinematically equivalent to the COM frame formed by the two external momenta. Therefore, when the same basis is used, the same coefficient $F_r(q^2)$ is obtained in both descriptions. 
The explicit form of $F_r(q^2)$ depends on the chosen tensor basis, but the physical information contained in the full matrix element is basis independent. 

In the Condon-Shortley phase convention, crossing the initial spin state
$|p_3;j a_3\rangle$ to the final side converts its spin ket into the corresponding bra state,
\begin{equation}
| j,a_3\rangle
\quad\longleftrightarrow\quad
(-1)^{j-a_3}\langle j,-a_3| .
\end{equation}
To make the spin structure explicit, we rewrite the matrix element in a crossed-channel form by moving the initial particle to the final side. Keeping the crossing phase explicitly, the matrix element can be written as
\begin{equation}
    \mathcal{M}^{i_1i_2\cdots i_n}(\mathbf{q};a_1,a_3)
    =
    (-1)^{j-a_3}\,
    \langle p_1,\bar p_3;j a_1,j(-a_3)|
    \mathcal{O}^{i_1i_2\cdots i_n}|0\rangle .
\end{equation}
Next, the final state and the crossed initial state can be coupled through CGCs to obtain the total spin part $S$. In the $\mathrm{SO}(3)$ representation, two particles with spin $j$ are decomposed by the standard angular momentum decomposition into irreducible $\mathrm{SO}(3)$ representations. The corresponding decomposition and CGC can be written as
\begin{equation}
j\otimes j=\bigoplus_{S=0}^{2j}S,
\qquad
(-1)^{j-a_3}C^{S,a_S}_{j,a_1;j,-a_3},
\label{eq:crossed_spin_decomp}
\end{equation}
where the phase from crossing the initial spin state to the final side is kept in front of the CGC. 

Equivalently, this crossed-channel coupling can also be expressed in terms of the rotation generators associated with the spin of the external particles. The spin coupling part $\Sigma^{(S)}_{a_S}$ is defined as
\begin{equation}
\Sigma^{(S)}_{a_S}(a_1,a_3)
\equiv
\mathcal N_{jS}\,
(-1)^{j-a_3}C^{S,a_S}_{j,a_1;j,-a_3},
\qquad
S=0,1,\ldots,2j ,
\label{eq:spin_tensor_cg_relation}
\end{equation}
where the constants $\mathcal N_{jS}$ are used for the normalization matched to the rotation generators associated with the spin of the external particles.

Next, the corresponding spin generators are given. For the external particles with spin-$\frac{1}{2}$,
the matrix elements of the generators are
\begin{equation}
(J^i)_{a_1a_3}=\frac{1}{2}(\sigma^i)_{a_1a_3},
\end{equation}
where
\begin{equation}
\sigma^x=
\begin{pmatrix}
0&1\\
1&0
\end{pmatrix},
\qquad
\sigma^y=
\begin{pmatrix}
0&-i\\
i&0
\end{pmatrix},
\qquad
\sigma^z=
\begin{pmatrix}
1&0\\
0&-1
\end{pmatrix}.
\end{equation}
Their spherical components\footnote{The detailed relation between the spherical basis and the tensor basis is given in Appendix~\ref{app:Spherical and tensor basis}.} are
\begin{equation}
\sigma_0=\sigma^z,\qquad
\sigma_{+1}=-\frac{1}{\sqrt{2}}(\sigma^x+i\sigma^y),\qquad
\sigma_{-1}=\frac{1}{\sqrt{2}}(\sigma^x-i\sigma^y).
\end{equation}

For the external particles with spin-$1$, the matrix elements of the generators are
\begin{equation}\label{Eq:spin1gener}
S^x
=
\frac{1}{\sqrt{2}}
\begin{pmatrix}
0&1&0\\
1&0&1\\
0&1&0
\end{pmatrix},
\qquad
S^y
=
\frac{1}{\sqrt{2}}
\begin{pmatrix}
0&-i&0\\
i&0&-i\\
0&i&0
\end{pmatrix},
\qquad
S^z
=
\begin{pmatrix}
1&0&0\\
0&0&0\\
0&0&-1
\end{pmatrix}.
\end{equation}
The corresponding spherical components are
\begin{equation}
S_0=S^z,\qquad
S_{+1}=-\frac{1}{\sqrt{2}}(S^x+iS^y),\qquad
S_{-1}=\frac{1}{\sqrt{2}}(S^x-iS^y).
\end{equation}

With these conventions, for the spin-$\frac{1}{2}$ case, the corresponding coupling gives total spin $S=0$ and $S=1$. The corresponding spin coupling structures can be written as
\begin{equation}
\begin{aligned}
    S&=0:\qquad
\Sigma^{(0)}_0(a_1,a_3)= \delta_{a_1a_3},\\
S&=1:\qquad
\Sigma^{(1)}_m(a_1,a_3)= (\sigma_m)_{a_1a_3},
\qquad
m=0,\pm1,
\end{aligned}
\end{equation}
where $\Sigma^{(1)}_m$ is the spherical spin component. The Cartesian spin vector is
\begin{equation}
\Sigma^{(1),i}_{\rm cart}(a_1,a_3)
\equiv
\sum_{m=-1}^{1}\bar e^i_m\Sigma^{(1)}_m(a_1,a_3)
=
(\sigma^i)_{a_1a_3},
\label{eq:spinhalf_cartesian_spin_tensor}
\end{equation}
where $\bar e^i_m$ denotes the dual spherical basis vector used to convert spherical components into tensor components, and it can be obtained from Eq.~\eqref{eq:dual_spherical_basis_vector}.

For spin-$1$, the corresponding coupling gives total spin $S=0$, $S=1$ and $S=2$. The corresponding spin coupling structures can be written as
\begin{equation}
\begin{aligned}
    S&=0:\qquad
\Sigma^{(0)}_0(a_1,a_3)= \delta_{a_1a_3},\\
S&=1:\qquad
\Sigma^{(1)}_m(a_1,a_3)= (S_m)_{a_1a_3},\\
S&=2:\qquad
\Sigma^{(2)}_M(a_1,a_3)= Q^{(2)}_{M,a_1a_3},
\end{aligned}
\end{equation}
where $(S_m)_{a_1a_3}$ and $Q^{(2)}_M$ is the spherical spin component.  The Cartesian form are
\begin{equation}
\Sigma^{(1),i}_{\rm cart}(a_1,a_3)
\equiv
\sum_{m=-1}^{1}\bar e^i_m\Sigma^{(1)}_m(a_1,a_3)
=
(S^i)_{a_1a_3}.
\label{eq:spinone_cartesian_dipole}
\end{equation}
and
\begin{equation}\label{eq:spinone_spherical_to_cartesian_quadrupole}
\begin{aligned}
\Sigma^{(2),ij}_{\rm cart}(a_1,a_3)
&\equiv
\sum_{M=-2}^{2}\bar e^{ij}_{2M}\Sigma^{(2)}_M(a_1,a_3)\\
&=\left[
\frac{1}{2}\{S^i,S^j\}
-\frac{1}{3}\delta^{ij}\mathbf S^2
\right]_{a_1a_3}\\
&=Q^{ij}_{a_1a_3},
\end{aligned}
\end{equation}
where $\bar e^{ij}_{2M}$ denotes the dual spherical basis tensor used to convert spherical components into tensor components.
Therefore for spin-$1$ the three crossed-channel sectors $S=0,1,2$ correspond to the monopole, dipole, and quadrupole spin structures.

The orbital part is considered next. In the Breit frame, the external kinematic dependence is encoded in the magnitude $q$ and the direction $\hat q$. For a fixed $\mathrm{SO}(3)$ component of the operator, the angular dependence can be expanded in spherical harmonics $Y^L_m(\hat q)$, which carry orbital angular momentum $L$. In Cartesian form, a rank-$L$ spherical harmonic is equivalent to the symmetric traceless part of $L$ factors of $\hat q^i$ projected onto the spherical-component tensor basis,
\begin{equation}
Y^L_M(\hat q)
=
\sqrt{\frac{(2L+1)!!}{4\pi L!}}\,
e^{i_1\cdots i_L}_{LM}\,
\hat q^{\langle i_1}\cdots \hat q^{i_L\rangle},
\label{eq:spherical_to_cartesian_symmetric_traceless_general}
\end{equation}
where angular brackets denote the symmetric traceless projection. Thus the spherical label $(L,M)$ and the Cartesian symmetric traceless tensor carry the same irreducible $\mathrm{SO}(3)$ representation. 

For $L=0$,
\begin{equation}
Y^0_0(\hat q)=\frac{1}{\sqrt{4\pi}}.
\end{equation}
For $L=1$, the spherical harmonics are the spherical components of the vector
$\hat q^i$, up to the standard normalization,
\begin{equation}
Y^1_M(\hat q)
=
\sqrt{\frac{3}{4\pi}}\,e^i_M\hat q^i .
\end{equation}
For $L=2$, the relevant Cartesian object is the symmetric traceless quadrupole
\begin{equation}
Q^{ij}(\hat q)
\equiv
\hat q^i\hat q^j-\frac{1}{3}\delta^{ij}.
\label{eq:rank2_symmetric_traceless_Qij}
\end{equation}
The five $L=2$ spherical harmonics are the five spherical components of this
symmetric traceless tensor, equivalently the contractions of $Q^{ij}$ with the
basis tensors in Eq.~\eqref{eq:rank2_spherical_tensor_basis}. Explicitly,
\begin{equation}
Y^2_0(\hat q)
=
\sqrt{\frac{5}{16\pi}}\,(3(\hat q^z)^2-1)
=
\sqrt{\frac{45}{16\pi}}\,Q^{zz},
\label{eq:Y20_cartesian_symmetric_traceless}
\end{equation}
\begin{equation}
Y^2_{+1}(\hat q)
=
-\sqrt{\frac{15}{8\pi}}\,
\hat q^z(\hat q^x+i\hat q^y)
=
-\sqrt{\frac{15}{8\pi}}\,
(Q^{zx}+iQ^{zy}),
\label{eq:Y21p_cartesian_symmetric_traceless}
\end{equation}
\begin{equation}
Y^2_{-1}(\hat q)
=
\sqrt{\frac{15}{8\pi}}\,
\hat q^z(\hat q^x-i\hat q^y)
=
\sqrt{\frac{15}{8\pi}}\,
(Q^{zx}-iQ^{zy}),
\label{eq:Y21m_cartesian_symmetric_traceless}
\end{equation}
and
\begin{equation}
Y^2_{+2}(\hat q)
=
\sqrt{\frac{15}{32\pi}}\,
(\hat q^x+i\hat q^y)^2
=
\sqrt{\frac{15}{32\pi}}\,
\left(Q^{xx}-Q^{yy}+2iQ^{xy}\right),
\label{eq:Y22p_cartesian_symmetric_traceless}
\end{equation}
\begin{equation}
Y^2_{-2}(\hat q)
=
\sqrt{\frac{15}{32\pi}}\,
(\hat q^x-i\hat q^y)^2
=
\sqrt{\frac{15}{32\pi}}\,
\left(Q^{xx}-Q^{yy}-2iQ^{xy}\right).
\label{eq:Y22m_cartesian_symmetric_traceless}
\end{equation}

Having identified the spin and orbital parts, the multipole expansion is organized by coupling the crossed-channel spin $S$ to the orbital angular momentum $L$ according to $L\otimes S \to J$. For a given component with definite total angular momentum $J$, the corresponding crossed-channel three-point structure is introduced as
\begin{equation}
\mathcal A^{J,M}(\mathbf{q};a_1,a_3)
=
\sum_{L,S}F^{(J)}_{LS}(q^2)
\sum_{m,a_S}
C^{J,M}_{L,m;S,a_S}
Y^L_m(\hat q)
\Sigma^{(S)}_{a_S}(a_1,a_3).
\label{eq:threepoint_amp}
\end{equation}
The tensor $\Sigma^{(S)}_{a_S}$ is the irreducible spin tensor defined in Eq.~\eqref{eq:spin_tensor_cg_relation}. It is a spherical component in spin space. The coefficients $F^{(J)}_{LS}(q^2)$ are the form factors in the $LS$ basis. They are scalar functions of $q^2$ and contain the dynamical information associated with the operator insertion in a channel of definite total angular momentum $J$, orbital angular momentum $L$, and crossed-channel spin $S$. By construction, all nontrivial angular dependence is carried by the spherical harmonics. 
The CGC $C^{J,M}_{L,m;S,a_S}$ couples the orbital part and the spin part into total angular momentum $J$ with projection $M$.

Once the spherical amplitude $\mathcal A^{J,M}$ is obtained, the corresponding Cartesian tensor matrix element is reconstructed by attaching the appropriate dual Cartesian reconstruction tensor,
\begin{equation}
\mathcal M^{i_1\cdots i_n}(\mathbf{q};a_1,a_3)
=
\sum_{J,M}
\mathbb T^{\,i_1\cdots i_n}_{J,M}\,
\mathcal A^{J,M}(\mathbf{q};a_1,a_3).
\label{eq:tensor_reconstruction}
\end{equation}
where the tensor $\mathbb T^{\,i_1\cdots i_n}_{J,M}$ converts the spherical component $\mathcal A^{J,M}$ into the Cartesian indices carried by the matrix element, which can be obtained from Eqs.~\eqref{eq:spherical_tensor_basis_general} and \eqref{eq:dual_spherical_tensor_basis}.
For a scalar operator component, there is no Cartesian index and only $J=0$ contributes, so that $\mathbb T_{0,0}=1$. For a vector operator component, $J=1$, and the reconstruction tensor is the dual spherical basis vector $\mathbb T^i_{1,M}=\bar e^i_M$. The Cartesian vector matrix element is therefore reconstructed as
\begin{equation}
\mathcal M^i(\mathbf q;a_1,a_3)
=
\sum_{M=-1}^{1}\bar e^i_M\,
\mathcal A^{1,M}(\mathbf q;a_1,a_3).
\end{equation}
Therefore, Eq.~\eqref{eq:tensor_reconstruction} is simply the conversion from irreducible spherical components to Cartesian tensor components.

\subsection{Zemach tensor expansion}

We now rewrite the multipole expansion in the Zemach tensor formulation~\cite{Zemach:1963bc,Zemach:1965ycj}. In the Breit frame, the orbital dependence is described by symmetric traceless tensors built from the direction $\hat q$, while the spin dependence is described by tensors constructed from the rest frame wave functions of the external states. The corresponding $LS$ decomposition is therefore realized entirely at the tensor level.

For the matrix element of an operator component with definite total angular momentum $J$ under spatial rotations, the orbital dependence is carried by the direction $\hat q$, and the rank-$L$ Zemach tensor is defined by
\begin{equation}
Y^{(L)}_{i_1\cdots i_L}(\hat q)
\equiv
\hat q_{i_1}\cdots \hat q_{i_L}-\text{trace part},
\end{equation}
where $Y^{(L)}_{i_1\cdots i_L}$ denotes the Cartesian symmetric traceless tensor built from $\hat q$, in contrast to the spherical harmonic $Y^L_m(\hat q)$ in spherical basis.
For the first few values of $L$, one has
\begin{equation}
Y^{(0)}=1,
\qquad
Y^{(1)}_i=\hat q_i,
\qquad
Y^{(2)}_{ij}=\hat q_i\hat q_j-\frac{1}{3}\delta_{ij}.
\end{equation}

We now turn to the spin part. For a particle of spin-$j$, let $\chi^{(j)}_{A}(a)$ denote the rest frame wave function with canonical spin projection $a=-j,-j+1,\dots,j$, where $A$ collectively denotes the spin indices in the chosen tensor realization. For integer spin, $\chi^{(j)}_{A}$ may be realized as a symmetric traceless Cartesian tensor of rank $j$. For half-integer spin, it is realized as the corresponding spinor or spinor-tensor wave function.

The Cartesian spin tensors are obtained by coupling the initial and final rest-frame wave functions into irreducible $\mathrm{SO}(3)$ tensors. For a given total spin $S$, we define
\begin{equation}
\Sigma^{(S),k_1\cdots k_S}_{\rm cart}(a_1,a_3)
=
\big[
\chi^{(j)\dagger}(a_1)\otimes\chi^{(j)}(a_3)
\big]^{(S),k_1\cdots k_S},
\label{eq:general_spin_tensor_def}
\end{equation}
namely the irreducible Cartesian rank-$S$ tensor obtained by coupling the initial and final rest-frame wave functions. Since $j\otimes j=\bigoplus\limits_{S=0}^{2j}S$, the allowed spin tensors are labeled by $S=0,1,\dots,2j$.

Within the Zemach tensor method, the $LS$ multipole expansion of the matrix element of an operator component with definite total angular momentum $J$ can be written as
\begin{equation}
\mathcal M^{i_1i_2\cdots i_J}(\mathbf q;a_1,a_3)
=
\sum_{L,S}
F^{(J)}_{LS}(q^2)
\big[
Y^{(L)}(\hat q)\otimes \Sigma^{(S)}_{\rm cart}(a_1,a_3)
\big]^{i_1i_2\cdots i_J}.
\label{eq:zemach_amp_general_rewrite}
\end{equation}
The functions $F^{(J)}_{LS}(q^2)$ are the form factors in the $LS$ basis. All angular dependence is carried by the Zemach tensors, while the dynamical information is encoded in the scalar coefficients $F^{(J)}_{LS}(q^2)$.

We now specialize to the spin-$\frac{1}{2}\to\frac{1}{2}$ case.
For spin-$\frac{1}{2}$, the rest-frame wave functions reduce to the two-component Pauli spinors
\begin{equation}
\chi_{+\frac{1}{2}}
=
\begin{pmatrix}
1\\
0
\end{pmatrix},
\qquad
\chi_{-\frac{1}{2}}
=
\begin{pmatrix}
0\\
1
\end{pmatrix}.
\end{equation}
Since $\frac{1}{2}\otimes\frac{1}{2}=0\oplus1$, the corresponding spin tensors are
\begin{equation}
\Sigma^{(0)}_0(a_1,a_3)
=
\chi^\dagger_{a_1}\chi_{a_3},\qquad \Sigma^{(1),i}_{\rm cart}(a_1,a_3)=\chi^\dagger_{a_1}\sigma^i\chi_{a_3},
\end{equation}
where $\sigma^i$ are the Pauli matrices.

\subsection{Comparison among three methods}
\subsubsection{Relation between multipoles and $LS$ expansion}

We now keep the same conventional electromagnetic multipole expansion, but rewrite it as a matrix element between the initial and final one-particle states.  Thus the current components $J^0(\mathbf q)$ and $J^i(\mathbf q)$ are replaced by $\mathcal M^0(\mathbf q;a_1,a_3)$ and $\mathcal M^i(\mathbf q;a_1,a_3)$, and the multipole coefficients acquire the little-group indices of the external states.  No new multipole basis is being introduced here; this is only a change of notation that keeps track of the spin labels and makes the comparison with the crossed-channel $LS$ method direct.  In the Breit frame, the time component is an $\mathrm{SO}(3)$ scalar and is therefore expanded in ordinary spherical harmonics,
\begin{align}
    \mathcal M^0(\mathbf q;a_1,a_3)&=\sum_{k=0}^{2j}\sum_{m=-k}^{k}C^k_{m}(q^2;a_1,a_3)\,Y^{*k}_m(\hat q)\\
    &=\sum_{k=0}^{2j}\sum_{m=-k}^{k}(-1)^mC^k_{m}(q^2;a_1,a_3)\,Y^{k}_{-m}(\hat q),
\label{eq:J0_general_multipole}
\end{align}
where $k$ denotes the conventional multipole rank.  The spatial component of the matrix element transforms as an $\mathrm{SO}(3)$ vector and is expanded in the same vector spherical harmonics used for the spatial current,
\begin{align}
\mathcal M^i(\mathbf q;a_1,a_3)
&=\sum_{k=0}^{2j}\sum_{m=-k}^{k}
\Big[
L^k_{m}(q^2;a_1,a_3)\,Y^{(\mathrm L)*\,i}_{km}(\hat q)
+
E^k_{m}(q^2;a_1,a_3)\,Y^{(\mathrm E)*\,i}_{km}(\hat q)
+
M^k_{m}(q^2;a_1,a_3)\,Y^{(\mathrm M)*\,i}_{km}(\hat q)
\Big]
\notag\\
&=\sum_{k=0}^{2j}\sum_{m=-k}^{k}(-1)^m
\Big[
L^k_{m}(q^2;a_1,a_3)\,Y^{(\mathrm L)\,i}_{k,-m}(\hat q)
+
E^k_{m}(q^2;a_1,a_3)\,Y^{(\mathrm E)\,i}_{k,-m}(\hat q)
+
M^k_{m}(q^2;a_1,a_3)\,Y^{(\mathrm M)\,i}_{k,-m}(\hat q)
\Big].
\label{eq:Jvec_general_multipole}
\end{align}
Here $C^k_m$, $L^k_m$, $E^k_m$, and $M^k_m$ are the same conventional charge,
longitudinal, electric, and magnetic multipoles defined above.  The vector
spherical harmonics are also the same ones introduced in
Eqs.~\eqref{eq:YL_def_current}, \eqref{eq:YE_def_current} and \eqref{eq:YM_def_current}; in particular,
for $k=0$ only the longitudinal branch survives.

For the spin-$\frac{1}{2}$ example below, only $k=0$ and $k=1$ can appear.  We
therefore first write the ordinary spherical harmonics in terms of
$q^i/q$.  For $k=0$,
\begin{equation}
Y^0_0(\hat q)=\frac{1}{\sqrt{4\pi}} .
\label{eq:Y00_qhat_explicit}
\end{equation}
For $k=1$,
\begin{equation}
Y^1_0(\hat q)
=
\sqrt{\frac{3}{4\pi}}\frac{q^z}{q}
=
\sqrt{\frac{3}{4\pi}}\hat q^z,
\label{eq:Y10_qhat_explicit}
\end{equation}
\begin{equation}
Y^1_{+1}(\hat q)
=
-\sqrt{\frac{3}{8\pi}}
\frac{q^x+iq^y}{q}
=
-\sqrt{\frac{3}{8\pi}}(\hat q^x+i\hat q^y),
\label{eq:Y1p_qhat_explicit}
\end{equation}
and
\begin{equation}
Y^1_{-1}(\hat q)
=
\sqrt{\frac{3}{8\pi}}
\frac{q^x-iq^y}{q}
=
\sqrt{\frac{3}{8\pi}}(\hat q^x-i\hat q^y).
\label{eq:Y1m_qhat_explicit}
\end{equation}

Thus $q\nabla^i$ removes the radial factor and projects onto the tangent plane of the sphere.

Now insert these results into the three vector harmonics
Eqs.~\eqref{eq:YL_def_current}, \eqref{eq:YE_def_current} and \eqref{eq:YM_def_current}.  For $k=0$, the ordinary harmonic
is constant, so
\begin{equation}
q\nabla^iY^0_0=0.
\end{equation}
The three vector harmonics are therefore
\begin{equation}
Y^{(\mathrm L)i}_{00}
=
\frac{1}{\sqrt{4\pi}}\,\hat q^i,
\qquad
Y^{(\mathrm E)i}_{00}=0,
\qquad
Y^{(\mathrm M)i}_{00}=0.
\label{eq:k0_vector_harmonics_explicit}
\end{equation}
This is why the $k=0$ spatial-current sector contains only the longitudinal
monopole $L_0$.

For $k=1$, it is compact to write
\begin{equation}
Y^1_m(\hat q)=a^j_m\hat q^j,
\end{equation}
where
\begin{equation}
a^j_0=\sqrt{\frac{3}{4\pi}}\delta^{jz},
\qquad
a^j_{+1}=-\sqrt{\frac{3}{8\pi}}(\delta^{jx}+i\delta^{jy}),
\qquad
a^j_{-1}=+\sqrt{\frac{3}{8\pi}}(\delta^{jx}-i\delta^{jy}).
\end{equation}
The three $k=1$ vector harmonics are
\begin{equation}
Y^{(\mathrm L)i}_{1m}
=
\hat q^i\,a^j_m\hat q^j,
\label{eq:k1_L_general_m}
\end{equation}
\begin{equation}
Y^{(\mathrm E)i}_{1m}
=
\frac{1}{\sqrt{2}}\,
a^j_m\left(\delta^{ij}-\hat q^i\hat q^j\right),
\label{eq:k1_E_general_m}
\end{equation}
and
\begin{equation}
Y^{(\mathrm M)i}_{1m}
=
\frac{1}{\sqrt{2}}\,
\epsilon^{il j}\hat q^l a^j_m .
\label{eq:k1_M_general_m}
\end{equation}
In particular, for $m=0$,
\begin{equation}
Y^{(\mathrm L)i}_{10}
=
\sqrt{\frac{3}{4\pi}}\,\hat q^i\hat q^z,
\qquad
Y^{(\mathrm E)i}_{10}
=
\sqrt{\frac{3}{8\pi}}
\left(\delta^{iz}-\hat q^i\hat q^z\right),
\label{eq:k1_LE_m0_explicit}
\end{equation}
\begin{equation}
Y^{(\mathrm M)i}_{10}
=
\sqrt{\frac{3}{8\pi}}\,
\epsilon^{il z}\hat q^l .
\label{eq:k1_M_m0_explicit}
\end{equation}
For $m=+1$,
\begin{align}
Y^{(\mathrm L)i}_{1,+1}
&=
-\sqrt{\frac{3}{8\pi}}\,
\hat q^i(\hat q^x+i\hat q^y),
\notag\\
Y^{(\mathrm E)i}_{1,+1}
&=
-\sqrt{\frac{3}{16\pi}}\,
\left[
\delta^{ix}+i\delta^{iy}
-\hat q^i(\hat q^x+i\hat q^y)
\right],
\notag\\
Y^{(\mathrm M)i}_{1,+1}
&=
-\sqrt{\frac{3}{16\pi}}\,
\epsilon^{il j}\hat q^l
(\delta^{jx}+i\delta^{jy}) .
\label{eq:k1_LEM_p_explicit}
\end{align}
For $m=-1$,
\begin{align}
Y^{(\mathrm L)i}_{1,-1}
&=
+\sqrt{\frac{3}{8\pi}}\,
\hat q^i(\hat q^x-i\hat q^y),
\notag\\
Y^{(\mathrm E)i}_{1,-1}
&=
+\sqrt{\frac{3}{16\pi}}\,
\left[
\delta^{ix}-i\delta^{iy}
-\hat q^i(\hat q^x-i\hat q^y)
\right],
\notag\\
Y^{(\mathrm M)i}_{1,-1}&=+\sqrt{\frac{3}{16\pi}}\,\epsilon^{il j}\hat q^l(\delta^{jx}-i\delta^{jy}) .
\label{eq:k1_LEM_m_explicit}
\end{align}

Next, we consider the spin part in front of the vector spherical harmonic functions obtained from the matrix element decomposition. This part contains the spin information of the initial and final states. Therefore, it can be written as
\begin{equation}
    X^k_m(q^2;a_1,a_3)=\langle p_1;ja_1 |X^k_m|p_3;ja_3\rangle,\qquad X=C,L,E,M,
\end{equation}
where $X^k_m$ is an operator with angular momentum $k$. By using the Wigner-Eckart theorem, the above expression can be decomposed as
\begin{align}
    \langle p_1;ja_1 |X^k_m|p_3;ja_3\rangle&=\frac{1}{\sqrt{2j+1}}X_k(q^2)\mathcal{S}^k_m(a_1,a_3)\\&=C^{j,a_1}_{j,a_3;k,m}\frac{1}{\sqrt{2j+1}}X_{k}(q^2)\\
    &=(-1)^{j-a_3}
\left[
\frac{2j+1}{2k+1}
\right]^{\frac{1}{2}}
C^{k,-m}_{j,a_3;j,-a_1}\frac{1}{\sqrt{2j+1}}
X_k(q^2).
\end{align}
where the standard symmetry relation of the CGCs has
been used. This formula is simply a CGC multiplied by a scalar coefficient.
Therefore, we only need to identify the corresponding spin tensor basis.

For example, for the particle spin-$\frac{1}{2}$ and $k=0$, it can be written as
\begin{equation}
\mathcal S^0_0(a_1,a_3)\propto\delta_{a_1a_3},
\label{eq:spinhalf_S00_def}
\end{equation}

For $k=1$, it can be written as
\begin{equation}
\mathcal S^1_0(a_1,a_3)\propto(\sigma^z)_{a_1a_3},
\qquad
\mathcal S^1_{+1}(a_1,a_3)
\propto
-\frac{1}{\sqrt{2}}(\sigma^x+i\sigma^y)_{a_1a_3},
\qquad
\mathcal S^1_{-1}(a_1,a_3)
\propto
\frac{1}{\sqrt{2}}(\sigma^x-i\sigma^y)_{a_1a_3}.
\label{eq:spinhalf_S1m_def}
\end{equation}

Because the spin tensors in Eqs.~\eqref{eq:spinhalf_S00_def} and \eqref{eq:spinhalf_S1m_def} have only been fixed up to normalization, the following spin-dependent structures are written up to the corresponding absorbed scalar coefficients.  Keeping only $k=0$ and $k=1$, the charge component is therefore
\begin{equation}
\mathcal M^0(\mathbf q;a_1,a_3)
\propto
C_0(q^2)\,
\mathcal S^0_0(a_1,a_3)Y^0_0(\hat q)
+
C_1(q^2)
\sum_{m=-1}^{1}(-1)^{m}
\mathcal S^1_m(a_1,a_3)Y^1_{-m}(\hat q).
\label{eq:spinhalf_relevant_C_expansion}
\end{equation}
The $k=0$ structure is
\begin{equation}
\mathcal S^0_0Y^0_0
\propto
\frac{1}{\sqrt{4\pi}}\delta_{a_1a_3},
\label{eq:C0_km_structure}
\end{equation}
whereas the $k=1$ structure is
\begin{equation}
\sum_{m=-1}^{1}(-1)^m\mathcal S^1_{m}Y^1_{-m}
\ \propto\
(\boldsymbol{\sigma}\cdot\hat q)_{a_1a_3},
\label{eq:C1_km_structure}
\end{equation}
up to the normalization convention for the spherical components.  These are the $km$-dependent parts multiplying the reduced form factors $C_0(q^2)$ and $C_1(q^2)$.

For the spatial current, the same factorization gives
\begin{align}
\mathcal M^i(\mathbf q;a_1,a_3)
&\propto
L_0(q^2)\,
\mathcal S^0_0(a_1,a_3)Y^{(\mathrm L)i}_{00}(\hat q)
\notag\\
&\quad
+
\sum_{m=-1}^{1}(-1)^{m}\mathcal S^1_{m}(a_1,a_3)
\Big[
L_1(q^2)Y^{(\mathrm L)i}_{1,-m}(\hat q)
+
E_1(q^2)Y^{(\mathrm E)i}_{1,-m}(\hat q)
+
M_1(q^2)Y^{(\mathrm M)i}_{1,-m}(\hat q)
\Big].
\label{eq:spinhalf_relevant_LEM_expansion}
\end{align}
The $km$-dependent structures multiplying the reduced form factors are
\begin{equation}
L_0:\quad
\mathcal S^0_0Y^{(\mathrm L)i}_{00}
\propto
\frac{1}{\sqrt{4\pi}}\hat q^i\delta_{a_1a_3},
\label{eq:L0_km_structure}
\end{equation}
\begin{equation}
L_1:\quad
\sum_{m=-1}^{1}
(-1)^{m}\mathcal S^1_{m}Y^{(\mathrm L)i}_{1,-m}
\ \propto\
\hat q^i(\boldsymbol{\sigma}\cdot\hat q)_{a_1a_3},
\label{eq:L1_km_structure}
\end{equation}
\begin{equation}
E_1:\quad
\sum_{m=-1}^{1}
(-1)^{m}\mathcal S^1_{m}Y^{(\mathrm E)i}_{1,-m}
\ \propto\
\left[\sigma^i-\hat q^i(\boldsymbol{\sigma}\cdot\hat q)\right]_{a_1a_3},
\label{eq:E1_km_structure}
\end{equation}
and
\begin{equation}
M_1:\quad
\sum_{m=-1}^{1}
(-1)^{m}\mathcal S^1_{m}Y^{(\mathrm M)i}_{1,-m}
\ \propto\
\bigl(i\boldsymbol{\sigma}\times\hat q\bigr)^i_{a_1a_3}.
\label{eq:M1_km_structure}
\end{equation}
The complete conventional spin-$\frac{1}{2}$ multipole content before imposing parity or current conservation is
\begin{equation}
J^0:\quad C_0,\ C_1,
\qquad
J^i:\quad L_0,\ L_1,\ E_1,\ M_1,
\label{eq:spinhalf_conventional_content}
\end{equation}
where each symbol denotes a reduced form factor multiplying the corresponding $km$ structure displayed above.

\paragraph{Spin-$\frac{1}{2}$ example}
For elastic spin-$\frac{1}{2}$ matrix elements, the crossed-channel spin can take only $S=0,1$. Therefore the possible spin tensors, after undoing the crossing convention, are the scalar spin structure and the vector spin structure,
\begin{equation}
S=0:\quad \delta_{a_1a_3},
\qquad
S=1:\quad (\sigma^i)_{a_1a_3}.
\end{equation}
In the $LS$ basis, the orbital harmonic $Y^L_m(\hat q)$ is coupled directly to these spin tensors.  In the conventional basis, one first builds $C/L/E/M$ harmonics and only then restores the spin dependence.

First consider the charge density $J^0$.  Since it is a scalar component, the
total angular momentum is $J=0$ and the $LS$ condition is $L=S$.  The two
allowed channels are
\begin{equation}
(L,S)=(0,0),\qquad (1,1),
\end{equation}
which give
\begin{equation}
(0,0):\quad \delta_{a_1a_3},
\qquad
(1,1):\quad
(\boldsymbol{\sigma}\cdot\hat q)_{a_1a_3}.
\end{equation}
Thus the one-to-one identification in the charge sector is
\begin{equation}
C_0\ \longleftrightarrow\ (L,S)=(0,0)
\ \longleftrightarrow\ \delta_{a_1a_3},
\end{equation}
\begin{equation}
C_1\ \longleftrightarrow\ (L,S)=(1,1)
\ \longleftrightarrow\ (\boldsymbol{\sigma}\cdot\hat q)_{a_1a_3}.
\end{equation}
Here $C_0$ uses the constant harmonic $Y^0_0$ and the spin scalar
$\delta_{a_1a_3}$; $C_1$ uses the dipole harmonic $Y^1_m\sim \hat q^m$ and
the spin vector $\sigma^m$, coupled to a scalar.  This is exactly the
$(1,1)$ channel found in Eq.~\eqref{eq:scalar_amp_spinhalf}.

Now consider the spatial current $J^i$.  In the $LS$ basis its vector index is
restored only at the end by $\mathbb T^i_{1,M}=\bar e^i_M$, so the relevant
component has total angular momentum $J=1$.  The allowed spin-$\frac{1}{2}$
$LS$ channels are therefore
\begin{equation}
(L,S)=(1,0),\qquad (0,1),\qquad (1,1),\qquad (2,1).
\end{equation}
After the Cartesian vector index is restored with
$\mathbb T^i_{1,M}=\bar e^i_M$, these four channels become
\begin{equation}
(1,0):\quad \hat q^i\delta_{a_1a_3},
\end{equation}
\begin{equation}
(0,1):\quad (\sigma^i)_{a_1a_3},
\end{equation}
\begin{equation}
(1,1):\quad
\bigl(i\boldsymbol{\sigma}\times\hat q\bigr)^i_{a_1a_3},
\end{equation}
\begin{equation}
(2,1):\quad
\bigl(3\hat q^i(\hat q\cdot\boldsymbol{\sigma})-\sigma^i\bigr)_{a_1a_3}.
\end{equation}
The conventional multipoles give the same four-dimensional vector space, but
they reach it in a different order.  Let us spell out each case.

The longitudinal monopole $L_0$ comes from the $k=0$ vector harmonic computed
in Eq.~\eqref{eq:k0_vector_harmonics_explicit},
\begin{equation}
Y^{(\mathrm L)i}_{00}\propto \hat q^i.
\end{equation}
Since $k=0$ carries no spin tensor beyond $S=0$, this is precisely the
$LS$ channel $(L,S)=(1,0)$:
\begin{equation}
L_0\ \longleftrightarrow\ (L,S)=(1,0)
\ \longleftrightarrow\ \hat q^i\delta_{a_1a_3},
\label{eq:L0_spinhalf_direct}
\end{equation}

The magnetic dipole $M_1$ comes from the $k=1$ magnetic vector harmonic.  From
Eq.~\eqref{eq:YM_def_current}, it is the cross product of $\hat q$ with the angular
gradient of the dipole harmonic.  When the spin-$1$ tensor
$\sigma^j$ is restored, the only rank-one antisymmetric coupling is
$i\epsilon^{ijk}\hat q_j\sigma_k$.  This is the $LS$ channel $(L,S)=(1,1)$:
\begin{equation}
M_1\ \longleftrightarrow\ (L,S)=(1,1)
\ \longleftrightarrow\
\bigl(i\boldsymbol{\sigma}\times\hat q\bigr)^i_{a_1a_3},
\label{eq:M1_spinhalf_direct}
\end{equation}

The remaining two conventional dipoles, $E_1$ and $L_1$, are more subtle and
this is where the comparison with the $LS$ basis is most useful.  The
electric and longitudinal $k=1$ vector harmonics are not single $LS$ channels.
They are the two orthogonal combinations of the two symmetric spin-dependent
channels
\begin{equation}
(0,1):\quad \sigma^i,
\qquad
(2,1):\quad
3\hat q^i(\hat q\cdot\boldsymbol{\sigma})-\sigma^i .
\end{equation}
Explicitly,
\begin{equation}
E_1\ \longleftrightarrow\
\sqrt{\frac{2}{3}}\,(0,1)-\sqrt{\frac{1}{3}}\,(2,1)
\ \longleftrightarrow\
\sqrt{\frac{2}{3}}\,\sigma^i
-
\sqrt{\frac{1}{3}}\,
\bigl(3\hat q^i(\hat q\cdot\boldsymbol{\sigma})-\sigma^i\bigr),
\label{eq:E1_spinhalf_direct}
\end{equation}
\begin{equation}
L_1\ \longleftrightarrow\
\sqrt{\frac{1}{3}}\,(0,1)+\sqrt{\frac{2}{3}}\,(2,1)
\ \longleftrightarrow\
\sqrt{\frac{1}{3}}\,\sigma^i
+
\sqrt{\frac{2}{3}}\,
\bigl(3\hat q^i(\hat q\cdot\boldsymbol{\sigma})-\sigma^i\bigr).
\label{eq:L1_spinhalf_direct}
\end{equation}
Therefore the precise statement is the following.  The conventional
$C/L/E/M$ multipoles and the $LS$ coefficients do not label the tensors in the
same order, but they span the same space.  $C_0$, $C_1$, $L_0$, and $M_1$
each select one $LS$ channel, while $E_1$ and $L_1$ are the two fixed
orthogonal linear combinations of $(0,1)$ and $(2,1)$.

\begin{table}[t]
\centering
\small
\renewcommand{\arraystretch}{1.25}
\begin{tabularx}{\textwidth}{>{\centering\arraybackslash}p{2.7cm}>{\centering\arraybackslash}p{2.3cm}>{\centering\arraybackslash}p{3.2cm}>{\raggedright\arraybackslash}X}
\toprule
Conventional multipole
&
Current component
&
$LS$ channel
&
Cartesian structure
\\
\midrule
$C_0$
&
$J^0$
&
$(L,S)=(0,0)$
&
$\delta_{a_1a_3}$
\\
$C_1$
&
$J^0$
&
$(L,S)=(1,1)$
&
$(\boldsymbol{\sigma}\cdot\hat q)_{a_1a_3}$
\\
$L_0$
&
$J^i$
&
$(L,S)=(1,0)$
&
$\hat q^i\,\delta_{a_1a_3}$
\\
$M_1$
&
$J^i$
&
$(L,S)=(1,1)$
&
$\bigl(i\boldsymbol{\sigma}\times\hat q\bigr)^i_{a_1a_3}$
\\
\addlinespace
$E_1$
&
$J^i$
&
$\sqrt{\frac{2}{3}}\,(0,1)-\sqrt{\frac{1}{3}}\,(2,1)$
&
fixed combination of $\sigma^i$ and
$3\hat q^i(\hat q\cdot\boldsymbol{\sigma})-\sigma^i$
\\
$L_1$
&
$J^i$
&
$\sqrt{\frac{1}{3}}\,(0,1)+\sqrt{\frac{2}{3}}\,(2,1)$
&
orthogonal combination of the same two spin-dependent structures
\\
\bottomrule
\end{tabularx}
\caption{Explicit relation between the conventional charge, longitudinal,
electric, and magnetic multipoles and the crossed-channel $LS$ basis for
spin-$\frac{1}{2}$ matrix elements.  The entries are given up to normalization
and phase conventions.}
\label{tab:spinhalf_explicit_multipole_LS_map}
\end{table}

Equivalently, the spin-$\frac{1}{2}$ matrix elements may be written as
\begin{equation}
\mathcal M^0(\mathbf q;a_1,a_3)
=
f_0(q^2)\,\delta_{a_1a_3}
+
g_0(q^2)\,(\boldsymbol{\sigma}\cdot\hat q)_{a_1a_3},
\label{eq:J0_pauli_general_recap}
\end{equation}
and
\begin{align}
\mathcal M^i(\mathbf q;a_1,a_3)
&=
f_1(q^2)\,\hat q^i\,\delta_{a_1a_3}
+g_1(q^2)\,(\sigma^i)_{a_1a_3}
+h_1(q^2)\,\bigl(i(\boldsymbol{\sigma}\times \hat q)^i\bigr)_{a_1a_3}
\notag\\
&\quad
+k_1(q^2)\,\bigl(3\hat q^i(\hat q\cdot\boldsymbol{\sigma})-\sigma^i\bigr)_{a_1a_3}.
\label{eq:Jvec_pauli_general_recap}
\end{align}
The charge multipoles $C_0$ and $C_1$ span the two-dimensional scalar space in
Eq.~\eqref{eq:J0_pauli_general_recap}.  Likewise, $L_0$, $L_1$, $E_1$, and $M_1$
span the four-dimensional vector space in
Eq.~\eqref{eq:Jvec_pauli_general_recap}.

\paragraph{General result}

The spin-$\frac12$ example shows the essential relation between the traditional electromagnetic multipoles and the crossed-channel $LS$ basis. The two bases do not use the same coupling order, but they span the same irreducible $\mathrm{SO}(3)$ tensor space. This statement extends directly to arbitrary external spin-$j$.

For the spin $j$, the crossed-channel total spin can take
\begin{equation}
S=0,1,\ldots,2j .
\end{equation}
Thus the conventional multipole rank-$k$ is identified with the crossed-channel spin rank,
\begin{equation}
S=k,
\qquad
0\le k\le 2j .
\end{equation}
The charge multipole is a scalar component of the current. Therefore the orbital part and the spin part must be coupled to total angular momentum zero. This fixes
\begin{equation}
C_k
\quad\longleftrightarrow\quad
(L,S)=(k,k).
\label{eq:Ck_general_LS_result}
\end{equation}

For the spatial current, the Cartesian vector index is treated differently in the two bases. In the conventional basis, the orbital harmonic is first combined with the vector index to form the longitudinal, electric, and magnetic vector harmonics. In the $LS$ basis, the orbital harmonic is first coupled with the crossed-channel spin tensor, and the Cartesian vector index is restored only at the end. Therefore the difference between the two descriptions is only a recoupling of the same $\mathrm{SO}(3)$ tensors.

The magnetic multipole selects the middle orbital branch and corresponds to one $LS$ channel,
\begin{equation}
M_k
\quad\longleftrightarrow\quad
(L,S)=(k,k).
\label{eq:Mk_general_LS_result}
\end{equation}
By contrast, the longitudinal and electric multipoles do not correspond to single $LS$ channels. For each $k\ge1$, they span the two-dimensional space generated by
\begin{equation}
(L,S)=(k-1,k),
\qquad
(L,S)=(k+1,k).
\end{equation}
More explicitly, up to the same phase and normalization conventions used above,
\begin{equation}
L_k
\quad\longleftrightarrow\quad
\sqrt{\frac{k}{2k+1}}\,(k-1,k)
+
\sqrt{\frac{k+1}{2k+1}}\,(k+1,k),
\label{eq:Lk_general_LS_result}
\end{equation}
and
\begin{equation}
E_k
\quad\longleftrightarrow\quad
\sqrt{\frac{k+1}{2k+1}}\,(k-1,k)
-
\sqrt{\frac{k}{2k+1}}\,(k+1,k).
\label{eq:Ek_general_LS_result}
\end{equation}
The relative signs depend on the phase convention for the vector spherical harmonics, but the invariant statement is that $L_k$ and $E_k$ are two orthogonal linear combinations of the same two $LS$ channels.

For $k=0$, the branch $(k-1,k)$ is absent. Therefore the spatial current has only the longitudinal monopole,
\begin{equation}
L_0
\quad\longleftrightarrow\quad
(L,S)=(1,0),
\label{eq:L0_general_LS_result}
\end{equation}
while $E_0$ and $M_0$ are absent.

The general correspondence is summarized in Table~\ref{tab:general_multipole_LS_map}.

\begin{table}[H]
\centering
\small
\renewcommand{\arraystretch}{1.25}
\begin{tabularx}{\textwidth}
{
>{\centering\arraybackslash}p{2.3cm}
>{\centering\arraybackslash}p{2.0cm}
>{\raggedright\arraybackslash}X
>{\raggedright\arraybackslash}X
}
\toprule
Conventional multipole
&
Current component
&
Conventional structure
&
Crossed-channel $LS$ content
\\
\midrule
$C_k$
&
$J^0$
&
charge multipole of rank $k$
&
single scalar channel $(L,S)=(k,k)$
\\
$L_0$
&
$J^i$
&
longitudinal monopole
&
single vector channel $(L,S)=(1,0)$
\\
$M_k$
&
$J^i$
&
magnetic multipole of rank $k$
&
single vector channel $(L,S)=(k,k)$
\\
$L_k$
&
$J^i$
&
longitudinal multipole of rank $k$
&
linear combination of $(k-1,k)$ and $(k+1,k)$
\\
$E_k$
&
$J^i$
&
electric multipole of rank $k$
&
orthogonal linear combination of $(k-1,k)$ and $(k+1,k)$
\\
\bottomrule
\end{tabularx}
\caption{Relation between the conventional electromagnetic multipoles and the
crossed-channel $LS$ basis for arbitrary external spin $j$. The allowed
multipole ranks satisfy $0\le k\le 2j$. For $k=0$, only $C_0$ and $L_0$ are
present.}
\label{tab:general_multipole_LS_map}
\end{table}

Therefore, the spin-$\frac12$ result is not a special feature of Pauli
matrices. It is the lowest-spin realization of a general recoupling relation.
The conventional electromagnetic multipoles classify the current by charge,
longitudinal, electric, and magnetic components, while the $LS$ basis
classifies the same structures by orbital angular momentum and
crossed-channel total spin.

\subsubsection{\texorpdfstring{Spin-$\frac{1}{2}$}{Spin-one-half to one-half} matrix elements}

The general $LS$ construction is now illustrated for spin $j=\frac{1}{2}$ particles. In this case, the crossed-channel total spin can only take the values $S=0,1$. The corresponding spin tensors follow from Eq.~\eqref{eq:spin_tensor_cg_relation}. For spin $j=\frac{1}{2}$, the spherical spin components reduce to
\begin{equation}
\Sigma^{(0)}_0=\delta_{a_1a_3},\qquad\Sigma^{(1)}_m=(\sigma_m)_{a_1a_3}.
\end{equation}

\paragraph{Scalar component}
For a scalar operator component, $J=0$ and
\begin{equation}
\mathcal M_{(0)}(\mathbf{q};a_1,a_3)=\langle p_1;\frac{1}{2}a_1|\mathcal O_{(0)}|p_3;\frac{1}{2}a_3\rangle.
\label{eq:scalar_me_def}
\end{equation}
The condition $L\otimes S\to0$ requires $L=S$, so the allowed channels are
\begin{equation}
(L,S)=(0,0),\qquad (1,1).
\end{equation}
Substituting these two channels into the general $LS$ formula Eq.~\eqref{eq:threepoint_amp} gives
\begin{equation}
\mathcal A^{0,0}(\mathbf q;a_1,a_3)=F^{(0)}_{00}(q^2)\,Y^0_0(\hat q)\,\Sigma^{(0)}_0(a_1,a_3)+F^{(0)}_{11}(q^2)\sum_{m,a_S}C^{0,0}_{1,m;1,a_S}Y^1_m(\hat q)\,\Sigma^{(1)}_{a_S}(a_1,a_3).
\label{eq:scalar_amp_spinhalf}
\end{equation}
For a scalar component, Eq.~\eqref{eq:tensor_reconstruction} has $\mathbb T_{0,0}=1$, so the matrix element is obtained directly from $\mathcal A^{0,0}$.  The first channel is immediate: 
\begin{equation}
Y^0_0\,\Sigma^{(0)}_0=\frac{1}{\sqrt{4\pi}}\delta_{a_1a_3}.
\end{equation}
Here the $S=0$ spin tensor is the spin-$\frac{1}{2}$ specialization of Eq.~\eqref{eq:spin_tensor_cg_relation}. For the second channel, the CGC coupling $1\otimes1\to0$ is
\begin{equation}
C^{0,0}_{1,m;1,a_S}=\frac{(-1)^{1-m}}{\sqrt3}\,\delta_{m,-a_S}.
\end{equation}
Therefore
\begin{align}
\sum_{m,a_S}C^{0,0}_{1,m;1,a_S}Y^1_m(\hat q)\,\Sigma^{(1)}_{a_S}&=\frac{1}{\sqrt3}\sum_{m=-1}^{1}(-1)^{1-m}Y^1_m(\hat q)\,\Sigma^{(1)}_{-m}\notag\\
&=-\frac{1}{\sqrt{4\pi}}\,(\boldsymbol{\sigma}\cdot\hat q)_{a_1a_3},
\end{align}
where in the last step we used the rank-one case of Eq.~\eqref{eq:spherical_to_cartesian_symmetric_traceless_general} for the orbital part and
Eq.~\eqref{eq:spin_tensor_cg_relation} for the rank-one spin part.  Hence the scalar component has the form
\begin{equation}
\mathcal M_{(0)}(\mathbf{q};a_1,a_3)
=
f_{(0)}(q^2)\,\delta_{a_1a_3}
+
g_{(0)}(q^2)\,(\boldsymbol{\sigma}\cdot\hat q)_{a_1a_3}.
\label{eq:scalar_pauli_final}
\end{equation}
The exact coefficient relation is therefore
\begin{equation}
f_{(0)}(q^2)=\frac{1}{\sqrt{4\pi}}F^{(0)}_{00}(q^2),
\qquad
g_{(0)}(q^2)=-\frac{1}{\sqrt{4\pi}}F^{(0)}_{11}(q^2).
\label{eq:spinhalf_scalar_F_to_f}
\end{equation}
If parity is imposed, the first structure is parity even, while the second is parity odd.

\paragraph{Vector component}

For a vector operator component, $J=1$ and
\begin{equation}
\mathcal M^i_{(1)}(\mathbf q;a_1,a_3)=\langle p_1;\frac{1}{2}a_1|\mathcal O^i_{(1)}|p_3;\frac{1}{2}a_3\rangle.
\label{eq:vector_me_def}
\end{equation}
The condition $L\otimes S\to1$ in Eq.~\eqref{eq:threepoint_amp} gives four allowed channels,
\begin{equation}
(L,S)=(1,0),\qquad (0,1),\qquad (1,1),\qquad (2,1).
\end{equation}
Keeping only these channels, Eq.~\eqref{eq:threepoint_amp} becomes
\begin{equation}
\mathcal A^{1,M}(\mathbf q;a_1,a_3)=\sum_{(L,S)}F^{(1)}_{LS}(q^2)\sum_{m,a_S}C^{1,M}_{L,m;S,a_S}Y^L_m(\hat q)\,\Sigma^{(S)}_{a_S}(a_1,a_3),
\label{eq:vector_amp_spinhalf}
\end{equation}
where the sum runs over the four channels above. The Cartesian vector is obtained as the special case of Eq.~\eqref{eq:tensor_reconstruction},
\begin{equation}
\mathcal M^i_{(1)}(\mathbf q;a_1,a_3)=\sum_{M=-1}^{1}\bar e^i_M\,\mathcal A^{1,M}(\mathbf q;a_1,a_3),
\label{eq:vector_spherical_to_cartesian}
\end{equation}
with $\bar e^i_M$ defined in Eq.~\eqref{eq:dual_spherical_basis_vector}. The four channels are now reduced separately.

For $(L,S)=(1,0)$, the spin tensor is the scalar tensor in Eq.~\eqref{eq:spin_tensor_cg_relation}, and $C^{1,M}_{1,m;0,0}=\delta_{Mm}$. Using Eq.~\eqref{eq:vector_spherical_to_cartesian} to attach the vector basis $\bar e^i_M$ gives
\begin{equation}
\sum_{M,m}\bar e^i_M C^{1,M}_{1,m;0,0}Y^1_m(\hat q)\Sigma^{(0)}_0\ =\sqrt{\frac{3}{4\pi}}\,\hat q^i\,\delta_{a_1a_3},
\end{equation}
where the last equality uses Eq.~\eqref{eq:spherical_to_cartesian_symmetric_traceless_general} at $L=1$. For $(L,S)=(0,1)$, the orbital part is a scalar and $C^{1,M}_{0,0;1,a_S}=\delta_{M a_S}$. Therefore,
\begin{equation}
\sum_{M,a_S}\bar e^i_M C^{1,M}_{0,0;1,a_S}Y^0_0(\hat q)\Sigma^{(1)}_{a_S}\ =\frac{1}{\sqrt{4\pi}}\,(\sigma^i)_{a_1a_3},
\end{equation}
where the spherical components of $\Sigma^{(1)}$, defined in Eq.~\eqref{eq:spin_tensor_cg_relation}, reconstruct the Cartesian Pauli matrix after contraction with the dual basis vector in Eq.~\eqref{eq:dual_spherical_basis_vector}.

For $(L,S)=(1,1)$, the coupling of two vectors to total $J=1$ gives the antisymmetric vector product. Applying the CGC coupling in Eq.~\eqref{eq:threepoint_amp} and then using Eq.~\eqref{eq:vector_spherical_to_cartesian} gives
\begin{align}
    \sum_{M,m,a_S}\bar e^i_M C^{1,M}_{1,m;1,a_S}Y^1_m(\hat q)\Sigma^{(1)}_{a_S}
\ &=-\sqrt{\frac{3}{8\pi}}\,i\epsilon^{ijk}\hat q_j(\sigma_k)_{a_1a_3}\\
&=\sqrt{\frac{3}{8\pi}}i(\boldsymbol\sigma\times\hat q)^i.
\end{align}

For $(L,S)=(2,1)$, the $L=2$ harmonic is represented by the symmetric traceless tensor $Q^{ij}=\hat q^i\hat q^j-\frac{1}{3}\delta^{ij}$ defined in Eq.~\eqref{eq:rank2_symmetric_traceless_Qij}, and its spherical components are obtained from the rank-two basis in Eq.~\eqref{eq:rank2_spherical_tensor_basis}. Coupling this rank-two tensor with the rank-one spin tensor to total $J=1$ gives the vector obtained by one contraction,
\begin{equation}
\sum_{M,m,a_S}\bar e^i_M C^{1,M}_{2,m;1,a_S}Y^2_m(\hat q)\Sigma^{(1)}_{a_S}\ =-\frac{1}{\sqrt{12\pi}}\,\bigl(3\hat q^i(\hat q\cdot\boldsymbol{\sigma})-\sigma^i\bigr)_{a_1a_3}.
\end{equation}
This is the same structure as the one-contraction form, since
\begin{equation}
\left(\hat q^i\hat q^j-\frac{1}{3}\delta^{ij}\right)\sigma_j=\frac{1}{3}\left[3\hat q^i(\hat q\cdot\boldsymbol{\sigma})-\sigma^i\right].
\end{equation}
Therefore, the vector component matrix element can be written as
\begin{equation}
\mathcal M^i_{(1)}(\mathbf q;a_1,a_3)=f_{(1)}(q^2)\,\hat q^i\,\delta_{a_1a_3}+g_{(1)}(q^2)\,(\sigma^i)_{a_1a_3}+h_{(1)}(q^2)\,\bigl(i(\boldsymbol{\sigma}\times\hat q)^i\bigr)_{a_1a_3}+k_{(1)}(q^2)\,\bigl(3\hat q^i(\hat q\cdot\boldsymbol{\sigma})-\sigma^i\bigr)_{a_1a_3},
\label{eq:vector_tensor_basis}
\end{equation}
where
\begin{equation}
f_{(1)}(q^2)=\sqrt{\frac{3}{4\pi}}F^{(1)}_{10}(q^2),\qquad
g_{(1)}(q^2)=\frac{1}{\sqrt{4\pi}}F^{(1)}_{01}(q^2),
\end{equation}
\begin{equation}
h_{(1)}(q^2)=\sqrt{\frac{3}{8\pi}}F^{(1)}_{11}(q^2),\qquad
k_{(1)}(q^2)=-\frac{1}{\sqrt{12\pi}}F^{(1)}_{21}(q^2).
\label{eq:spinhalf_vector_F_to_f}
\end{equation}
The four terms in Eq.~\eqref{eq:vector_tensor_basis} correspond respectively to the channels $(1,0)$, $(0,1)$, $(1,1)$, and $(2,1)$.

\paragraph{Tensor component}
We finally consider a general rank-two Cartesian tensor component,
\begin{equation}
\mathcal M^{ij}(\mathbf q;a_1,a_3)=\langle p_1;\frac{1}{2}a_1|\mathcal O^{ij}|p_3;\frac{1}{2}a_3\rangle .
\label{eq:tensor_me_def_spinhalf}
\end{equation}
A general rank-two spatial tensor is reducible under rotations and decomposes as
\begin{equation}
1\otimes1=0\oplus1\oplus2.
\end{equation}
In Cartesian form, this decomposition can be written as
\begin{equation}
\mathcal M^{ij}=\frac{1}{3}\delta^{ij}\mathcal M_{(0)}+\epsilon^{ijk}\mathcal M^k_{(1)}+\mathcal M^{ij}_{(2)}.
\label{eq:rank2_cartesian_decomposition}
\end{equation}
The three irreducible components are
\begin{equation}
\mathcal M_{(0)}=\delta_{ij}\mathcal M^{ij},
\qquad
\mathcal M^k_{(1)}=\frac{1}{2}\epsilon^{kij}\mathcal M^{ij},
\qquad
\mathcal M^{ij}_{(2)}=\frac{1}{2}(\mathcal M^{ij}+\mathcal M^{ji})
-\frac{1}{3}\delta^{ij}\mathcal M_{(0)}.
\label{eq:rank2_irreducible_parts}
\end{equation}
Thus, the trace gives the $J=0$ part, the antisymmetric part gives the $J=1$ part, and the symmetric traceless part gives the $J=2$ part. These three components are evaluated below in the $LS$ basis.

The $J=0$ trace part is exactly the scalar component already computed in Eq.~\eqref{eq:scalar_pauli_final},
\begin{equation}
\mathcal M_{(0)}=f_{(0)}(q^2)\,\delta_{a_1a_3}+g_{(0)}(q^2)(\boldsymbol{\sigma}\cdot\hat q)_{a_1a_3}.
\label{eq:tensor_trace_part_spinhalf}
\end{equation}
According to Eq.~\eqref{eq:rank2_cartesian_decomposition}, its contribution to the original tensor is therefore
\begin{equation}
\mathcal M^{ij}_{J=0}
=
\frac{1}{3}\delta^{ij}
\left[
f_{(0)}(q^2)\,\delta_{a_1a_3}
+
g_{(0)}(q^2)(\boldsymbol{\sigma}\cdot\hat q)_{a_1a_3}
\right].
\label{eq:tensor_J0_spinhalf}
\end{equation}

The $J=1$ part is the antisymmetric component. From Eq.~\eqref{eq:rank2_irreducible_parts}, its Cartesian tensor contribution is $\epsilon^{ijk}\mathcal M^k_{(1)}$. Using the vector result in Eq.~\eqref{eq:vector_tensor_basis} gives
\begin{align}
\mathcal M^{ij}_{J=1}
&=
\epsilon^{ijk}
\Big[
f_{(1)}(q^2)\hat q^k\,\delta_{a_1a_3}
+
g_{(1)}(q^2)(\sigma^k)_{a_1a_3}
\notag\\
&\qquad\qquad
+
h_{(1)}(q^2)\bigl(i(\boldsymbol{\sigma}\times\hat q)^k\bigr)_{a_1a_3}
+
k_{(1)}(q^2)
\bigl(3\hat q^k(\hat q\cdot\boldsymbol{\sigma})-\sigma^k\bigr)_{a_1a_3}
\Big].
\label{eq:tensor_J1_spinhalf}
\end{align}
This term is antisymmetric in $i,j$, as required.

It remains to compute the $J=2$ symmetric traceless part. With $S=0,1$, the condition $L\otimes S\to2$ in Eq.~\eqref{eq:threepoint_amp} gives
\begin{equation}
(L,S)=(2,0),\qquad (1,1),\qquad (2,1),\qquad (3,1).
\end{equation}
Therefore,
\begin{equation}
\mathcal A^{2,M}(\mathbf q;a_1,a_3)
=
\sum_{(L,S)}
F^{(2)}_{LS}(q^2)
\sum_{m,a_S}
C^{2,M}_{L,m;S,a_S}
Y^L_m(\hat q)\,
\Sigma^{(S)}_{a_S}(a_1,a_3),
\label{eq:tensor_amp_spinhalf}
\end{equation}
where the sum runs over these four channels. The Cartesian symmetric traceless tensor is obtained from Eq.~\eqref{eq:tensor_reconstruction} using the dual rank-two basis in Eq.~\eqref{eq:dual_spherical_tensor_basis},
\begin{equation}
\mathcal M^{ij}_{(2)}(\mathbf q;a_1,a_3)
=
\sum_{M=-2}^{2}
\bar e^{ij}_{2M}\,
\mathcal A^{2,M}(\mathbf q;a_1,a_3).
\label{eq:tensor_spherical_to_cartesian_spinhalf}
\end{equation}

For $(L,S)=(2,0)$, the spin tensor is the scalar tensor in Eq.~\eqref{eq:spin_tensor_cg_relation}, and the CGC reduces to
\begin{equation}
C^{2,M}_{2,m;0,0}=\delta_{M m}.
\end{equation}
Thus Eq.~\eqref{eq:tensor_spherical_to_cartesian_spinhalf} gives
\begin{align}
\sum_{M,m}
\bar e^{ij}_{2M}C^{2,M}_{2,m;0,0}
Y^2_m(\hat q)\Sigma^{(0)}_0
&=
\sum_{M}
\bar e^{ij}_{2M}Y^2_M(\hat q)\Sigma^{(0)}_0
\notag\\
&=
\sqrt{\frac{15}{8\pi}}\,
Q^{ij}(\hat q)\delta_{a_1a_3},
\end{align}
where the last equality uses Eqs.~\eqref{eq:rank2_spherical_tensor_basis} and \eqref{eq:rank2_symmetric_traceless_Qij}. Hence
\begin{equation}
(2,0):\qquad
\big[Y^{(2)}\otimes\Sigma^{(0)}\big]^{(2)ij}
\ =
\sqrt{\frac{15}{8\pi}}\,
\left(\hat q^i\hat q^j-\frac{1}{3}\delta^{ij}\right)\delta_{a_1a_3}.
\end{equation}

For $(L,S)=(1,1)$, the required CGCs are those for $1\otimes1\to2$. They are
\begin{equation}
C^{2,M}_{1,m;1,a}
=
\begin{cases}
\sqrt{\frac{(1+M)(2+M)}{12}}, & a=1,\quad m=M-1,\\[6pt]
\sqrt{\frac{(2-M)(2+M)}{6}}, & a=0,\quad m=M,\\[6pt]
\sqrt{\frac{(1-M)(2-M)}{12}}, & a=-1,\quad m=M+1,\\[6pt]
0, & \text{otherwise}.
\end{cases}
\label{eq:CG_11_to_2}
\end{equation}
Substitution of Eq.~\eqref{eq:CG_11_to_2} into Eq.~\eqref{eq:tensor_spherical_to_cartesian_spinhalf}, together with the rank-one relations in Eqs.~\eqref{eq:spherical_to_cartesian_symmetric_traceless_general} and \eqref{eq:spin_tensor_cg_relation}, gives the standard symmetric traceless product of two Cartesian vectors,
\begin{align}
\sum_{M,m,a}
\bar e^{ij}_{2M}C^{2,M}_{1,m;1,a}
Y^1_m(\hat q)\Sigma^{(1)}_{a}
&=
\sqrt{\frac{3}{4\pi}}\,
\left[
\frac{1}{2}\left(\hat q^i\sigma^j+\hat q^j\sigma^i\right)
-\frac{1}{3}\delta^{ij}(\hat q\cdot\boldsymbol{\sigma})
\right]_{a_1a_3}.
\end{align}
Therefore,
\begin{equation}
(1,1):\qquad
\big[Y^{(1)}\otimes\Sigma^{(1)}\big]^{(2)ij}
\ =
\sqrt{\frac{3}{4\pi}}\,
\left[
\frac{1}{2}\left(\hat q^i\sigma^j+\hat q^j\sigma^i\right)
-\frac{1}{3}\delta^{ij}(\hat q\cdot\boldsymbol{\sigma})
\right]_{a_1a_3}.
\label{eq:tensor_11_spinhalf_structure}
\end{equation}

For $(L,S)=(2,1)$, the orbital tensor is the quadrupole $Q^{ij}$ in Eq.~\eqref{eq:rank2_symmetric_traceless_Qij}. The CGCs for $2\otimes1\to2$ can be written as
\begin{equation}
C^{2,M}_{2,m;1,a}
=
\begin{cases}
-\sqrt{\frac{(3-M)(2+M)}{12}}, & a=1,\quad m=M-1,\\[6pt]
\frac{M}{\sqrt6}, & a=0,\quad m=M,\\[6pt]
\sqrt{\frac{(3+M)(2-M)}{12}}, & a=-1,\quad m=M+1,\\[6pt]
0, & \text{otherwise}.
\end{cases}
\label{eq:CG_21_to_2}
\end{equation}
These antisymmetric coefficients project the product of the rank-two orbital tensor and the spin vector onto
\begin{align}
\sum_{M,m,a}
\bar e^{ij}_{2M}C^{2,M}_{2,m;1,a}
Y^2_m(\hat q)\Sigma^{(1)}_{a}
&=
-\sqrt{\frac{5}{16\pi}}\,
i\left[
\epsilon^{ikl}Q^{jk}(\hat q)\sigma^l
+\epsilon^{jkl}Q^{ik}(\hat q)\sigma^l
\right]_{a_1a_3}
\notag\\
&=
-\sqrt{\frac{5}{16\pi}}\,
i\left[
\hat q^i(\boldsymbol{\sigma}\times\hat q)^j
+
\hat q^j(\boldsymbol{\sigma}\times\hat q)^i
\right]_{a_1a_3},
\end{align}
where the $\frac{1}{3}\delta^{ij}$ part of $Q^{ij}$ drops out after the contraction with $\epsilon^{ijk}$. Thus,
\begin{equation}
(2,1):\qquad
\big[Y^{(2)}\otimes\Sigma^{(1)}\big]^{(2)ij}
\ =
-\sqrt{\frac{5}{16\pi}}\,
i\left[
\hat q^i(\boldsymbol{\sigma}\times\hat q)^j
+
\hat q^j(\boldsymbol{\sigma}\times\hat q)^i
\right]_{a_1a_3}.
\label{eq:tensor_21_spinhalf_structure}
\end{equation}
This tensor is symmetric, and it is traceless because $\hat q\cdot(\boldsymbol{\sigma}\times\hat q)=0$.

For $(L,S)=(3,1)$, the orbital part is the rank-three symmetric traceless tensor $\hat q^{\langle i}\hat q^j\hat q^{k\rangle}$, obtained from the general formula in Eq.~\eqref{eq:spherical_to_cartesian_symmetric_traceless_general}. The CGCs for $3\otimes1\to2$ are
\begin{equation}
C^{2,M}_{3,m;1,a}
=
\begin{cases}
\sqrt{\frac{(4-M)(3-M)}{42}}, & a=1,\quad m=M-1,\\[6pt]
-\sqrt{\frac{(3-M)(3+M)}{21}}, & a=0,\quad m=M,\\[6pt]
\sqrt{\frac{(4+M)(3+M)}{42}}, & a=-1,\quad m=M+1,\\[6pt]
0, & \text{otherwise}.
\end{cases}
\label{eq:CG_31_to_2}
\end{equation}
Substitution into Eq.~\eqref{eq:tensor_spherical_to_cartesian_spinhalf} selects the unique contraction of the rank-three symmetric traceless orbital tensor with the spin vector:
\begin{align}
\sum_{M,m,a}
\bar e^{ij}_{2M}C^{2,M}_{3,m;1,a}
Y^3_m(\hat q)\Sigma^{(1)}_{a}
&=
-\frac{5}{\sqrt{8\pi}}\,
\hat q^{\langle i}\hat q^j\hat q^{k\rangle}
(\sigma_k)_{a_1a_3}.
\end{align}
Therefore,
\begin{equation}
(3,1):\qquad
\big[Y^{(3)}\otimes\Sigma^{(1)}\big]^{(2)ij}
\ =
-\frac{5}{\sqrt{8\pi}}\,
\hat q^{\langle i}\hat q^j\hat q^{k\rangle}(\sigma_k)_{a_1a_3}.
\label{eq:tensor_31_spinhalf_symmetric_traceless}
\end{equation}
Equivalently,
\begin{equation}
\hat q^{\langle i}\hat q^j\hat q^{k\rangle}\sigma_k
=
\hat q^i\hat q^j(\hat q\cdot\boldsymbol{\sigma})
-\frac15\left[
\hat q^i\sigma^j+\hat q^j\sigma^i
+\delta^{ij}(\hat q\cdot\boldsymbol{\sigma})
\right],
\label{eq:tensor_31_spinhalf_explicit}
\end{equation}
which is symmetric and traceless in $i,j$.

The $J=2$ part is therefore
\begin{align}
\mathcal M^{ij}_{(2)}(\mathbf q;a_1,a_3)
&=
f_{(2)}(q^2)
\left(\hat q^i\hat q^j-\frac{1}{3}\delta^{ij}\right)\delta_{a_1a_3}
\notag\\
&\quad
+
g_{(2)}(q^2)
\left[
\frac{1}{2}\left(\hat q^i\sigma^j+\hat q^j\sigma^i\right)
-\frac{1}{3}\delta^{ij}(\hat q\cdot\boldsymbol{\sigma})
\right]_{a_1a_3}
\notag\\
&\quad
+
h_{(2)}(q^2)
\left[
i\hat q^i(\boldsymbol{\sigma}\times\hat q)^j
+i\hat q^j(\boldsymbol{\sigma}\times\hat q)^i
\right]_{a_1a_3}
\notag\\
&\quad
+
k_{(2)}(q^2)
\left[
\hat q^i\hat q^j(\hat q\cdot\boldsymbol{\sigma})
-\frac15\left(
\hat q^i\sigma^j+\hat q^j\sigma^i
+\delta^{ij}(\hat q\cdot\boldsymbol{\sigma})
\right)
\right]_{a_1a_3}.
\label{eq:tensor_symmetric_traceless_spinhalf_final}
\end{align}
The four terms correspond respectively to the $LS$ channels $(2,0)$, $(1,1)$, $(2,1)$, and $(3,1)$. The corresponding exact coefficient map is
\begin{equation}
f_{(2)}(q^2)=\sqrt{\frac{15}{8\pi}}F^{(2)}_{20}(q^2), \qquad g_{(2)}(q^2)=\sqrt{\frac{3}{4\pi}}F^{(2)}_{11}(q^2),
\end{equation}
\begin{equation}
h_{(2)}(q^2)=-\sqrt{\frac{5}{16\pi}}F^{(2)}_{21}(q^2),
\qquad
k_{(2)}(q^2)=-\frac{5}{\sqrt{8\pi}}F^{(2)}_{31}(q^2).
\label{eq:spinhalf_tensor_F_to_f}
\end{equation}

Combining Eqs.~\eqref{eq:tensor_J0_spinhalf},
\eqref{eq:tensor_J1_spinhalf}, and \eqref{eq:tensor_symmetric_traceless_spinhalf_final}, the most general rank-two spin-$\frac{1}{2}$ tensor matrix element is
\begin{equation}
\mathcal M^{ij}=\mathcal M^{ij}_{J=0}+\mathcal M^{ij}_{J=1}+\mathcal M^{ij}_{(2)}.
\label{eq:tensor_spinhalf_final}
\end{equation}

\subsubsection{\texorpdfstring{Spin-$1$}{Spin-1 to 1} matrix elements}

Matrix elements between spin-$1$ states are considered next. The construction is parallel to the spin $j=\frac{1}{2}$ case, while the crossed-channel spin space contains one additional irreducible sector. In the crossed channel,
\begin{equation}
1\otimes1=0\oplus1\oplus2 .
\end{equation}
The label $S$ denotes the total spin obtained by coupling the final spin-$1$ state to the crossed initial spin-$1$ state,
\begin{equation}
(-1)^{1-a_3}C^{S,a_S}_{1,a_1;1,-a_3}.
\end{equation}
The three possibilities have the direct interpretation
\begin{equation}
S=0:\ \text{Monopole},\qquad
S=1:\ \text{Dipole},\qquad
S=2:\ \text{Quadrupole}.
\end{equation}
The the rotation generators in the spin-$1$ representation are denoted by $\mathbf S$, with matrix elements $(S^i)_{a_1a_3}$ in the canonical spin basis $a_1,a_3=-1,0,1$. The explicit matrices can be obtained from Eq.~\eqref{Eq:spin1gener}.

\paragraph{Scalar component}
For the scalar component $\mathcal O_{(0)}$, the matrix element is
\begin{equation}
\mathcal M_{(0)}(\mathbf q;a_1,a_3)=\langle p_1;1a_1|\mathcal O_{(0)}|p_3;1a_3\rangle .
\label{eq:scalar_me_def_spinone}
\end{equation}
Since this operator component is a rotational scalar, the total angular momentum is $J=0$. The crossed-channel amplitude is
\begin{equation}
\mathcal A^{0,0}(\mathbf q;a_1,a_3)=\sum_{L,S}F^{(0)}_{LS}(q^2)\sum_{m,a_S}C^{0,0}_{L,m;S,a_S}Y^L_m(\hat q)\Sigma^{(S)}_{a_S}(a_1,a_3).
\label{eq:scalar_amp_general_spinone}
\end{equation}
For spin $j=1$, the crossed-channel spin can take the values
\begin{equation}
S=0,\ 1,\ 2.
\end{equation}
The scalar condition $L\otimes S\to0$ again requires $L=S$, so that the allowed channels are
\begin{equation}
(L,S)=(0,0),\qquad (1,1),\qquad (2,2).
\end{equation}
The amplitude therefore becomes
\begin{align}
\mathcal A^{0,0}(\mathbf q;a_1,a_3)&=F^{(0)}_{00}(q^2)\,C^{0,0}_{0,0;0,0}Y^0_0(\hat q)\Sigma^{(0)}_0(a_1,a_3)\notag\\
&\quad+F^{(0)}_{11}(q^2)\sum_{m,a_S}C^{0,0}_{1,m;1,a_S}Y^1_m(\hat q)\Sigma^{(1)}_{a_S}(a_1,a_3)\notag\\
&\quad+F^{(0)}_{22}(q^2)\sum_{m,a_S}C^{0,0}_{2,m;2,a_S}Y^2_m(\hat q)\Sigma^{(2)}_{a_S}(a_1,a_3).
\label{eq:scalar_amp_spinone}
\end{align}
The three channels reduce to
\begin{equation}
(0,0):\qquad Y^0_0\Sigma^{(0)}_0=\frac{1}{\sqrt{4\pi}}\delta_{a_1a_3},
\end{equation}
\begin{equation}
(1,1):\qquad\sum_{m,a_S}C^{0,0}_{1,m;1,a_S}Y^1_m\Sigma^{(1)}_{a_S}=-\frac{1}{\sqrt{4\pi}}(\mathbf S\cdot\hat q)_{a_1a_3},
\end{equation}
and
\begin{equation}
(2,2):\qquad\sum_{m,a_S}C^{0,0}_{2,m;2,a_S}Y^2_m\Sigma^{(2)}_{a_S}=\sqrt{\frac{3}{8\pi}}\,\hat q_i\hat q_j Q^{ij}_{a_1a_3}.
\end{equation}
After the spherical tensors are converted to Cartesian form, the scalar matrix element can be written as
\begin{equation}
\mathcal M_{(0)}(\mathbf q;a_1,a_3)=f_{(0)}(q^2)\,\delta_{a_1a_3}+g_{(0)}(q^2)\,(\mathbf S\cdot\hat q)_{a_1a_3}+h_{(0)}(q^2)\,\hat q_i\hat q_j Q^{ij}_{a_1a_3},
\label{eq:scalar_spinone_final}
\end{equation}
where
\begin{equation}
f_{(0)}(q^2)=\frac{1}{\sqrt{4\pi}}F^{(0)}_{00}(q^2),\qquad g_{(0)}(q^2)=-\frac{1}{\sqrt{4\pi}}F^{(0)}_{11}(q^2),\qquad h_{(0)}(q^2)=\sqrt{\frac{3}{8\pi}}F^{(0)}_{22}(q^2).
\label{eq:spinone_scalar_F_to_f}
\end{equation}

\paragraph{Vector component}
For a vector component $\mathcal O^i_{(1)}$, the matrix element is
\begin{equation}
\mathcal M^i_{(1)}(\mathbf q;a_1,a_3)=\langle p_1;1a_1|\mathcal O^i_{(1)}|p_3;1a_3\rangle .
\label{eq:vector_me_def_spinone}
\end{equation}
Since $\mathcal O^i_{(1)}$ transforms as a vector under rotations, the total
angular momentum is $J=1$.  The corresponding crossed-channel amplitude is
\begin{equation}
\mathcal A^{1,M}(\mathbf q;a_1,a_3)=\sum_{L,S}F^{(1)}_{LS}(q^2)\sum_{m,a_S}C^{1,M}_{L,m;S,a_S}Y^L_m(\hat q)\Sigma^{(S)}_{a_S}(a_1,a_3).
\label{eq:vector_amp_general_spinone}
\end{equation}
For spin-$1$, $S=0,1,2$.  The condition
\begin{equation}
L\otimes S\to1
\end{equation}
then gives the seven allowed channels
\begin{equation}
\begin{gathered}
(L,S)=(1,0),\\
(L,S)=(0,1),\qquad (1,1),\qquad (2,1),\\
(L,S)=(1,2),\qquad (2,2),\qquad (3,2).
\end{gathered}
\label{eq:vector_channels_spinone}
\end{equation}
Correspondingly,
\begin{align}
\mathcal A^{1,M}(\mathbf q;a_1,a_3)&=F^{(1)}_{10}\sum_m C^{1,M}_{1,m;0,0}Y^1_m\Sigma^{(0)}_0(a_1,a_3)\notag\\
&\quad+F^{(1)}_{01}\sum_{a_S}C^{1,M}_{0,0;1,a_S}Y^0_0\Sigma^{(1)}_{a_S}(a_1,a_3)\notag\\
&\quad+F^{(1)}_{11}\sum_{m,a_S}C^{1,M}_{1,m;1,a_S}Y^1_m\Sigma^{(1)}_{a_S}(a_1,a_3)\notag\\
&\quad+F^{(1)}_{21}\sum_{m,a_S}C^{1,M}_{2,m;1,a_S}Y^2_m\Sigma^{(1)}_{a_S}(a_1,a_3)\notag\\
&\quad+F^{(1)}_{12}\sum_{m,a_S}C^{1,M}_{1,m;2,a_S}Y^1_m\Sigma^{(2)}_{a_S}(a_1,a_3)\notag\\
&\quad+F^{(1)}_{22}\sum_{m,a_S}C^{1,M}_{2,m;2,a_S}Y^2_m\Sigma^{(2)}_{a_S}(a_1,a_3)\notag\\
&\quad+F^{(1)}_{32}\sum_{m,a_S}C^{1,M}_{3,m;2,a_S}Y^3_m\Sigma^{(2)}_{a_S}(a_1,a_3),
\label{eq:vector_amp_spinone}
\end{align}
where the $q^2$ of the form factors and the $\hat q$ of the spherical harmonics have been suppressed on the right-hand
side.  The Cartesian matrix element is again obtained from
\begin{equation}
\mathcal M^i_{(1)}(\mathbf q;a_1,a_3)=\sum_{M=-1}^{1}\bar e^i_M\,\mathcal A^{1,M}(\mathbf q;a_1,a_3).
\end{equation}
After this spherical-to-Cartesian conversion, using the Cartesian spin tensors
in Eq.~\eqref{Eq:spin1gener}, the
seven channels in Eq.~\eqref{eq:vector_channels_spinone} become
\begin{equation}
(1,0):\qquad\big[Y^{(1)}\otimes\Sigma^{(0)}\big]^{(1)i}\ =\sqrt{\frac{3}{4\pi}}\,\hat q^i\,\delta_{a_1a_3},
\end{equation}
\begin{equation}
(0,1):\qquad\big[Y^{(0)}\otimes\Sigma^{(1)}\big]^{(1)i}\ =\frac{1}{\sqrt{4\pi}}\,(S^i)_{a_1a_3},
\end{equation}
\begin{equation}
(1,1):\qquad\big[Y^{(1)}\otimes\Sigma^{(1)}\big]^{(1)i}\ =\sqrt{\frac{3}{8\pi}}\,i\epsilon^{ijk}\hat q_j(S_k)_{a_1a_3},
\end{equation}
\begin{equation}
(2,1):\qquad\big[Y^{(2)}\otimes\Sigma^{(1)}\big]^{(1)i}\ =-\frac{1}{\sqrt{12\pi}}\,\bigl(3\hat q^i(\hat q\cdot\mathbf S)-S^i\bigr)_{a_1a_3},
\end{equation}
\begin{equation}
(1,2):\qquad\big[Y^{(1)}\otimes\Sigma^{(2)}\big]^{(1)i}\ =-\frac{3}{\sqrt{20\pi}}\,Q^{ij}_{a_1a_3}\hat q_j,
\end{equation}
\begin{equation}
(2,2):\qquad\big[Y^{(2)}\otimes\Sigma^{(2)}\big]^{(1)i}\ =-\sqrt{\frac{3}{4\pi}}\,i\epsilon^{ijk}\hat q_j\hat q_l Q^{kl}_{a_1a_3},
\end{equation}
and
\begin{equation}
(3,2):\qquad\big[Y^{(3)}\otimes\Sigma^{(2)}\big]^{(1)i}\ =\sqrt{\frac{15}{8\pi}}\,\left[\hat q^i\hat q_j\hat q_k-\frac15\left(\delta^i_j\hat q_k+\delta^i_k\hat q_j\right)\right]Q^{jk}_{a_1a_3}.
\end{equation}
In the last line the possible term proportional to
$\hat q^i\delta_{jk}Q^{jk}$ has vanished because $Q^{jk}$ is traceless.
The spin-$1$ vector matrix element is therefore
\begin{align}
\mathcal M^i_{(1)}(\mathbf q;a_1,a_3)
&=f_{(1)}(q^2)\,\hat q^i\delta_{a_1a_3}+g_{(1)}(q^2)\,(S^i)_{a_1a_3}+h_{(1)}(q^2)\,\bigl(i(\mathbf S\times\hat q)^i\bigr)_{a_1a_3}\notag\\
&\quad+k_{(1)}(q^2)\,\bigl(3\hat q^i(\hat q\cdot\mathbf S)-S^i\bigr)_{a_1a_3}+u_{(1)}(q^2)\,Q^{ij}_{a_1a_3}\hat q_j\notag\\
&\quad+v_{(1)}(q^2)\,i\epsilon^{ijk}\hat q_j\hat q_l Q^{kl}_{a_1a_3}+w_{(1)}(q^2)\,\left[\hat q^i\hat q_j\hat q_k-\frac15\left(\delta^i_j\hat q_k+\delta^i_k\hat q_j\right)\right]Q^{jk}_{a_1a_3}.
\label{eq:vector_spinone_final}
\end{align}
The exact coefficient map is
\begin{equation}
f_{(1)}(q^2)=\sqrt{\frac{3}{4\pi}}F^{(1)}_{10}(q^2),\qquad
g_{(1)}(q^2)=\frac{1}{\sqrt{4\pi}}F^{(1)}_{01}(q^2),\qquad
h_{(1)}(q^2)=-\sqrt{\frac{3}{8\pi}}F^{(1)}_{11}(q^2),
\end{equation}
\begin{equation}
k_{(1)}(q^2)=-\frac{1}{\sqrt{12\pi}}F^{(1)}_{21}(q^2),\qquad
u_{(1)}(q^2)=-\frac{3}{\sqrt{20\pi}}F^{(1)}_{12}(q^2),\qquad
v_{(1)}(q^2)=-\sqrt{\frac{3}{4\pi}}F^{(1)}_{22}(q^2),
\end{equation}
\begin{equation}
w_{(1)}(q^2)=\sqrt{\frac{15}{8\pi}}F^{(1)}_{32}(q^2).
\label{eq:spinone_vector_F_to_f}
\end{equation}
The first four structures are the direct spin-$1$ analogues of the spin-$\frac{1}{2}$ vector structures, with $\boldsymbol\sigma$ replaced by the spin-$1$ generators $\mathbf S$.  The last three are new quadrupole structures, which arise because the spin-$1$ crossed channel contains the additional $S=2$ sector.

\section{Canonical spinor formalism}
\label{sec:Canonical-spinor formulation of the multipole expansion}

In the previous section, the multipole expansion and the $LS$ coupling were formulated in the Breit frame. Only in this special frame does the momentum transfer become purely spatial, so that the matrix element can be organized directly under the three dimensional rotation group and the orbital and spin degrees of freedom admit the usual nonrelativistic $LS$ interpretation. In a general reference frame, Lorentz boosts mix the time and spatial components of tensor operators and also rotate the spin degrees of freedom through Wigner rotations. As a result, different multipole structures are generally mixed, and the simple nonrelativistic $LS$ decomposition is no longer manifest. Helicity form factors provide another useful parametrization of matrix elements, but the helicity basis ties the spin projection to the momentum direction and therefore does not display an explicit separation between orbital angular momentum and spin. A covariant tensor construction can restore Lorentz covariance, but the $LS$ interpretation is not always transparent in that form.

The purpose of this section is to introduce a canonical spinor formulation of the covariant $LS$ construction. In this formulation, Lorentz covariance is carried by the spinor indices, while the spin coupling is performed in the little group space. This separation allows one to keep the $LS$ organization explicit while constructing Lorentz covariant matrix element bases. More concretely, a matrix element of a local operator in a Lorentz representation $(j_L,j_R)$ can be converted into an auxiliary massive three point amplitude, and the complete canonical spinor $LS$ basis of that amplitude can then be translated back into a complete basis of matrix elements. We first introduce the spinor variables and wave function conventions, then describe the direct $LS$ coupling form and illustrate it with low spin examples. Finally, we compare the resulting structures with the nonrelativistic $LS$ and multipole bases discussed in section~\ref{sec:Non-relativistic multipole expansion}.

\subsection{Spinor helicity formalism and relativistic wave functions}

Before discussing the direct $LS$ coupling form and the low spin examples, let us briefly summarize the  spinor formalism with canonical particle states \cite{Arkani-Hamed:2017jhn,Huang:2026egv}. In this construction, the massive spinor helicity variables are taken as the basic building blocks, and the amplitude is decomposed into orbital and spin parts in the little group space.

For a massive particle, the four-momentum can be written in spinorial form as
\begin{equation}
p_{\alpha\dot\alpha}=|p^Q_\alpha\rangle [p_{Q\dot\alpha}|.
\end{equation}
The two-component canonical Wyel spinor variables are defined as
\begin{equation}
|p^Q\rangle_\alpha \equiv \lambda^Q_\alpha,
\qquad
|p_Q]_{\dot\alpha} \equiv \tilde\lambda_{Q\dot\alpha},
\qquad
\langle p^Q|^\alpha \equiv \lambda^{Q\alpha},
\qquad
[p_Q|^{\dot\alpha} \equiv \tilde\lambda_Q^{\dot\alpha},
\end{equation}
where $\alpha$ and $\dot\alpha$ are the undotted and dotted $\mathrm{SL}(2,\mathbb C)$ Lorentz spinor indices, while $Q$ is the $\mathrm{SU}(2)$ little group index. The angle spinor $|p^Q\rangle_\alpha$ is left handed, and the square spinor $|p_Q]_{\dot\alpha}$ is right handed.

The Lorentz spinor indices are raised and lowered by the antisymmetric tensors $\epsilon^{\alpha\beta}$, $\epsilon_{\alpha\beta}$ and $\epsilon^{\dot\alpha\dot\beta}$, $\epsilon_{\dot\alpha\dot\beta}$, while the little group indices are raised and lowered by $\epsilon^{QK}$ and $\epsilon_{QK}$. Explicitly,
\begin{equation}
\lambda^{Q\alpha}=\epsilon^{\alpha\beta}\lambda^Q_\beta,
\qquad
\tilde\lambda_Q^{\dot\alpha}=\epsilon^{\dot\alpha\dot\beta}\tilde\lambda_{Q\dot\beta},
\qquad
\lambda_{Q\alpha}=\epsilon_{KQ}\lambda^K_\alpha,
\qquad
\tilde\lambda^Q_{\dot\alpha}=\epsilon^{QK}\tilde\lambda_{K\dot\alpha}.
\end{equation}
Accordingly, the basic spinor contractions are written as
\begin{equation}
\langle p^Q q^K\rangle,
\qquad
[p_Q q_K],
\qquad
\langle q^Q|p|k_K].
\end{equation}

Using the canonical spinor variables introduced above, the wave function of a massive particle in the Lorentz representation $(s_L,s_R)$ with spin-$s$ can be written in a compact form. Before giving the general expression, it is helpful to begin with a few simple low spin examples:
\begin{equation}
\begin{alignedat}{2}
u^I_{\alpha}(p) &\sim |p^I\rangle_{\alpha},
&\qquad& s=\tfrac{1}{2}, \quad (s_L,s_R)=\left(\tfrac{1}{2},0\right), \\
u^I_{\dot\beta}(p) &\sim |p^I]_{\dot\beta},
&& s=\tfrac{1}{2}, \quad (s_L,s_R)=\left(0,\tfrac{1}{2}\right), \\
u_{\alpha,\dot\beta}(p) &\sim p_{\alpha\dot\beta},
&& s=0, \quad (s_L,s_R)=\left(\tfrac{1}{2},\tfrac{1}{2}\right), \\
u^{(I_1 I_2)}_{\alpha,\dot\beta}(p) &\sim |p^{(I_1}\rangle_{\alpha}\,[p^{I_2)}|_{\dot\beta},
&& s=1, \quad (s_L,s_R)=\left(\tfrac{1}{2},\tfrac{1}{2}\right), \\
u^{(I_1 I_2)}_{\alpha_1\alpha_2}(p) &\sim |p^{(I_1}\rangle_{(\alpha_1}\, |p^{I_2)}\rangle_{\alpha_2)},
&& s=1, \quad (s_L,s_R)=(1,0), \\
u^{(I_1 I_2)}_{\dot\beta_1\dot\beta_2}(p) &\sim |p^{(I_1}]_{(\dot\beta_1}\, |p^{I_2)}]_{\dot\beta_2)},
&& s=1, \quad (s_L,s_R)=(0,1), \\
u^{(I_1 I_2)}_{\alpha_1\alpha_2,\dot\beta_1\dot\beta_2}(p)
&\sim |p^{(I_1}\rangle_{(\alpha_1}\, p_{\alpha_2)(\dot\beta_1}\, [p^{I_2)}|_{\dot\beta_2)},
&& s=1, \quad (s_L,s_R)=(1,1).
\end{alignedat}
\end{equation}
These examples already show how the canonical spinor wave functions are built from angle spinors, square spinors, and, when needed, explicit factors of the momentum bispinor $p_{\alpha\dot\beta}$. The general canonical spinor wave function of a massive particle with spin-$s$ in the Lorentz representation $(s_L,s_R)$ is then given by
\begin{equation}
    \begin{aligned}
    u^{\{Q\}}_{\{\alpha\}, \{\dot\beta\}}(p)
    \sim\;& | p^{(Q_1 \dots Q_x} \rangle^x_{(\alpha_1 \dots \alpha_x} \, p^{(z)}_{\alpha_{x+1} \dots \alpha_{2s_L}), (\dot{\beta}_1 \dots \dot{\beta}_z} \, [ p^{Q_{x+1} \dots Q_{2s})} |^y_{\dot{\beta}_{z+1} \dots \dot{\beta}_{2s_R})} \\
    \sim\;&
    C^{s,\{Q\}}_{s_L,\{J\};\,s_R,\{K\}}\,
    | p^{\{J\}} \rangle^{2s_L}_{\{\alpha\}}\,
    [ p^{\{K\}} |^{2s_R}_{\{\dot\beta\}} \,,
    \end{aligned}
\end{equation}
where we use the brace notation $\{\alpha\}\equiv(\alpha_1 \dots \alpha_{2s_L})$ and $\{\dot\beta\}\equiv(\dot{\beta}_1 \dots \dot{\beta}_{2s_R})$ for totally symmetrized Lorentz multiindices, with analogous notations such as $\{\beta\}$ and $\{\dot\alpha\}$ understood similarly. The integers are $x = s_L - s_R + s$, $y = -s_L + s_R + s$, and $z = s_L + s_R - s$. In the second line, $\{J\}\equiv(J_1 \dots J_{2s_L})$ and $\{K\}\equiv(K_1 \dots K_{2s_R})$ denote totally symmetrized little group multiindices, while
\begin{equation}
    | p^{\{J\}} \rangle^{2s_L}_{\{\alpha\}} \equiv | p^{(J_1} \rangle_{(\alpha_1} \cdots | p^{J_{2s_L})} \rangle_{\alpha_{2s_L})},
    \qquad
    [ p^{\{K\}} |^{2s_R}_{\{\dot\beta\}} \equiv [ p^{(K_1} |_{(\dot{\beta}_1} \cdots [ p^{K_{2s_R})} |_{\dot{\beta}_{2s_R})}.
\end{equation}
This formula applies uniformly to both integer spin and half integer spin. The distinction between the two cases is encoded covariantly by the numbers $(x,y,z)$ and by the totally symmetrized little group indices $\{Q\}$, so the same canonical spinor construction already contains the complete set of general spin wave functions needed below.

For integer spin, the most important case for the later general construction is $2s_L=2s_R=L$ with $s=L$. When $p$ is taken to its rest frame, the wave function is equivalent to the spherical tensor basis introduced in Eq.~\eqref{eq:spherical_tensor_basis_general}. More precisely, after converting each spinor pair $(\alpha_a,\dot\beta_a)$ into a vector index $i_a$ and rewriting the little group indices $\{Q\}$ in the standard magnetic basis $M$, one may write schematically
\begin{equation*}
    u^{\{Q\}}_{\{\alpha\},\{\dot\beta\}}(p) \sim u^M_{i_1\cdots i_L}(p) \sim e^{i_1\cdots i_L}_{LM},
\end{equation*}
where $u^M_{i_1\cdots i_L}(p)$ denotes the same wave function expressed with spatial vector indices and magnetic quantum number $M$. In our construction, the spinor wave function $u^{\{Q\}}_{\{\alpha\},\{\dot\beta\}}(p)$ with $2s_L=2s_R=L$ and $s=L$ should be regarded as the Lorentz covariant form of $e^{i_1\cdots i_L}_{LM}$. Correspondingly, $u^{\{Q\}}_{\{\alpha\},\{\dot\beta\}}(p)$ also provides the Lorentz covariant form of the reconstruction tensor $\mathbb T$ appearing in Eq.~\eqref{eq:tensor_reconstruction}. The simplest example is the $L=1$ case, for which after converting the spinor indices into a spatial vector index one has
\begin{equation*}
    |p^1\rangle_\alpha [p^1|_{\dot\beta} \sim e^i_{-1},
    \qquad
    |p^{(1}\rangle_\alpha [p^{2)}|_{\dot\beta} \sim e^i_0,
    \qquad
    |p^2\rangle_\alpha [p^2|_{\dot\beta} \sim e^i_{+1}.
\end{equation*}
For half integer spin, one uses the same canonical expression above with half integer $s$. The resulting wave functions again carry totally symmetrized little group indices and provide the covariant spin tensors needed for arbitrary spin external states. These wave functions therefore supply the basic building blocks for the constructions below, while the more systematic general spin discussion is deferred to section~\ref{sec:Covariant Bases from Method II}.

For an outgoing particle with momentum $p_1$ and little group indices $\{I\}$, the corresponding state $\langle p_1, \{I\} |$ involves the complex conjugate of the wave function. Let us denote this conjugated wave function by applying the complex conjugation directly to the wave function written above. To derive its explicit form, we use the detailed complex conjugation properties of massive spinors and momenta:
\begin{align}
    (|p^Q \rangle)^* &= [p_Q |, & (|p_Q \rangle)^* &= -[p^Q |, \nonumber \\
    (|p^Q ])^* &= -\langle p_Q |, & (|p_Q ])^* &= \langle p^Q |, \nonumber \\
    (p_{\alpha\dot{\beta}})^* &= p_{\beta\dot{\alpha}}, & (p^{\dot{\beta}\alpha})^* &= p^{\dot{\alpha}\beta}.
\end{align}
Applying these operations step by step to the three components of the wave function and ignoring the overall sign $(-1)^y$ (which can be absorbed into the coupling constant), we obtain:
\begin{align}
    &\left( u^{\{I\}}_{\{\alpha\}, \{\dot\beta\}}(p_1) \right)^* \nonumber \\
    &\sim \left( | p_1^{(I_1 \dots I_x} \rangle^x_{(\alpha_1 \dots \alpha_x} \, p^{(z)}_{1,\alpha_{x+1} \dots \alpha_{2s_L}), (\dot{\beta}_1 \dots \dot{\beta}_z} \, [ p_1^{I_{x+1} \dots I_{2s})} |^y_{\dot{\beta}_{z+1} \dots \dot{\beta}_{2s_R})} \right)^* \nonumber \\
    &\sim [ p_{1, (I_1 \dots I_x} |^x_{(\dot{\alpha}_1 \dots \dot{\alpha}_x} \, p_{1, \dot{\alpha}_{x+1} \dots \dot{\alpha}_{2s_L}), (\beta_1 \dots \beta_z}^{(z)} \, | p_{1, I_{x+1} \dots I_{2s})} \rangle^y_{(\beta_{z+1} \dots \beta_{2s_R})}.
\end{align}
Rearranging the terms into the standard order (left handed spinors, momentum matrices, right handed spinors) yields:
\begin{equation}
    \left( u^{\{I\}}_{\{\alpha\}, \{\dot\beta\}}(p_1) \right)^* \sim | p_{1, (I_{x+1} \dots I_{2s}} \rangle^y_{(\beta_{z+1} \dots \beta_{2s_R}} \, p_{1, \dot{\alpha}_{x+1} \dots \dot{\alpha}_{2s_L}), (\beta_1 \dots \beta_z}^{(z)} \, [ p_{1, I_1 \dots I_x)} |^x_{\dot{\alpha}_1 \dots \dot{\alpha}_x)}.
\end{equation}
Observing the indices, this expression has $y+z = 2s_R$ left handed undotted indices $\beta$, $x+z = 2s_L$ right handed dotted indices $\dot{\alpha}$, and lowered little group indices $\{I\}$. This exactly defines an incoming wave function with lowered little group indices:
\begin{equation}
    \left( u^{\{I\}}_{\{\alpha\}, \{\dot\beta\}}(p_1) \right)^* \sim u(p_1)_{\{I\}; \{\beta\}, \{\dot\alpha\}}.
\end{equation}
This derivation proves that the wave function of an outgoing particle with momentum $p_1$ and raised little group indices is equivalent to the wave function of an incoming particle with momentum $p_1$ and lowered indices. Consequently, in the matrix element or the amplitude, the outgoing particle 1 can be effectively treated as an incoming particle. The matrix element of an operator $\hat{\mathcal O}$ is thus formally written as follows, where $\{\alpha\}$ and $\{\dot\beta\}$ now denote the totally symmetrized Lorentz indices with $2j_L$ and $2j_R$ entries, respectively:
\begin{equation}\label{eq:AGeneralFormFactor}
    \mathcal M_{\{I\}\,\{\alpha\},\,\{\dot\beta\}}^{\{A\}} = \langle p_1, \{I\} | \hat{\mathcal O} | p_3, \{A\} \rangle.
\end{equation}
Furthermore, we define $P = p_1 + p_3$. This $P$ represents the momentum of an effective outgoing particle, which is equivalent to an incoming particle with momentum $-P$. Because the spinors satisfy $|-p^Q \rangle \sim |p^Q \rangle$ and $|-p^Q ] \sim |p^Q ]$, the wave function respects $u(-P) \sim u(P)$.

For the three point process $3\to1+2$, the canonical spinor amplitude can be decomposed into $LS$ partial waves as
\begin{equation}
\mathcal A^{\{Q^{(1)}\}\{Q^{(2)}\}}_{\{Q^{(3)}\}}(p_3,p_1,p_2)
\sim
\sum_{L,S}
\mathcal A^{\{Q^{(1)}\}\{Q^{(2)}\}}_{\{Q^{(3)}\}}(p_3,p_1,p_2;L,S),
\end{equation}
where and in the following, the notation $\{Q\}=(Q_1,\dots,Q_{2s})$ denotes the completely symmetrized $\mathrm{SU}(2)$ little group index associated with a spin-$s$ representation. 
Little group indices for a particle with spin-$s$ are grouped as $\{Q\} = (Q_1, Q_2, \dots, Q_{2s})$. The polarization states are constructed from massive spinors:
\begin{equation}
    |p_{\{Q\}} \rangle^{2s} = |p_{(Q_1}\rangle |p_{Q_2}\rangle \dots |p_{Q_{2s})}\rangle, \quad
    |p_{\{Q\}} ]^{2s} = |p_{(Q_1}] |p_{Q_2}] \dots |p_{Q_{2s})}].
\end{equation}

The orbital part is denoted by
\begin{equation}
\mathcal Y^L_{\{Q\}}(p_3,p_1,p_2)\sim(\langle \mathbf{3} | p_{1} | \mathbf{3} ]^L)_{\{Q\}},
\end{equation}
with
\begin{equation}
(\langle \mathbf{3} | p_{1} | \mathbf{3} ]^L)_{\{Q\}}
=
\langle3_{(Q_{1}}|p_{1}|3_{Q_{2}}]\cdots\langle3_{Q_{2L-1}}|p_{1}|3_{Q_{2L})}]
=(\langle \mathbf{3} | p_{1} | \mathbf{3} ]^L)_{Q_1 \dots Q_{2L}},
\end{equation}
which carries $2L$ totally symmetric little group indices of particle 3. The basic building block is the tensor
\begin{equation}
\langle 3^Q|p_1|3^K].
\end{equation}

The spin part is denoted by
\begin{equation}
\mathcal S^{\{Q^{(1)}\}\{Q^{(2)}\}}_{\{K\}}(p_3,p_1,p_2),
\end{equation}
which couples the daughter spins $s_1$ and $s_2$ into a total spin-$S$ representation. Its general form is
\begin{equation}
\mathcal S^{\{Q^{(1)}\}\{Q^{(2)}\}}_{\{K\}}
\sim
(\tau_1^{2s_1})^{\{Q^{(1)}\}}_{\{L^{(1)}\}}
(\tau_2^{2s_2})^{\{Q^{(2)}\}}_{\{L^{(2)}\}}
\,C^{S,\{K\}}_{s_1,\{L^{(1)}\};\,s_2,\{L^{(2)}\}},
\end{equation}
which carries the total spin-$S$ representation in the little group space of particle 3. It is obtained by first transporting the little group indices of particles 1 and 2 into the parent little group space through the covariant maps $\tau_1$ and $\tau_2$, and then coupling the daughter spins $s_1$ and $s_2$ into a total spin-$S$ representation. Its general form is
\begin{equation}
(\tau_1)^Q_K(p_3,p_1)
\sim
\left(
\langle 1^Q 3_K\rangle-[1^Q 3_K]
\right),
\end{equation}
\begin{equation}
(\tau_2)^Q_K(p_3,p_2)
\sim
\left(
\langle 2^Q 3_K\rangle-[2^Q 3_K]
\right),
\end{equation}
where only the kinematic structure is kept here, while the normalization factor is suppressed.
The tensors $\tau_1^{2s_1}$ and $\tau_2^{2s_2}$ are obtained by taking tensor products of the basic $\tau_1$ and $\tau_2$ and then completely symmetrizing the corresponding little group indices.

For later use, and in particular for the spin part introduced below, we make the corresponding symmetrized shorthands explicit:
\begin{align}
    (\langle \mathbf{1} \mathbf{3} \rangle - [\mathbf{1} \mathbf{3}])_I^A
    &= \langle 1_I 3^A \rangle - [ 1_I 3^A ], \\[3ex]
    (\langle \mathbf{1} \mathbf{3} \rangle - [\mathbf{1} \mathbf{3}])_{I_1 \dots I_{2s_1}}^{2s_1;\, A_1 \dots A_{2s_1}}
    &= (\langle \mathbf{1} \mathbf{3} \rangle - [\mathbf{1} \mathbf{3}])_{(I_1}^{(A_1} (\langle \mathbf{1} \mathbf{3} \rangle - [\mathbf{1} \mathbf{3}])_{I_2}^{A_2} \dots
    (\langle \mathbf{1} \mathbf{3} \rangle - [\mathbf{1} \mathbf{3}])_{I_{2s_1})}^{A_{2s_1})}.
\end{align}

Finally, the orbital and spin parts are coupled to the parent little group indices by another $\mathrm{SU}(2)$ CGC, which can be written as
\begin{equation}
\mathcal A^{\{Q^{(1)}\}\{Q^{(2)}\}}_{\{Q^{(3)}\}}(p_3,p_1,p_2;L,S)
=
\mathcal Y^L_{\{Q\}}(p_3,p_1,p_2)\,
\mathcal S^{\{Q^{(1)}\}\{Q^{(2)}\}}_{\{K\}}(p_3,p_1,p_2)\,
C_{s_3,\{Q^{(3)}\}}^{L,\{Q\};\,S,\{K\}}.
\label{eq:canonical_spinor_LS_basic}
\end{equation}
Therefore, the canonical spinor three point amplitude is obtained by first constructing the orbital part with definite $L$, then constructing the spin part with definite $S$, and finally coupling them to the total spin-$s_3$ of the parent particle through the $\mathrm{SU}(2)$ CGC.

\subsection{Covariant $LS$ form factor}

The construction in this subsection furnishes a complete basis for the matrix elements in an arbitrary Lorentz irreducible representation (irrep), and each basis element carries definite quantum numbers $(S,L,J)$. Here $S$ is obtained by coupling the spin degrees of freedom of particles 1 and 3 in the auxiliary three point amplitude, $L$ is the orbital angular momentum in the covariant three point construction, and $J$ is the $\mathrm{SO}(3)$ angular momentum obtained from the Lorentz representation $(j_L,j_R)$ of the operator after restricting to the rotation subgroup. Accordingly, $J$ takes values in the range $J \in \{ |j_L - j_R|, |j_L - j_R| + 1, \dots, j_L + j_R \}$. Once the spins and momenta of the external particles are fixed, every allowed triple $(S,L,J)$ labels exactly one basis element, meaning there is no degeneracy for a fixed $(S,L,J)$.

The direct $LS$ coupling form is the direct covariant generalization of the traditional $LS$ construction. When specialized to the Breit frame of the original matrix element, equivalently the Breit frame of the corresponding auxiliary amplitude with $P=p_1+p_3$, the labels $(S,L,J)$ have exactly the same physical meaning as in the traditional nonrelativistic construction. For this reason, the direct $LS$ coupling form can be directly compared with the nonrelativistic $LS$ and multipole bases discussed in section~\ref{sec:Non-relativistic multipole expansion}.

The basic idea of the direct $LS$ coupling form is to convert a matrix element into an auxiliary three point amplitude and then use the complete canonical spinor $LS$ basis of that amplitude. By contracting the matrix element with the wave functions of $P$, one converts the Lorentz representation $(j_L,j_R)$ carried by the operator into little group indices associated with $P$, thereby projecting the matrix element onto a set of effective three point amplitudes. In this form, the three legs of the auxiliary amplitude correspond to momenta $p_1$, $p_3$, and $P$, carrying spins $s_1$, $s_3$, and $J$, respectively. Writing the little group indices of particle 1 as lowered indices $\{I\}$ since it is an outgoing particle, and those of particle 3 as $\{A\}$ since it is an incoming particle, the corresponding canonical spinor amplitude can be written in the explicit shorthand notation introduced above as
\begin{align}
    \mathcal A_{\{I\}}^{\{A\}\{Q\}}(P, J; p_1,s_1;p_3,s_3;L,S)
    &=
    (\langle \mathbf{P} | p_1 | \mathbf{P} ]^L)^{\{N\}}\,
    (\langle \mathbf{1} \mathbf{P} \rangle - [\mathbf{1} \mathbf{P}])_{\{I\}}^{2s_1;\,\{B\}}\,
    (\langle \mathbf{3} \mathbf{P} \rangle - [\mathbf{3} \mathbf{P}])^{2s_3;\,\{A\},\{C\}} \nonumber \\
    &\quad \times
    C^{S,\{K\}}_{s_1,\{B\};\,s_3,\{C\}}\,
    C^{J,\{Q\}}_{L,\{N\};\,S,\{K\}}.
\end{align}
Here $\{Q\}$ denotes the little group indices associated with $P$ after projection onto definite spin $J$. This is precisely the three point amplitude that will be used in the direct $LS$ coupling form below.

To make this map explicit, we define a tensor $T_{1\{I\}}^{\{A\}\{C\}\{D\}}$ by contracting the Lorentz indices of the matrix element with the spinors of the intermediate momentum $P$:
\begin{equation}
    T_{1\{I\}}^{\{A\}\{C\}\{D\}} = \mathcal M_{\{I\}\,\{\alpha\},\,\{\dot\beta\}}^{\{A\}} \langle P^{C_1} |^{\alpha_1} \dots \langle P^{C_{2j_L}} |^{\alpha_{2j_L}} | P^{D_1} ]^{\dot{\beta}_1} \dots | P^{D_{2j_R}} ]^{\dot{\beta}_{2j_R}}.
\end{equation}
In this way, all Lorentz indices carried by the operator are traded for little group indices associated with $P$. The tensor $T_{1\{I\}}^{\{A\}\{C\}\{D\}}$ therefore contains only little group information together with the external state labels.

Furthermore we also have
\begin{equation}
    \mathcal{M}_{\{I\}\,\{\alpha\},\,\{\dot\beta\}}^{\{A\}} \sim T_{1\{I\}}^{\{A\}\{C\}\{D\}} | P_{\{C\}} \rangle^{2j_L}_{\{\alpha\}} [ P_{\{D\}} |^{2j_R}_{\{\dot\beta\}}\,,
\end{equation}
which makes the transformation reversible. We then use CGCs to take the direct product of the little group indices carried by the wave functions of the particle with momentum $P$:
\begin{align}
    \mathcal{M}_{\{I\}\,\{\alpha\},\,\{\dot\beta\}}^{\{A\}} &\sim \sum_{J, \{Q\}} T_{1\{I\}}^{\{A\}\{C\}\{D\}} C^{J, \{Q\}}_{j_L, \{C\}; j_R, \{D\}} \nonumber \\
    &\quad \times \left( C_{J, \{Q\}}^{j_L, \{C'\}; j_R, \{D'\}} | P_{\{C'\}} \rangle^{2j_L}_{\{\alpha\}} [ P_{\{D'\}} |^{2j_R}_{\{\dot\beta\}} \right) \nonumber \\
    &\sim \sum_{J, \{Q\}} \left( T_{1\{I\}}^{\{A\}\{C\}\{D\}} C^{J, \{Q\}}_{j_L, \{C\}; j_R, \{D\}} \right) u_{J, \{Q\}; \{\alpha\}, \{\dot\beta\}}(P)\,,
\end{align}
where $u_{J, \{Q\}}(P)$ is the relativistic wave function introduced above. The other part is exactly an effective three point amplitude $\mathcal A$:
\begin{equation}
    \mathcal A^{\{I\}\{A\}\{Q\}}(p_1, s_1; -P, J; p_3, s_3) \equiv T_{1}^{\{I\}\{A\}\{C\}\{D\}} C^{J, \{Q\}}_{j_L, \{C\}; j_R, \{D\}}= \mathcal{M}_{\{\alpha\},\{\dot\beta\}}^{\{I\}\{A\}}u_{J, }^{\{Q\}\,\{\alpha\}, \{\dot\beta\}}(P),
\end{equation}
with all momenta incoming, as induced by the projection with the wave function. Hence, the complete structural basis of the matrix element is generated by
\begin{equation}
    \{ \mathcal{A}_{\{I\}}^{\{A\}\{Q\}} u_{J, \{Q\}; \{\alpha\}, \{\dot\beta\}}(P) \} = \{ \mathcal M_{\{I\}\,\{\alpha\},\,\{\dot\beta\}}^{\{A\}} \},
\end{equation}
which is the direct spinor analogue of Eq.~\eqref{eq:tensor_reconstruction}. The relativistic wave function $u_{J,\{Q\};\{\alpha\},\{\dot\beta\}}(P)$ plays the role of the reconstruction tensor, while the $\mathrm{SO}(3)$ labels $(J,M)$ are replaced here by little group indices $\{Q\}$. In the rest frame of $P$, after converting the spinor indices into Cartesian vector indices and rewriting the little group indices in the magnetic basis, one has schematically
\begin{equation}
    u_{J,\{Q\};\{\alpha\},\{\dot\beta\}}(P) \sim u^{\,i_1\cdots i_n}_{J,M}(P) \sim \mathbb T^{\,i_1\cdots i_n}_{J,M},
\end{equation}
so that the present construction reduces directly to the reconstruction tensor appearing in Eq.~\eqref{eq:tensor_reconstruction}. In this sense, the orbital and spin angular momentum degrees of freedom are encoded in the spinor formalism through little group indices. Moreover, this relation is valid in an arbitrary reference frame, so it provides the Lorentz covariant generalization of Eq.~\eqref{eq:tensor_reconstruction}.

This relation shows that the complete kinematic structures of the matrix element are generated once the complete set of three point amplitudes is known. Since the canonical spinor three point amplitudes have already been organized into a complete $LS$ basis, the above construction immediately provides a complete basis for the matrix element in the Lorentz representation $(j_L,j_R)$.


\subsection{Form factors of vector operator}

This subsection collects the explicit spin-$\frac{1}{2}$ matrix element bases with irrep $(1/2,1/2)$ generated in by $LS$ coupling and extends to the general spin cases, in which we derive their forms in the Breit frame for the comparison with the results we get in section~\ref{sec:Non-relativistic multipole expansion}. We will show that the $LS$ coupling form is the covariant version of the traditional $LS$ construction, which means that the resulting basis elements reduce directly in the Breit frame to the $LS$ results discussed in section~\ref{sec:Non-relativistic multipole expansion}. 

\paragraph{Basis of spin-$\tfrac{1}{2}$ external particles.}
For the matrix elements with irrep $(1/2,1/2)$ and the external particle state with $s_1=s_3=1/2$, when we restrict the irrep $(1/2,1/2)$ to the rotation subgroup, it gives the $\mathrm{SO}(3)$ angular momenta $J=0$ and $J=1$. Since the spin of particles 1 and 3 couple to $S=0$ or $S=1$, the allowed channels in the direct $LS$ coupling form are
\begin{equation}
    (S,L,J)=(0,0,0),\quad (1,1,0),\quad (1,0,1),\quad (0,1,1),\quad (1,1,1).
\end{equation}
To make it clear, the matrix elements of the operator with $(\frac{1}{2}, \frac{1}{2})$ Lorentz irrep and spin-$\frac{1}{2}$ external particles can be expanded as
\begin{equation}
    \mathcal{M}=\mathcal{M}_{(0,0,0)}+\mathcal{M}_{(1,1,0)}+\mathcal{M}_{(1,0,1)}+\mathcal{M}_{(0,1,1)}+\mathcal{M}_{(1,1,1)}.
\end{equation}
These five channels are exactly the low spin basis elements in the direct $LS$ coupling form that match the vector multipoles of section~\ref{sec:Non-relativistic multipole expansion}.

For later use we introduce the shorthand
\begin{equation}
    \tau_{1\,I}{}^{B} \equiv (\langle \mathbf{1} \mathbf{P} \rangle - [\mathbf{1} \mathbf{P}])_I^{\ B},
    \qquad
    \tau_3{}^{A}{}_{B} \equiv (\langle \mathbf{3} \mathbf{P} \rangle - [\mathbf{3} \mathbf{P}])^A_{\ B}\,,
\end{equation}
where $I$ and $A$ denote the little group indices of particle 1 and particle 3, respectively, and $B$ is the little group index of the particle with momentum $P$ in the $LS$ 3-point amplitude.
Before turning to the explicit Breit frame basis elements, let us make the lowered little group notation for particle 3 explicit. When particle 3 is written with lowered little group indices, its raised little group indices $\{A\}$ are converted into lowered indices $\{\bar A\}$, which are obtained from $\{A\}$ by exchanging $1\leftrightarrow 2$ in $\{A\}$. Accordingly, we write
\begin{equation}
    \tau_{3\,\bar A B} \sim \tau_3{}^{A}{}_{B},
    \qquad
    \tau_{3\,\bar A}{}^{B} \sim \tau_3^{AB},
    \qquad
    \mathcal S_{\{I\}\{\bar A\}}{}_{\{K\}} \sim \mathcal S_{\{I\}}^{\{A\}}{}_{\{K\}},
\end{equation}
so that, to facilitate comparison with the earlier text, throughout this subsection the spin tensor written in this lowered index notation may be regarded as $\mathcal S_{\{I\}\{\bar A\}}{}^{\{K\}} \sim C^{S,\{K\}}_{s_1,\{I\};\,s_3,\{\bar A\}}$.

In the rest frame of $P$, the canonical-spinor tensors have the simple forms:
\begin{equation}
\begin{aligned}
    \tau_3{}^{A}{}_{B} &\sim \delta^A_B,
    \qquad
    \tau_3^{AB} \equiv \tau_3{}^{A}{}_{C}\epsilon^{BC} \sim \epsilon^{AB}, \\
    \tau_{1\,I}{}^{B} &\sim \delta_I^B,
    \qquad
    \tau_{1\,IB} \equiv \tau_{1\,I}{}^{C}\epsilon_{CB} \sim \epsilon_{IB}.
\end{aligned}
\end{equation}
To make the hierarchy transparent, we first discuss the two $J=0$ channels of the vector operator and only afterwards turn to the three $J=1$ channels. In the explicit intermediate steps below, we keep the little group components $\mathcal Y^{(1)}_{11}$, $\mathcal Y^{(1)}_{12}$, and $\mathcal Y^{(1)}_{22}$ manifest, and only afterwards compare them with the conventional spherical harmonics labeled by $M=0,\pm1$ through
\begin{equation}
    \mathcal Y^{(1)}_{11} \sim \mathcal Y^{(1)}_{-1},
    \qquad
    \mathcal Y^{(1)}_{12} \sim \mathcal Y^{(1)}_{0},
    \qquad
    \mathcal Y^{(1)}_{22} \sim \mathcal Y^{(1)}_{+1},
\end{equation}
\begin{equation}
    |P^1\rangle_\alpha [P^1|_{\dot\beta} \sim e_{-1}^\mu,
    \qquad
    |P^{(1}\rangle_\alpha [P^{2)}|_{\dot\beta} \sim e_{0}^\mu,
    \qquad
    |P^2\rangle_\alpha [P^2|_{\dot\beta} \sim e_{+1}^\mu.
\end{equation}
That is, once the spinor indices on the left hand side are converted into the vector index $\mu$, the resulting objects correspond precisely to the polarization vectors on the right hand side. In the Breit frame, they are exactly the polarization vectors introduced earlier in Eq.~\eqref{eq:spherical_basis_vector}, now viewed as four-vectors with vanishing time component, namely $e^\mu_M=(0,e^i_M)$. Hence only the spatial components are nonzero, and one may equivalently use the spatial index $i$ instead of the vector index $\mu$. This is the direct covariant realization of the vector $LS$ basis of section~\ref{sec:Non-relativistic multipole expansion}.

\paragraph{$J=0$ channels for $(j_L,j_R)=(1/2,1/2)$.}
The restriction of the vector operator to $J=0$ gives two channels in the direct $LS$ coupling form, namely $(S,L,J)=(0,0,0)$ and $(1,1,0)$. In the Breit frame these become the two timelike structures multiplying the unique $J=0$ wave function $P^\mu$.

We first consider the channel $(S,L,J)=(0,0,0)$. Since both the orbital tensor and the spin coupling are trivial, the corresponding 3-point amplitude is simply
\begin{equation}
    (\langle \mathbf{3} \mathbf{P} \rangle - [\mathbf{3} \mathbf{P}])^A_{\ B}
    (\langle \mathbf{1} \mathbf{P} \rangle - [\mathbf{1} \mathbf{P}])_I^{\ B}
    =
    \tau_3{}^{A}{}_{B}\tau_{1\,I}{}^{B}.
\end{equation}
After dressing it with the $J=0$ wave function in the $(1/2,1/2)$ representation, the basis element of the matrix element generated in the direct $LS$ coupling form becomes
\begin{equation}
    (\langle \mathbf{3} \mathbf{P} \rangle - [\mathbf{3} \mathbf{P}])^A_{\ B}
    (\langle \mathbf{1} \mathbf{P} \rangle - [\mathbf{1} \mathbf{P}])_I^{\ B}\,
    P_{\alpha \dot{\beta}}.
\end{equation}
Converting the spinor indices into a Lorentz vector index then gives
\begin{equation}\label{eq:FFin000channel}
    \mathcal{M}_{(0,0,0)}^\mu\propto\langle \mathbf{3} \mathbf{P} \rangle - [\mathbf{3} \mathbf{P}])^A_{\ B}
    (\langle \mathbf{1} \mathbf{P} \rangle - [\mathbf{1} \mathbf{P}])_I^{\ B}\,
    P^{\mu}.
\end{equation}

To compare with the results in Breit frame constructed in section~\ref{sec:Non-relativistic multipole expansion}, we replace the momentum in Eq.~\eqref{eq:FFin000channel} with that in the Breit frame, find 
\begin{equation}
    (\tau_3{}^{A}{}_{B}\tau_{1\,I}{}^{B}) P^\mu
    \sim
    \delta_I^A P^\mu.
\end{equation}
In this lowered index notation, the spin tensor with $S=0$ can be rewritten as
\begin{equation}
    \mathcal S^{(0)}_{I\bar A}
    \sim
    \tau_{3\,\bar A }{}^C\tau_{1\,I}{}^{D}
    C^{0}_{\frac{1}{2},D;\,\frac{1}{2},C}
    \sim
    C^{0}_{\frac{1}{2},I;\,\frac{1}{2},\bar A}.
\end{equation}
Accordingly, the basis obtained in the direct $LS$ coupling form for $(S,L,J)=(0,0,0)$ is precisely $\mathcal S^{(0)}_{I\bar A} P^\mu$.
In the rest frame of $P$, namely $P^\mu=(M,0,0,0)$, only the zeroth component is nonvanishing, so this basis is the purely timelike $J=0$ basis in the vector representation.

We next consider the channel $(S,L,J)=(1,1,0)$. The spin part is now the symmetric triplet tensor
\begin{equation}
    \tau_3^{A(B_1}\tau_{1\,I}^{B_2)}.
\end{equation}
Coupling it to the orbital tensor and then dressing it with the $J=0$ wave function gives the corresponding basis element
\begin{equation}
\begin{aligned}
&\mathcal{M}_{(1,1,0)}(p_1,p_2,P)\sim\bigl(\tau_3^{A(B_1}\tau_{1\,I}^{B_2)}\bigr) \mathcal Y^{(1)}_{B_1B_2}\, P_{\alpha\dot{\beta}} \\
\sim\;& \Big[
\tau_3^{A1}\tau_{1\,I}^{1}\, \mathcal Y^{(1)}_{11}
+ 2\tau_3^{A(1}\tau_{1\,I}^{2)}\, \mathcal Y^{(1)}_{12}
+ \tau_3^{A2}\tau_{1\,I}^{2}\, \mathcal Y^{(1)}_{22}
\Big] P_{\alpha\dot{\beta}}.
\end{aligned}
\end{equation}
Equivalently, after converting the spinor indices into a Lorentz vector index, one finds
\begin{equation}
    \mathcal{M}_{(1,1,0)}^\mu(p_1,p_2,P)\propto\Big[
    \tau_3^{A1}\tau_{1\,I}^{1}\, \mathcal Y^{(1)}_{11}
    + 2\tau_3^{A(1}\tau_{1\,I}^{2)}\, \mathcal Y^{(1)}_{12}
    + \tau_3^{A2}\tau_{1\,I}^{2}\, \mathcal Y^{(1)}_{22}
    \Big] P^\mu.
\end{equation}

Again, working in the Breit frame we find 
\begin{equation}
    \mathcal S^{(1)}_{I\bar A}{}^{B_1B_2}
    \sim
    \tau_{3\,\bar A}{}^{C_1}\tau_{1\,I}{}^{C_2}
    C^{1,(B_1B_2)}_{\frac{1}{2},C_1;\,\frac{1}{2},C_2}
    \sim
    C^{1,(B_1B_2)}_{\frac{1}{2},I;\,\frac{1}{2},\bar A}.
\end{equation}
With this lowered index notation, the basis obtained in the direct $LS$ coupling form for $(S,L,J)=(1,1,0)$ is precisely $\bigl(\mathcal S^{(1)}_{I\bar A}{}^{B_1B_2}\mathcal Y^{(1)}_{B_1B_2}\bigr)P^\mu$.
In components, this means
\begin{equation}
    \Big[
    \mathcal S^{(1)}_{I\bar A}{}^{11}\mathcal Y^{(1)}_{11}
    + 2\mathcal S^{(1)}_{I\bar A}{}^{12}\mathcal Y^{(1)}_{12}
    + \mathcal S^{(1)}_{I\bar A}{}^{22}\mathcal Y^{(1)}_{22}
    \Big] P^\mu,
\end{equation}
\begin{equation}
    \mathcal S^{(1)}_{I\bar A}{}^{11}\mathcal Y^{(1)}_{11} P^\mu
    \sim
    \mathcal S^{(1)}_{I\bar A}{}^{11}\mathcal Y^{(1)}_{-1} P^\mu,
    \qquad
    2\mathcal S^{(1)}_{I\bar A}{}^{12}\mathcal Y^{(1)}_{12} P^\mu
    \sim
    \mathcal S^{(1)}_{I\bar A}{}^{12}\mathcal Y^{(1)}_{0} P^\mu,
    \qquad
    \mathcal S^{(1)}_{I\bar A}{}^{22}\mathcal Y^{(1)}_{22} P^\mu
    \sim
    \mathcal S^{(1)}_{I\bar A}{}^{22}\mathcal Y^{(1)}_{+1} P^\mu.
\end{equation}
Just as in the previous example, only the timelike component survives in the rest frame of $P$, although the coefficient is now built from the spin triplet structure rather than the spin singlet one.

Now we come to the $J=1$ channels for $(j_L,j_R)=(1/2,1/2)$. 
The restriction of the vector operator to $J=1$ gives the three channels in the direct $LS$ coupling form $(S,L,J)=(1,0,1)$, $(0,1,1)$, and $(1,1,1)$. These are the spatial vector structures that reduce in the Breit frame directly to the familiar nonrelativistic multipole basis. 

The first instructive example is the channel $(S,L,J)=(1,0,1)$. Since $L=0$, there is no orbital tensor, and the spin triplet structure is dressed directly by the $J=1$ wave function. It is convenient to denote $\mathcal S_I^{A,Q_1Q_2} \equiv \tau_3^{A(Q_1}\tau_{1\,I}^{Q_2)}$. The corresponding basis element obtained in the direct $LS$ coupling form is then $\mathcal S_I^{A,Q_1Q_2}\, |P_{(Q_1}\rangle_\alpha [P_{Q_2)}|_{\dot\beta}$.
Expanding the symmetrized little group indices gives
\begin{equation}
\begin{aligned}
&\mathcal{M}_{(1)}^\mu\sim\mathcal S_I^{A,Q_1Q_2}\,
|P_{(Q_1}\rangle_\alpha [P_{Q_2)}|_{\dot\beta} \\
=\;&
\tau_3^{A1}\tau_{1\,I}^{1}\, |P_1\rangle_\alpha [P_1|_{\dot\beta}
+ 2\tau_3^{A(1}\tau_{1\,I}^{2)}\, |P_{(1}\rangle_\alpha [P_{2)}|_{\dot\beta}
+ \tau_3^{A2}\tau_{1\,I}^{2}\, |P_2\rangle_\alpha [P_2|_{\dot\beta}.
\end{aligned}
\end{equation}
Using the lowered index notation for particle 3, we may equally write
\begin{equation}
    \mathcal S^{(1)}_{I\bar A}{}^{Q_1Q_2}
    \sim
    \tau_{3\,\bar A}{}^{C_1}\tau_{1\,I}{}^{C_2}
    C^{1,(Q_1Q_2)}_{\frac{1}{2},C_1;\,\frac{1}{2},C_2}
    \sim
    C^{1,(Q_1Q_2)}_{\frac{1}{2},I;\,\frac{1}{2},\bar A}.
\end{equation}
In this notation, the basis obtained in the direct $LS$ coupling form for $(S,L,J)=(1,0,1)$ is precisely
\begin{equation}
    \mathcal S^{(1)}_{I\bar A}{}^{Q_1Q_2}\,
    |P_{(Q_1}\rangle_\alpha [P_{Q_2)}|_{\dot\beta},
\end{equation}
whose three components are equivalently written as
\begin{equation}
    \mathcal S^{(1)}_{I\bar A}{}^{11}|P_1\rangle_\alpha [P_1|_{\dot\beta}
    \sim
    \mathcal S^{(1)}_{I\bar A}{}^{11} e_{-1}^\mu,
    \qquad
    2\mathcal S^{(1)}_{I\bar A}{}^{12}|P_{(1}\rangle_\alpha [P_{2)}|_{\dot\beta}
    \sim
    \mathcal S^{(1)}_{I\bar A}{}^{12} e_0^\mu,
\end{equation}
\begin{equation}
    \mathcal S^{(1)}_{I\bar A}{}^{22}|P_2\rangle_\alpha [P_2|_{\dot\beta}
    \sim
    \mathcal S^{(1)}_{I\bar A}{}^{22} e_{+1}^\mu.
\end{equation}

We next consider the channel $(S,L,J)=(0,1,1)$. In this case, the corresponding amplitude reads
\begin{equation}
    \mathcal{M}_{(0,1,1)}^\mu\sim(\langle \mathbf{3} \mathbf{P} \rangle - [\mathbf{3} \mathbf{P}])^A_{\ B}
    (\langle \mathbf{1} \mathbf{P} \rangle - [\mathbf{1} \mathbf{P}])_I^{\ B}
    (\langle \mathbf{P} | p_1 | \mathbf{P} ]^1)_{Q_1 Q_2}
    =
    (\tau_3{}^{A}{}_{B}\tau_{1\,I}{}^{B}) \mathcal Y^{(1)}_{Q_1 Q_2}.
\end{equation}
We denote the orbital structure by
\begin{equation}
    \mathcal Y^{(1)}_{Q_1 Q_2} \equiv (\langle \mathbf{P} | p_1 | \mathbf{P} ])_{Q_1 Q_2}.
\end{equation}
In the rest frame of $P$, which is also the Breit frame of the initial and final states, the tensor $\mathcal Y^{(1)}_{Q_1Q_2}$ represents the $L=1$ spherical harmonics constructed from the spatial components of $p_1$. Its components are identified with the magnetic quantum number $M$ only up to normalization.
After contracting the little group indices $Q_1,Q_2$ with the fully symmetrized massive spin-1 polarization basis, the full $J=1$ basis element takes the form
\begin{equation}
\begin{aligned}
&(\tau_3{}^{A}{}_{B}\tau_{1\,I}{}^{B}) \mathcal Y^{(1)}_{Q_1 Q_2}
|P^{(Q_1}\rangle_\alpha [P^{Q_2)}|_{\dot{\beta}} \\
\sim\;& (\tau_3{}^{A}{}_{B}\tau_{1\,I}{}^{B}) \Big[
\mathcal Y^{(1)}_{11} |P^1\rangle_\alpha [P^1|_{\dot{\beta}}
+ 2\mathcal Y^{(1)}_{12} |P^{(1}\rangle_\alpha [P^{2)}|_{\dot{\beta}}
+ \mathcal Y^{(1)}_{22} |P^2\rangle_\alpha [P^2|_{\dot{\beta}}
\Big] \\
\sim\;&
(\tau_3{}^{A}{}_{B}\tau_{1\,I}{}^{B}) \mathcal Y^{(1)}_{11} |P^1\rangle_\alpha [P^1|_{\dot{\beta}}
+ 2(\tau_3{}^{A}{}_{B}\tau_{1\,I}{}^{B}) \mathcal Y^{(1)}_{12} |P^{(1}\rangle_\alpha [P^{2)}|_{\dot{\beta}}
+ (\tau_3{}^{A}{}_{B}\tau_{1\,I}{}^{B}) \mathcal Y^{(1)}_{22} |P^2\rangle_\alpha [P^2|_{\dot{\beta}} .
\end{aligned}
\end{equation}
Equivalently, in the lowered index notation,
\begin{equation}
    \mathcal S^{(0)}_{I\bar A}
    \Big[
    \mathcal Y^{(1)}_{11} |P^1\rangle_\alpha [P^1|_{\dot{\beta}}
    + 2\mathcal Y^{(1)}_{12} |P^{(1}\rangle_\alpha [P^{2)}|_{\dot{\beta}}
    + \mathcal Y^{(1)}_{22} |P^2\rangle_\alpha [P^2|_{\dot{\beta}}
    \Big],
\end{equation}
with the same spin tensor $\mathcal S^{(0)}_{I\bar A}\sim C^{0}_{\frac{1}{2},I;\,\frac{1}{2},\bar A}$. With this identification, the basis obtained in the direct $LS$ coupling form for $(S,L,J)=(0,1,1)$ is precisely the expression above, and the terms appearing in it satisfy the following equivalence relations.
\begin{equation}
    \mathcal S^{(0)}_{I\bar A}\mathcal Y^{(1)}_{11}|P^1\rangle_\alpha [P^1|_{\dot{\beta}}
    \sim
    \mathcal S^{(0)}_{I\bar A}\mathcal Y^{(1)}_{-1} e_{-1}^\mu,
    \qquad
    2\mathcal S^{(0)}_{I\bar A}\mathcal Y^{(1)}_{12}|P^{(1}\rangle_\alpha [P^{2)}|_{\dot{\beta}}
    \sim
    \mathcal S^{(0)}_{I\bar A}\mathcal Y^{(1)}_{0} e_{0}^\mu,
    \qquad
\end{equation}
\begin{equation}
    \mathcal S^{(0)}_{I\bar A}\mathcal Y^{(1)}_{22}|P^2\rangle_\alpha [P^2|_{\dot{\beta}}
    \sim
    \mathcal S^{(0)}_{I\bar A}\mathcal Y^{(1)}_{+1} e_{+1}^\mu.
\end{equation}

Finally, let us analyze the channel $(S,L,J)=(1,1,1)$. Following the notation introduced in the tables, we denote the spin triplet structure carrying one little group index associated with $P$ by $\mathcal S_{I,Q}^{A,B}$. The corresponding basis element is then written as $\mathcal S_{I,(Q_1}^{A,B} \mathcal Y^{(1)}_{Q_2)B}\, |P^{Q_1}\rangle_\alpha [P^{Q_2}|_{\dot{\beta}}$,
where $Q_1$ is the lower index of the spin tensor $\mathcal S$, $Q_2$ is the lower index of the orbital tensor $\mathcal Y^{(1)}$, and the pair $(Q_1Q_2)$ is fully symmetrized. Expanding first this symmetrization and then the contraction over $B=1,2$, and using the symmetry $\mathcal Y^{(1)}_{Q_2B}=\mathcal Y^{(1)}_{BQ_2}$, one finds
\begin{equation}
\begin{aligned}
&\mathcal{M}_{(1,1,1)}^\mu\sim\mathcal S_{I,(Q_1}^{A,B} \mathcal Y^{(1)}_{Q_2)B}\,
|P^{Q_1}\rangle_\alpha [P^{Q_2}|_{\dot{\beta}} \\
=\;&
\Big(
\mathcal S_{I,1}^{A,1} \mathcal Y^{(1)}_{11}
+ \mathcal S_{I,1}^{A,2} \mathcal Y^{(1)}_{12}
\Big)\, |P^1\rangle_\alpha [P^1|_{\dot{\beta}} \\
&+
\Big(
\mathcal S_{I,1}^{A,1} \mathcal Y^{(1)}_{12}
+ \mathcal S_{I,1}^{A,2} \mathcal Y^{(1)}_{22}
+ \mathcal S_{I,2}^{A,1} \mathcal Y^{(1)}_{11}
+ \mathcal S_{I,2}^{A,2} \mathcal Y^{(1)}_{12}
\Big)\, |P^{(1}\rangle_\alpha [P^{2)}|_{\dot{\beta}} \\
&+
\Big(
\mathcal S_{I,2}^{A,1} \mathcal Y^{(1)}_{12}
+ \mathcal S_{I,2}^{A,2} \mathcal Y^{(1)}_{22}
\Big)\, |P^2\rangle_\alpha [P^2|_{\dot{\beta}} .
\end{aligned}
\end{equation}
The same final expression can also be rewritten using the lowered index notation for particle 3, with
\begin{equation}
    \mathcal S^{(1)}_{I\bar A}{}^{BQ}
    \sim
    \tau_{3\,\bar A}{}^{C_1}\tau_{1\,I}{}^{C_2}
    C^{1,(BQ)}_{\frac{1}{2},C_1;\,\frac{1}{2},C_2}
    \sim
    C^{1,(QB)}_{\frac{1}{2},I;\,\frac{1}{2},\bar A},
\end{equation}
so that the terms appearing in this expression satisfy the following equivalence relations.
\begin{equation}
\begin{aligned}
&\Big(
\mathcal S_{I,1}^{A,1} \mathcal Y^{(1)}_{11}
+ \mathcal S_{I,1}^{A,2} \mathcal Y^{(1)}_{12}
\Big)|P^1\rangle_\alpha [P^1|_{\dot{\beta}}
\sim
\Big(
\mathcal S^{(1)}_{I\bar A,1}{}^{1} \mathcal Y^{(1)}_{-1}
+ \mathcal S^{(1)}_{I\bar A,1}{}^{2} \mathcal Y^{(1)}_{0}
\Big)e_{-1}^\mu, \\
&\Big(
\mathcal S_{I,1}^{A,1} \mathcal Y^{(1)}_{12}
+ \mathcal S_{I,1}^{A,2} \mathcal Y^{(1)}_{22}
+ \mathcal S_{I,2}^{A,1} \mathcal Y^{(1)}_{11}
+ \mathcal S_{I,2}^{A,2} \mathcal Y^{(1)}_{12}
\Big)|P^{(1}\rangle_\alpha [P^{2)}|_{\dot{\beta}} \\
&\hspace{1.5cm}\sim
\Big(
\mathcal S^{(1)}_{I\bar A,1}{}^{1} \mathcal Y^{(1)}_{0}
+ \mathcal S^{(1)}_{I\bar A,1}{}^{2} \mathcal Y^{(1)}_{+1}
+ \mathcal S^{(1)}_{I\bar A,2}{}^{1} \mathcal Y^{(1)}_{-1}
+ \mathcal S^{(1)}_{I\bar A,2}{}^{2} \mathcal Y^{(1)}_{0}
\Big)e_{0}^\mu, \\
&\Big(
\mathcal S_{I,2}^{A,1} \mathcal Y^{(1)}_{12}
+ \mathcal S_{I,2}^{A,2} \mathcal Y^{(1)}_{22}
\Big)|P^2\rangle_\alpha [P^2|_{\dot{\beta}}
\sim
\Big(
\mathcal S^{(1)}_{I\bar A,2}{}^{1} \mathcal Y^{(1)}_{0}
+ \mathcal S^{(1)}_{I\bar A,2}{}^{2} \mathcal Y^{(1)}_{+1}
\Big)e_{+1}^\mu .
\end{aligned}
\end{equation}
Thus, the basis obtained in the direct $LS$ coupling form for $(S,L,J)=(1,1,1)$ is precisely the vector expression above, or equivalently the same expression written with $\mathcal S^{(1)}_{I\bar A,Q}{}^{B}$.
This makes it explicit that, once the symmetrization between the spin index $Q_1$ and the orbital index $Q_2$ is handled correctly, the triplet $J=1$ basis element again decomposes into the three massive polarization vectors with coefficients built from the appropriate spherical harmonics and spin tensors.

\paragraph{Vector operator with general spin external particles}
The same decomposition persists more generally. Still in the $(1/2,1/2)$ representation, the general $J=1$ basis in the direct $LS$ coupling form for arbitrary $s_1$, $s_3$, and $S$ can be written as
\begin{equation}
\begin{aligned}
&\mathcal{M}_{(S,L,J)}^\mu\sim\mathcal S_{\{I\}}^{\{A\}}{}_{\{K\}}\, \mathcal Y^{(L)}_{\{N\}}\,
C^{S,\{K\};\,L,\{N\}}_{J,Q_1Q_2}\,
|P^{(Q_1}\rangle_\alpha [P^{Q_2)}|_{\dot{\beta}} \\
=\;& \mathcal S_{\{I\}}^{\{A\}}{}_{\{K\}}\, \mathcal Y^{(L)}_{\{N\}}
\left[
\begin{aligned}[t]
&C^{S,\{K\};\,L,\{N\}}_{J,11}\, |P^1\rangle_\alpha [P^1|_{\dot{\beta}}
+ C^{S,\{K\};\,L,\{N\}}_{J,12}\, |P^{(1}\rangle_\alpha [P^{2)}|_{\dot{\beta}} \\
&+ C^{S,\{K\};\,L,\{N\}}_{J,22}\, |P^2\rangle_\alpha [P^2|_{\dot{\beta}}
\end{aligned}
\right] \\
\end{aligned}
\end{equation}
Here one may identify $|P^1\rangle_\alpha [P^1|_{\dot{\beta}} \sim e_{-1}^\mu$, $|P^{(1}\rangle_\alpha [P^{2)}|_{\dot{\beta}} \sim e_{0}^\mu$, and $|P^2\rangle_\alpha [P^2|_{\dot{\beta}} \sim e_{+1}^\mu$ in Breit frame.
In this general $J=1$ result, one may likewise rewrite the spin factor using the lowered index notation for particle 3 as
\begin{equation}
\begin{aligned}
&\mathcal S_{\{I\}\{\bar A\}}{}_{\{K\}}\, \mathcal Y^{(L)}_{\{N\}}\,
C^{S,\{K\};\,L,\{N\}}_{J,Q_1Q_2}\,
|P^{(Q_1}\rangle_\alpha [P^{Q_2)}|_{\dot{\beta}} \\
\end{aligned}
\end{equation}
Similarly, the expression can again be identified with the three polarization vectors. This shows that the $J=1$ basis elements reproduce the multipole expansion of the vector current discussed in section~\ref{sec:Non-relativistic multipole expansion}. With $\mathcal S_{\{I\}\{\bar A\}}{}_{\{K\}}\sim C^{S,\{K\}}_{s_1,\{I\};\,s_3,\{\bar A\}}$, this gives the general $J=1$ basis in the direct $LS$ coupling form. Throughout this expression, $J=1$ is understood, and the result makes it clear that, for arbitrary spin content in the external legs, the $J=1$ basis element is always a linear combination of the three massive polarization vectors weighted by the spin structure, the orbital tensor, and the corresponding CGCs.
A complementary simplification occurs directly in the spin tensor. Using the same symmetrized shorthand as in the canonical-spinor amplitude, one may write for arbitrary $s_1$ and $s_3$
\begin{align}
    \mathcal S_{\{I\}}^{\{A\}}{}_{\{K\}}
    &\sim
    (\tau_3^{2s_3})^{\{A\}\{B\}}
    (\tau_1^{2s_1})_{\{I\}}{}^{\{C\}}
    C^{S,\{K\}}_{s_1,\{C\};\,s_3,\{B\}}, \\
    &\sim
    \epsilon^{2s_3,\{A\}\{B\}}
    C^{S,\{K\}}_{s_1,\{I\};\,s_3,\{B\}},
\end{align}
where $\epsilon^{2s_3,\{A\}\{B\}}$ denotes the symmetrized product of $2s_3$ little group epsilon tensors. If one further rewrites particle 3 with lowered little group indices, this epsilon tensor simply converts the raised multiindex $\{A\}$ into the corresponding lowered multiindex $\{\bar{A}\}$, obtained by exchanging $1\leftrightarrow 2$ on each little group index. In this lowered index rest frame description, the spin tensor takes the CGC form
\begin{equation}
    \mathcal S_{\{I\}}^{\{A\}}{}_{\{K\}}
    \sim
    C^{S,\{K\}}_{s_1,\{I\};\,s_3,\{\bar{A}\}}.
\end{equation}
In other words, in the lowered index notation introduced at the beginning of this subsection, the same statement is
\begin{equation}
    \mathcal S_{\{I\}\{\bar A\}}{}_{\{K\}}
    \sim
    C^{S,\{K\}}_{s_1,\{I\};\,s_3,\{\bar A\}}.
\end{equation}
\subsection{Comparison with canonical $LS$ and multipole methods}

For direct comparison with Eqs.~\eqref{eq:scalar_amp_spinhalf} and \eqref{eq:vector_spherical_to_cartesian}, it is useful to reorganize the basis elements in a more compact multipole form. After factoring out the unique $J=0$ wave function $P^\mu$, the remaining coefficient may be written as
\begin{equation}
    \mathcal M_{(0)}(P;\{I\},\{\bar A\})
    \sim
    F_{00}^{(0)}(P^2)\,\mathcal Y^{(0)}_0\,\mathcal S^{(0)}_{\{I\}\{\bar A\}}
    +
    F_{11}^{(0)}(P^2)\,
    C_{0}^{1,\{B\};\,1,\{C\}}\,
    \mathcal Y^{(1)}_{\{B\}}\,
    \mathcal S^{(1)}_{\{I\}\{\bar A\}}{}_{\{C\}}.
\end{equation}
The first term here is exactly the $(S,L,J)=(0,0,0)$ basis written above, while the second term is exactly the $(S,L,J)=(1,1,0)$ basis written above. Obviously this formula contains exactly the same information as Eq.~\eqref{eq:scalar_amp_spinhalf}.
    In details, for the spin-$\tfrac{1}{2}$ example, it is easy to get the following relationship between the corresponding object in traditional form of Eqs.~\eqref{eq:spin_tensor_cg_relation} and \eqref{eq:vector_spherical_to_cartesian} and the spinor form used here:
\begin{align}
    \mathcal Y^{(0)}_0 &\sim Y^0_0(\hat q), \notag\\
    \mathcal Y^{(1)}_{11} &\sim Y^1_{-1}(\hat q), \qquad
    \mathcal Y^{(1)}_{12} \sim Y^1_0(\hat q), \qquad
    \mathcal Y^{(1)}_{22} \sim Y^1_{+1}(\hat q), \notag\\
    \mathcal S^{(0)}_{\{I\}\{\bar A\}}
    &\sim C^{0}_{\frac{1}{2},I;\,\frac{1}{2},\bar A}
    \sim \Sigma^{(0)}_0(a_1,a_3)
    \sim \delta_{a_1a_3}, \notag\\
    \mathcal S^{(1)}_{\{I\}\{\bar A\}}{}_{\{C\}}
    &\sim C^{1,a_S}_{\frac{1}{2},I;\,\frac{1}{2},\bar A}
    \sim \Sigma^{(1)}_{a_S}(a_1,a_3)
    \sim (\sigma_{a_S})_{a_1a_3}, \qquad a_S=0,\pm1, \notag\\
    C_{0}^{1,\{B\};\,1,\{C\}}
    &\sim C^{0,0}_{1,m;1,a_S}, \qquad
    C_{1,(Q_1Q_2)}^{L,\{B\};\,S,\{C\}}
    \sim C^{1,M}_{L,m;S,a_S}.
\end{align}
Furthermore, we have
\begin{align}
    C^{1,11A}_{\frac{1}{2},I;\,\frac{1}{2}}
    &\sim \epsilon^{A(1}\delta^{1)}_I
    \sim \frac{1}{\sqrt{2}}(\sigma^x-i\sigma^y)
    = \sigma_{-1}
    \sim \Sigma^{(1)}_{-1}(a_1,a_3), \notag\\
    C^{1,22A}_{\frac{1}{2},I;\,\frac{1}{2}}
    &\sim \epsilon^{A(2}\delta^{2)}_I
    \sim \frac{1}{\sqrt{2}}(\sigma^x+i\sigma^y)
    = \sigma_{+1}
    \sim \Sigma^{(1)}_{+1}(a_1,a_3), \notag\\
    C^{1,12A}_{\frac{1}{2},I;\,\frac{1}{2}}
    &\sim \epsilon^{A(1}\delta^{2)}_I
    \sim \sigma^z
    = \sigma_0
    \sim \Sigma^{(1)}_0(a_1,a_3).
\end{align}
Using the relationships above, we get the exact corresponding relationship
\begin{align}
    F_{00}^{(0)}(P^2)\,\mathcal Y^{(0)}_0\,\mathcal S^{(0)}_{\{I\}\{\bar A\}}
    &\sim F_{00}^{(0)}(q^2)\,Y^0_0(\hat q)\,\Sigma^{(0)}_0(a_1,a_3)
    \sim
    f_{(0)}(q^2)\,\delta_{a_1a_3}\,,
    \\
    F_{11}^{(0)}(P^2)\,
    C_{0}^{1,\{B\};\,1,\{C\}}\,
    \mathcal Y^{(1)}_{\{B\}}\,
    \mathcal S^{(1)}_{\{I\}\{\bar A\}}{}_{\{C\}}
    &\sim
    F_{11}^{(0)}(q^2)\sum_{m,a_S}
    C^{0,0}_{1,m;1,a_S}Y^1_m(\hat q)\,\Sigma^{(1)}_{a_S}(a_1,a_3)\,,
    \notag\\
    &\sim
    g_{(0)}(q^2)\,(\boldsymbol{\sigma}\cdot\hat q)_{a_1a_3}.
\end{align}
The first line is exactly the $(S,L,J)=(0,0,0)$ basis written above, while the second line is exactly the $(S,L,J)=(1,1,0)$ basis written above.

Similarly, the same $J=1$ content may be reorganized as
\begin{equation}
\begin{aligned}
    \mathcal M^\mu_{(1)}(P;\{I\},\{\bar A\})
    \sim\;&
    \sum_{L,S}
    F_{LS}^{(1)}(P^2)\,
    \mathcal A^{(L,S)}_{\{I\}\{\bar A\}(Q_1Q_2)}\,
    |P^{(Q_1}\rangle_\alpha [P^{Q_2)}|_{\dot\beta} \\
    \sim\;&
    \sum_{L,S}
    F_{LS}^{(1)}(P^2)\,
    \mathcal Y^{(L)}_{\{B\}}\,
    \mathcal S^{(S)}_{\{I\}\{\bar A\}}{}_{\{C\}}\,
    C_{1,(Q_1Q_2)}^{L,\{B\};\,S,\{C\}}\,
    |P^{(Q_1}\rangle_\alpha [P^{Q_2)}|_{\dot\beta} \,,
\end{aligned}
\end{equation}
where $\mathcal A^{(L,S)}_{\{I\}\{\bar A\}(Q_1Q_2)}$ denotes the corresponding amplitude written in a compact form. In Breit frame, using $|P^1\rangle_\alpha [P^1|_{\dot\beta} \sim e_{-1}^\mu$, $|P^{(1}\rangle_\alpha [P^{2)}|_{\dot\beta} \sim e_{0}^\mu$, and $|P^2\rangle_\alpha [P^2|_{\dot\beta} \sim e_{+1}^\mu$, the second line may be identified directly with the polarization-vector reconstruction in Eq.~\eqref{eq:vector_spherical_to_cartesian}. In parallel, comparison with Eq.~\eqref{eq:vector_spherical_to_cartesian} identifies $\mathcal S^{(S)}_{\{I\}\{\bar A\}}{}_{\{C\}}$ with $\Sigma^{(S)}_{a_S}(a_1,a_3)$, the CGC $C_{1,(Q_1Q_2)}^{L,\{B\};\,S,\{C\}}$ with $C^{1,M}_{L,m;S,a_S}$, and the orbital tensor $\mathcal Y^{(L)}_{\{B\}}$ with the spherical harmonic $Y^L_m(\hat q)$. Thus the present expression is strictly parallel to the earlier formulas, with the same orbital and spin angular momentum content rewritten in terms of little group indices in the spinor formalism. 

In details, we have the corresponding relationship
\begin{align}
    F_{10}^{(1)}(P^2)\,
    \mathcal A^{(1,0)}_{\{I\}\{\bar A\}(Q_1Q_2)}\,
    |P^{(Q_1}\rangle_\alpha [P^{Q_2)}|_{\dot\beta}
    &\sim
    F_{10}^{(1)}(q^2)\sum_{M,m}
    \bar e^i_M C^{1,M}_{1,m;0,0}Y^1_m(\hat q)\,\Sigma^{(0)}_0(a_1,a_3)
    \notag\\
    &\sim
    f_{(1)}(q^2)\,\hat q^i\,\delta_{a_1a_3},
    \\
    F_{01}^{(1)}(P^2)\,
    \mathcal A^{(0,1)}_{\{I\}\{\bar A\}(Q_1Q_2)}\,
    |P^{(Q_1}\rangle_\alpha [P^{Q_2)}|_{\dot\beta}
    &\sim
    F_{01}^{(1)}(q^2)\sum_{M,a_S}
    \bar e^i_M C^{1,M}_{0,0;1,a_S}Y^0_0(\hat q)\,\Sigma^{(1)}_{a_S}(a_1,a_3)
    \notag\\
    &\sim
    g_{(1)}(q^2)\,(\sigma^i)_{a_1a_3},
    \\
    F_{11}^{(1)}(P^2)\,
    \mathcal A^{(1,1)}_{\{I\}\{\bar A\}(Q_1Q_2)}\,
    |P^{(Q_1}\rangle_\alpha [P^{Q_2)}|_{\dot\beta}
    &\sim
    F_{11}^{(1)}(q^2)\sum_{M,m,a_S}
    \bar e^i_M C^{1,M}_{1,m;1,a_S}Y^1_m(\hat q)\,\Sigma^{(1)}_{a_S}(a_1,a_3)
    \notag\\
    &\sim
    h_{(1)}(q^2)\,\bigl(i(\boldsymbol{\sigma}\times\hat q)^i\bigr)_{a_1a_3},
    \\
    F_{21}^{(1)}(P^2)\,
    \mathcal A^{(2,1)}_{\{I\}\{\bar A\}(Q_1Q_2)}\,
    |P^{(Q_1}\rangle_\alpha [P^{Q_2)}|_{\dot\beta}
    &\sim
    F_{21}^{(1)}(q^2)\sum_{M,m,a_S}
    \bar e^i_M C^{1,M}_{2,m;1,a_S}Y^2_m(\hat q)\,\Sigma^{(1)}_{a_S}(a_1,a_3)
    \notag\\
    &\sim
    k_{(1)}(q^2)\,
    \bigl(3\hat q^i(\hat q\cdot\boldsymbol{\sigma})-\sigma^i\bigr)_{a_1a_3}.
\end{align}

These examples show explicitly how the direct $LS$ coupling form reproduces the $J=0$ scalar structures as well as the spatial $J=1$ multipole structures.

\section{General spin construction}
\label{sec:Covariant Bases from Method II}
\subsection{Form factor for general spin particles}

To get the complete bases with simpler forms, we adopt a new coupling way to construct the bases of general spin. Instead of organizing the basis through the auxiliary three point amplitude with spins $(s_1,s_3,J)$, it introduces an auxiliary scalar leg with $s_2=0$ and rewrites the bases in a different but equivalent coupling scheme. Each basis element again carries definite quantum numbers, now labeled by $(L,S,J,s_{p_3})$, and for every allowed choice of these labels there is exactly one corresponding basis element, so the basis is again nondegenerate.

In this coupling scheme, the auxiliary leg with outgoing momentum $P$ has spin $0$. If one keeps the same $LS$ language as in the direct $LS$ coupling form, the spin label $S$ is therefore identified with the intrinsic spin $s_1$ of particle 1. The angular momentum coupling is then reorganized as follows: one couples $L$ with $S=s_1$ to obtain the intermediate spin $s_{p_3}$ in the auxiliary amplitude, while equivalently the same $s_{p_3}$ is obtained by coupling the spin $s_3$ of particle 3 with the operator angular momentum $J$. This is precisely an angular momentum recoupling of the direct $LS$ coupling basis. Because one leg is scalar, the resulting auxiliary amplitudes often take a simpler form than in the direct $LS$ coupling form. For this reason, the explicit covariant bases are tabulated later in the alternative coupling form.

The map between the matrix element and the auxiliary amplitude closely parallels the direct $LS$ coupling form. We now define $T_{2\{I\}}^{\{A\}\{C\}\{D\}}$ by contracting the matrix element with the spinors of $p_3$:
\begin{equation}
    T_{2\{I\}}^{\{A\}\{C\}\{D\}} = \mathcal{M}_{\{I\}\,\{\alpha\},\,\{\dot\beta\}}^{\{A\}} \langle 3^{C_1} |^{\alpha_1} \dots \langle 3^{C_{2j_L}} |^{\alpha_{2j_L}} | 3^{D_1} ]^{\dot{\beta}_1} \dots | 3^{D_{2j_R}} ]^{\dot{\beta}_{2j_R}}.
\end{equation}
Inserting the completeness relation:
\begin{equation}
    \mathcal{M}_{\{I\}\,\{\alpha\},\,\{\dot\beta\}}^{\{A\}} \sim T_{2\{I\}}^{\{A\}\{C\}\{D\}} | 3_{\{C\}} \rangle^{2j_L}_{\{\alpha\}} [ 3_{\{D\}} |^{2j_R}_{\{\dot\beta\}}.
\end{equation}
Applying the CGC completeness relation, we get
\begin{align}
    \mathcal{M}_{\{I\}\,\{\alpha\},\,\{\dot\beta\}}^{\{A\}} &\sim \sum_{J, \{Q\}} T_{2\{I\}}^{\{A\}\{C\}\{D\}} C^{J, \{Q\}}_{j_L, \{C\}; j_R, \{D\}} \left( C_{J, \{Q\}}^{j_L, \{C'\}; j_R, \{D'\}} | 3_{\{C'\}} \rangle^{2j_L}_{\{\alpha\}} [ 3_{\{D'\}} |^{2j_R}_{\{\dot\beta\}} \right) \nonumber \\ 
    &\sim \sum_{J, \{Q\}, \{Q'\}} T_{2\{I\}}^{\{A'\}\{C\}\{D\}} C^{J, \{Q'\}}_{j_L, \{C\}; j_R, \{D\}} \left( \sum_{s_{p_3}, \{E\}} C^{s_{p_3}, \{E\}}_{s_3, \{A'\}; J, \{Q'\}} C_{s_{p_3}, \{E\}}^{s_3, \{A\}; J, \{Q\}} \right) \nonumber \\
    &\quad \times \left( C_{J, \{Q\}}^{j_L, \{C'\}; j_R, \{D'\}} | 3_{\{C'\}} \rangle^{2j_L}_{\{\alpha\}} [ 3_{\{D'\}} |^{2j_R}_{\{\dot\beta\}} \right) \nonumber \\
    &\sim \sum_{J, \{Q\}, \{Q'\}, s_{p_3}, \{E\}} \left( T_{2\{I\}}^{\{A'\}\{C\}\{D\}} C^{J, \{Q'\}}_{j_L, \{C\}; j_R, \{D\}} C^{s_{p_3}, \{E\}}_{s_3, \{A'\}; J, \{Q'\}} \right) \nonumber \\
    &\quad \times C_{s_{p_3}, \{E\}}^{s_3, \{A\}; J, \{Q\}} u_{J, \{Q\}; \{\alpha\}, \{\dot\beta\}}(p_3).
\end{align}
This defines a new effective three point amplitude $\mathcal{A}_{\{I\}}^{\{E\}}$ representing the interaction of spins $s_1$ and $s_{p_3}$ with a scalar leg $-P$:
\begin{equation}
    \mathcal{A}_{\{I\}}^{\{E\}}(p_1, s_1; -P, 0; p_3, s_{p_3}) \equiv T_{2\{I\}}^{\{A'\}\{C\}\{D\}} C^{J, \{Q'\}}_{j_L, \{C\}; j_R, \{D\}} C^{s_{p_3}, \{E\}}_{s_3, \{A'\}; J, \{Q'\}}.
\end{equation}
Here the operator angular momentum $J$ ranges over $J \in \{ |j_L - j_R|, |j_L - j_R| + 1, \dots, j_L + j_R \}$. For each fixed $J$, the intermediate spin $s_{p_3}$ takes values
\begin{equation}
    s_{p_3} \in \{ |s_3 - J|, |s_3 - J| + 1, \dots, s_3 + J \}.
\end{equation}
Using the complete set of three point amplitudes $\{ \mathcal{A}_{\{I\}}^{\{E\}} \}$, the matrix element structures are completely given by
\begin{equation}
    \{ \mathcal{M}_{\{I\}\,\{\alpha\},\,\{\dot\beta\}}^{\{A\}} \} = \{ \mathcal{A}_{\{I\}}^{\{E\}} C_{s_{p_3}, \{E\}}^{s_3, \{A\}; J, \{Q\}} u_{J, \{Q\}; \{\alpha\}, \{\dot\beta\}}(p_3) \}.
\end{equation}
In this form, the underlying $LS$ labels are read directly from the amplitude. The orbital label is $L$, the spin label is $S=s_1$, and $J$ is the total angular momentum carried by the corresponding basis element. The alternative coupling form gives the same number of independent structures as the direct $LS$ coupling form because the two are different choices of angular momentum coupling within the same covariant $LS$ construction and span the same tensor space. Completeness and linear independence therefore follow from the same three point amplitude basis, merely written in a recoupled form. This is also why the alternative coupling form is especially convenient for explicit tabulation.

To illustrate this recoupled construction, let us again consider the Lorentz representation $(1/2,1/2)$ and take $s_1=s_3=1/2$. To begin with, in the alternative coupling form, the spin part is entirely carried by the basic tensor
\begin{equation}
    \mathcal S_I{}^A \equiv \tau_I{}^A \equiv (\langle \mathbf{1} \mathbf{3} \rangle - [\mathbf{1} \mathbf{3}])_I{}^A,
\end{equation}
while the $L=1$ orbital part is encoded in $\mathcal Y^{(1)}_{Q_1Q_2} \equiv (\langle \mathbf{3} | p_1 | \mathbf{3} ])_{Q_1Q_2}$.
For the illustrative low spin examples below, the tensor $\mathcal Y^{(1)}_{Q_1Q_2}$ is identified with the familiar $L=1$ spherical harmonic components, and we again keep the little group components explicit through the relations
\begin{equation}
    \mathcal Y^{(1)}_{11}\sim \mathcal Y^{(1)}_{-1},
    \qquad
    \mathcal Y^{(1)}_{12}\sim \mathcal Y^{(1)}_{0},
    \qquad
    \mathcal Y^{(1)}_{22}\sim \mathcal Y^{(1)}_{+1}.
\end{equation}
It is also convenient to introduce the mixed index version $\mathcal Y_Q{}^A \equiv \mathcal Y^{(1)}_{QB}\epsilon^{AB}$,
whose components are
\begin{equation}
    \mathcal Y_1{}^1\sim \mathcal Y^{(1)}_{0},
    \qquad
    \mathcal Y_1{}^2\sim \mathcal Y^{(1)}_{-1},
    \qquad
    \mathcal Y_2{}^1\sim \mathcal Y^{(1)}_{+1},
    \qquad
    \mathcal Y_2{}^2\sim \mathcal Y^{(1)}_{0}.
\end{equation}
We also use $|3_1\rangle_\alpha [3_1|_{\dot\beta}\sim e_{+1}^\mu$, $|3_{(1}\rangle_\alpha [3_{2)}|_{\dot\beta}\sim e_0^\mu$, and $|3_2\rangle_\alpha [3_2|_{\dot\beta}\sim e_{-1}^\mu$ when comparing with the conventional polarization basis.
The relevant structures can be read directly from Table~\ref{tab:spinhalf_irrep_half_half}.

We begin with the channel $L=1$, $J=0$, and $s_3=1/2$. The table gives the basis element
\begin{equation}
    \mathcal S_I{}^K \mathcal Y_K{}^A\, (p_3)_{\alpha\dot{\beta}}.
\end{equation}
Using the explicit $1,2$ components of $\mathcal Y_K{}^A$, this becomes
\begin{equation}
    \mathcal S_I{}^K \mathcal Y_K{}^A\, p_3^\mu
    \sim
    \Big[
    (\mathcal S_I{}^1 \mathcal Y_1{}^1 + \mathcal S_I{}^2 \mathcal Y_2{}^1) \delta_1^A
    +
    (\mathcal S_I{}^1 \mathcal Y_1{}^2 + \mathcal S_I{}^2 \mathcal Y_2{}^2) \delta_2^A
    \Big] p_3^\mu.
\end{equation}
The two terms satisfy the following equivalence relations.
\begin{equation}
    (\mathcal S_I{}^1 \mathcal Y_1{}^1 + \mathcal S_I{}^2 \mathcal Y_2{}^1)\delta_1^A p_3^\mu
    \sim
    (\mathcal S_I{}^1 \mathcal Y^{(1)}_{0} + \mathcal S_I{}^2 \mathcal Y^{(1)}_{+1}) \delta_1^A p_3^\mu,
    \qquad
    (\mathcal S_I{}^1 \mathcal Y_1{}^2 + \mathcal S_I{}^2 \mathcal Y_2{}^2)\delta_2^A p_3^\mu
    \sim
    (\mathcal S_I{}^1 \mathcal Y^{(1)}_{-1} + \mathcal S_I{}^2 \mathcal Y^{(1)}_{0}) \delta_2^A p_3^\mu.
\end{equation}
Accordingly, this channel again gives a purely timelike structure, now multiplied by the $L=1$ spherical harmonics.

Next consider the channel $L=1$, $J=1$, and $s_3=1/2$. From Table~\ref{tab:spinhalf_irrep_half_half}, the corresponding basis element is
\begin{equation}
    \mathcal S_I{}^K \mathcal Y_K{}^{(J_1}\epsilon^{J_2)A}
    |3_{J_1}\rangle_\alpha [3_{J_2}|_{\dot{\beta}}.
\end{equation}
Using $\epsilon^{12}=-1$, $\epsilon^{21}=1$, together with the explicit components of $\mathcal Y_K{}^A$, we may expand the symmetrized indices as
\begin{equation}
\begin{aligned}
&\mathcal S_I{}^K \mathcal Y_K{}^{(J_1}\epsilon^{J_2)A}
|3_{J_1}\rangle_\alpha [3_{J_2}|_{\dot{\beta}} \\
\sim\;&
\Big[
\bigl(\mathcal S_I{}^1 \mathcal Y_1{}^1 + \mathcal S_I{}^2 \mathcal Y_2{}^1\bigr)\delta_2^A
\Big] |3_1\rangle_\alpha [3_1|_{\dot{\beta}} \\
&-
\Big[
\bigl(\mathcal S_I{}^1 \mathcal Y_1{}^1 + \mathcal S_I{}^2 \mathcal Y_2{}^1\bigr)\delta_1^A
-
\bigl(\mathcal S_I{}^1 \mathcal Y_1{}^2 + \mathcal S_I{}^2 \mathcal Y_2{}^2\bigr)\delta_2^A
\Big]
|3_{(1}\rangle_\alpha [3_{2)}|_{\dot{\beta}} \\
&-
\Big[
\bigl(\mathcal S_I{}^1 \mathcal Y_1{}^2 + \mathcal S_I{}^2 \mathcal Y_2{}^2\bigr)\delta_1^A
\Big] |3_2\rangle_\alpha [3_2|_{\dot{\beta}} .
\end{aligned}
\end{equation}
The corresponding relationship are
\begin{equation}
\begin{aligned}
&\Big[
\bigl(\mathcal S_I{}^1 \mathcal Y_1{}^1 + \mathcal S_I{}^2 \mathcal Y_2{}^1\bigr)\delta_2^A
\Big] |3_1\rangle_\alpha [3_1|_{\dot{\beta}}
\sim
\Big[
\bigl(\mathcal S_I{}^1 \mathcal Y^{(1)}_{0} + \mathcal S_I{}^2 \mathcal Y^{(1)}_{+1}\bigr)\delta_2^A
\Big] e_{+1}^\mu, \\
&\Big[
\bigl(\mathcal S_I{}^1 \mathcal Y_1{}^1 + \mathcal S_I{}^2 \mathcal Y_2{}^1\bigr)\delta_1^A
-
\bigl(\mathcal S_I{}^1 \mathcal Y_1{}^2 + \mathcal S_I{}^2 \mathcal Y_2{}^2\bigr)\delta_2^A
\Big] |3_{(1}\rangle_\alpha [3_{2)}|_{\dot{\beta}} \\
&\hspace{1.5cm}\sim
\Big[
\bigl(\mathcal S_I{}^1 \mathcal Y^{(1)}_{0} + \mathcal S_I{}^2 \mathcal Y^{(1)}_{+1}\bigr)\delta_1^A
-
\bigl(\mathcal S_I{}^1 \mathcal Y^{(1)}_{-1} + \mathcal S_I{}^2 \mathcal Y^{(1)}_{0}\bigr)\delta_2^A
\Big] e_0^\mu, \\
&\Big[
\bigl(\mathcal S_I{}^1 \mathcal Y_1{}^2 + \mathcal S_I{}^2 \mathcal Y_2{}^2\bigr)\delta_1^A
\Big] |3_2\rangle_\alpha [3_2|_{\dot{\beta}}
\sim
\Big[
\bigl(\mathcal S_I{}^1 \mathcal Y^{(1)}_{-1} + \mathcal S_I{}^2 \mathcal Y^{(1)}_{0}\bigr)\delta_1^A
\Big] e_{-1}^\mu .
\end{aligned}
\end{equation}
Thus the $s_{p_3}=1/2$ channel decomposes into the three spatial polarization vectors, with coefficients determined by the spin factor $\mathcal S_I{}^A$ and the rank one spherical harmonics.

Finally, for $L=1$, $J=1$, and $s_{p_3}=3/2$, the table gives $\mathcal S_I{}^{(J_1} \mathcal Y^{J_2A)} |3_{J_1}\rangle_\alpha [3_{J_2}|_{\dot{\beta}}$,
where $\mathcal Y^{AB}\equiv (\langle \mathbf{3} | p_1 | \mathbf{3} ])^{AB}$ and
\begin{equation}
    \mathcal Y^{11}\sim \mathcal Y^{(1)}_{-1},
    \qquad
    \mathcal Y^{12}\sim \mathcal Y^{(1)}_{0},
    \qquad
    \mathcal Y^{22}\sim \mathcal Y^{(1)}_{+1}.
\end{equation}
Expanding the symmetrized indices gives
\begin{equation}
\begin{aligned}
&\mathcal S_I{}^{(J_1} \mathcal Y^{J_2A)}
|3_{J_1}\rangle_\alpha [3_{J_2}|_{\dot{\beta}} \\
\sim\;&
\Big[
\mathcal S_I{}^1\bigl(\mathcal Y^{11}\delta_1^A + \mathcal Y^{12}\delta_2^A\bigr)
\Big] |3_1\rangle_\alpha [3_1|_{\dot{\beta}} \\
&+
\Big[
\bigl(\mathcal S_I{}^1 \mathcal Y^{12} + \mathcal S_I{}^2 \mathcal Y^{11}\bigr)\delta_1^A
+
\bigl(\mathcal S_I{}^1 \mathcal Y^{22} + \mathcal S_I{}^2 \mathcal Y^{12}\bigr)\delta_2^A
\Big]
|3_{(1}\rangle_\alpha [3_{2)}|_{\dot{\beta}} \\
&+
\Big[
\mathcal S_I{}^2\bigl(\mathcal Y^{12}\delta_1^A + \mathcal Y^{22}\delta_2^A\bigr)
\Big] |3_2\rangle_\alpha [3_2|_{\dot{\beta}} .
\end{aligned}
\end{equation}
Using $\mathcal Y^{11}\sim \mathcal Y^{(1)}_{-1}$, $\mathcal Y^{12}\sim \mathcal Y^{(1)}_{0}$, and $\mathcal Y^{22}\sim \mathcal Y^{(1)}_{+1}$, the terms appearing here satisfy the following equivalence relations.
\begin{equation}
\begin{aligned}
&\Big[
\mathcal S_I{}^1\bigl(\mathcal Y^{11}\delta_1^A + \mathcal Y^{12}\delta_2^A\bigr)
\Big] |3_1\rangle_\alpha [3_1|_{\dot{\beta}}
\sim
\Big[
\mathcal S_I{}^1\bigl(\mathcal Y^{(1)}_{-1}\delta_1^A + \mathcal Y^{(1)}_{0}\delta_2^A\bigr)
\Big] e_{+1}^\mu, \\
&\Big[
\bigl(\mathcal S_I{}^1 \mathcal Y^{12} + \mathcal S_I{}^2 \mathcal Y^{11}\bigr)\delta_1^A
+
\bigl(\mathcal S_I{}^1 \mathcal Y^{22} + \mathcal S_I{}^2 \mathcal Y^{12}\bigr)\delta_2^A
\Big] |3_{(1}\rangle_\alpha [3_{2)}|_{\dot{\beta}} \\
&\hspace{1.5cm}\sim
\Big[
\bigl(\mathcal S_I{}^1 \mathcal Y^{(1)}_{0} + \mathcal S_I{}^2 \mathcal Y^{(1)}_{-1}\bigr)\delta_1^A
+
\bigl(\mathcal S_I{}^1 \mathcal Y^{(1)}_{+1} + \mathcal S_I{}^2 \mathcal Y^{(1)}_{0}\bigr)\delta_2^A
\Big] e_0^\mu, \\
&\Big[
\mathcal S_I{}^2\bigl(\mathcal Y^{12}\delta_1^A + \mathcal Y^{22}\delta_2^A\bigr)
\Big] |3_2\rangle_\alpha [3_2|_{\dot{\beta}}
\sim
\Big[
\mathcal S_I{}^2\bigl(\mathcal Y^{(1)}_{0}\delta_1^A + \mathcal Y^{(1)}_{+1}\delta_2^A\bigr)
\Big] e_{-1}^\mu .
\end{aligned}
\end{equation}
Therefore, the $s_{p_3}=3/2$ case has a similar vector decomposition pattern, but its coefficients now belong to the fully symmetrized spin-$3/2$ little group structure. These examples make it clear that the alternative coupling form organizes the basis elements through spherical harmonic coefficients multiplying either the vector $p_3^\mu$ or the polarization vectors $e_{+1}^\mu$, $e_0^\mu$, and $e_{-1}^\mu$.

The same recoupled construction also gives the following compact parametrizations for the scalar
part with $J=0$ and the spatial vector part with $J=1$ of the matrix elements:
\begin{equation}
    \mathcal M_{(0)}(p_1;\{I\},\{\bar A\})
    \sim
    F_{00}^{(0)}(p_1^2)\,\mathcal Y^{(0)}_0\,\mathcal S^{(0)}_{\{I\}\{\bar A\}}
    +
    F_{11}^{(0)}(p_1^2)\,
    C_{0}^{1,\{B\};\,1,\{C\}}\,
    \mathcal Y^{(1)}_{\{B\}}\,
    \mathcal S^{(1)}_{\{I\}\{\bar A\}}{}_{\{C\}} .
\end{equation}

\begin{equation}
\begin{aligned}
    \mathcal M^\mu_{(1)}(p_1;\{I\},\{\bar A\})
    \sim\;&
    \sum_{L,S}
    F_{LS}^{(1)}(p_1^2)\,
    \mathcal A^{(L,S)}_{\{I\}\{\bar A\}(Q_1Q_2)}\,
    |3^{(Q_1}\rangle_\alpha [3^{Q_2)}|_{\dot\beta}
    \\
    \sim\;&
    \sum_{L,S}
    F_{LS}^{(1)}(p_1^2)\,
    \mathcal Y^{(L)}_{\{B\}}\,
    \mathcal S^{(S)}_{\{I\}\{\bar A\}}{}_{\{C\}}\,
    C_{1,(Q_1Q_2)}^{L,\{B\};\,S,\{C\}}\,
    |3^{(Q_1}\rangle_\alpha [3^{Q_2)}|_{\dot\beta} .
\end{aligned}
\end{equation}
Here $\mathcal A^{(L,S)}_{\{I\}\{\bar A\}(Q_1Q_2)}$ again denotes the corresponding amplitude written in a compact form. Using the identifications $|3_1\rangle_\alpha [3_1|_{\dot\beta}\sim e_{+1}^\mu$, $|3_{(1}\rangle_\alpha [3_{2)}|_{\dot\beta}\sim e_0^\mu$, and $|3_2\rangle_\alpha [3_2|_{\dot\beta}\sim e_{-1}^\mu$, the second line may likewise be identified with the polarization-vector form.

\subsection{Counting}

The same covariant $LS$ construction admits two useful choices of angular momentum coupling:
\begin{itemize}
    \item In the \textbf{direct $LS$ coupling form}, the amplitude requires the spin $S$ obtained by coupling particles $1$ and $3$ to satisfy
    \begin{equation}
        |s_1-s_3| \leq S \leq s_1+s_3 .
    \end{equation}
    The orbital angular momentum $L$ then couples with $S$ to give an $\mathrm{SO}(3)$ angular momentum $J$ obtained by restricting the Lorentz representation of the operator to the rotation subgroup, so that
    \begin{equation}
        |L-S| \leq J \leq L+S .
    \end{equation}

    \item In the \textbf{alternative coupling form}, the amplitude includes an auxiliary scalar leg with $s_2=0$. The spin label $s_{p_3}$ is obtained by coupling the $\mathrm{SO}(3)$ angular momentum $J$ of the operator with the spin $s_3$ of particle $3$. The coupling constraints are therefore reorganized as
    \begin{equation}
        |s_1-s_{p_3}| \leq L \leq s_1+s_{p_3},
        \qquad
        |s_3-J| \leq s_{p_3} \leq s_3+J .
    \end{equation}
\end{itemize}

The counting can be performed most directly in the COM frame using
the direct $LS$ coupling form. For an operator transforming in the Lorentz representation
$(j_L,j_R)$, the possible $\mathrm{SO}(3)$ angular momenta appearing after restricting to the rotation subgroup are
\begin{equation}
    |j_L-j_R| \leq J \leq j_L+j_R .
\end{equation}
For fixed $J$ and $S$, the allowed values of the orbital angular momentum are equivalently
\begin{equation}
    |J-S| \leq L \leq J+S .
\end{equation}
Hence the number of independent basis elements is
\begin{equation}
\begin{aligned}
    N(s_1,s_3;j_L,j_R)
    &=
    \sum_{J=|j_L-j_R|}^{j_L+j_R}
    \sum_{S=|s_1-s_3|}^{s_1+s_3}
    \sum_{L=|J-S|}^{J+S} 1
    \\[2mm]
    &=
    \sum_{J=|j_L-j_R|}^{j_L+j_R}
    \sum_{S=|s_1-s_3|}^{s_1+s_3}
    \bigl(2\min(S,J)+1\bigr)
    \\[2mm]
    &=
    \sum_{J=|j_L-j_R|}^{j_L+j_R}
    \left[
        2\min(s_1,s_3)+1
        +2\sum_{S=|s_1-s_3|}^{s_1+s_3}\min(S,J)
    \right]
    \\[2mm]
    &=
    \bigl(2\min(s_1,s_3)+1\bigr)
    \bigl(2\min(j_L,j_R)+1\bigr)
    \\
    &\hspace{1.5cm}
    +2
    \sum_{J=|j_L-j_R|}^{j_L+j_R}
    \sum_{S=|s_1-s_3|}^{s_1+s_3}
    \min(S,J) .
\end{aligned}
\label{eq:method-I-COM-counting}
\end{equation}

Although this counting formula is derived in the COM frame using the direct $LS$ coupling form, the alternative coupling form gives the same total number of independent structures. This equality follows because the two are different choices of angular momentum coupling within the same covariant $LS$ construction and span the same tensor space. Here and in the tables below, the numbers refer to the basis before imposing $P$ or $T$ constraints. If one further imposes $P$ and $T$ conservation, the corresponding numbers can be obtained by applying the nonrelativistic counting method of Ref.~\cite{Sun:2026wqa}.

To make this counting more explicit, we now evaluate the formula for the cases used in the tables below. For equal external spins $s_1=s_3=\tfrac{1}{2}$ and $s_1=s_3=1$, the counting results for several low rank Lorentz representations are summarized in Tables~\ref{tab:counting-spin-half} and~\ref{tab:counting-spin-one}.
\begin{table}[htbp]
    \centering
    \renewcommand{\arraystretch}{1.2}
    \begin{tabular}{c c c}
        \hline
        Lorentz representation $(j_L,j_R)$ & Allowed $J$ & $N\!\left(\tfrac{1}{2},\tfrac{1}{2};j_L,j_R\right)$ \\
        \hline
        $(0,0)$ & $0$ & $2$ \\
        $\left(\tfrac{1}{2},\tfrac{1}{2}\right)$ & $0,1$ & $6$ \\
        $(1,0)$ & $1$ & $4$ \\
        $(0,1)$ & $1$ & $4$ \\
        $(1,1)$ & $0,1,2$ & $10$ \\
        \hline
    \end{tabular}
    \caption{Counting of independent structures for spin-$\tfrac{1}{2}$ external particles.}
    \label{tab:counting-spin-half}
\end{table}

\begin{table}[htbp]
    \centering
    \renewcommand{\arraystretch}{1.2}
    \begin{tabular}{c c c}
        \hline
        Lorentz representation $(j_L,j_R)$ & Allowed $J$ & $N(1,1;j_L,j_R)$ \\
        \hline
        $(0,0)$ & $0$ & $3$ \\
        $\left(\tfrac{1}{2},\tfrac{1}{2}\right)$ & $0,1$ & $10$ \\
        $(1,0)$ & $1$ & $7$ \\
        $(0,1)$ & $1$ & $7$ \\
        $(1,1)$ & $0,1,2$ & $19$ \\
        \hline
    \end{tabular}
    \caption{Counting of independent structures for spin-$1$ external particles.}
    \label{tab:counting-spin-one}
\end{table}
\newpage

\subsection{Explicit covariant bases}
For practical applications, the following tables present the Lorentz covariant bases of the matrix element $\langle p_1,\{I\}|\hat O|p_3,\{A\}\rangle$ written in the alternative coupling form, together with the corresponding auxiliary three point amplitudes. This form is especially convenient here because it leads to simpler auxiliary three point amplitudes. Each row gives one independent structure. The first five columns list $s_1$, $L$, $s_3$, $J$, and $s_{p_3}$. Here $s_1$ and $s_3$ are the spins of particles 1 and 3, $L$ is the orbital label, $J$ is the $\mathrm{SO}(3)$ spin carried by the basis element, or equivalently by the operator after restriction to the rotation subgroup, and $s_{p_3}$ is the intermediate spin obtained by coupling $J$ with $s_3$. The last two columns give the auxiliary $LS$ amplitude and the resulting Lorentz covariant basis element. In all the following tables, $\{I\}$ denote the little group indices of particle 1 with spin $s_1$, while $\{A\}$ denote the little group indices of particle 3 with spin $s_3$.

For convenience, we first collect a compact summary of the main explicit results.
We summarize our main explicit results here in the notation of Ref.~\cite{Sun:2026wqa}, before presenting the detailed tables in the alternative coupling form. We denote the scalar operator by $\mathcal{O}$, the vector operator by $\mathcal{J}^\mu$, the antisymmetric tensor operator by $\mathcal{A}^{[\mu\nu]}$, the symmetric rank-2 operator by $\mathcal{O}^{(\mu\nu)}$, the general rank-2 operator by $\mathcal{O}^{\mu\nu}$, and the totally symmetric traceless rank-3 operator by $\mathcal{O}^{\{\mu\nu\rho\}}$. More generally, for arbitrary external spin and arbitrary operator representation $(j_L,j_R)$, we give the universal construction formula that generates the complete basis from the corresponding massive three point amplitude basis. For completeness, the general construction for arbitrary spin and representations is also presented in the appendix.

For spin-$\tfrac{1}{2}$, spin-$1$, and spin-$\tfrac{3}{2}$ external particles, the explicit results in the main text can be summarized as follows.

{\small
For spin-$\tfrac{1}{2}$ particles,
\begin{align*}
\langle p_1,\{I\}| \mathcal{O} |p_3,\{A\}\rangle
&= \bigl\{\text{basis elements in Table~\ref{tab:spinhalf_irrep_0_0}}\bigr\}_{\{I\}}^{\{A\}}, \\
\langle p_1,\{I\}| \mathcal{J}^\mu |p_3,\{A\}\rangle
&= \bigl\{\text{basis elements in Table~\ref{tab:spinhalf_irrep_half_half}}\bigr\}_{\{I\}}^{\{A\}}{}_{\alpha\dot{\beta}}
\bar{\sigma}^{\mu,\dot{\beta}\alpha}, \\
\langle p_1,\{I\}| \mathcal{A}^{[\mu\nu]} |p_3,\{A\}\rangle
&= \bigl\{\text{basis elements in Tables~\ref{tab:spinhalf_irrep_1_0} and \ref{tab:spinhalf_irrep_0_1}}\bigr\}_{\{I\}}^{\{A\}}{}_{\alpha_1\alpha_2\dot{\beta}_1\dot{\beta}_2}
\bar{\sigma}^{\mu,\dot{\beta}_1\alpha_1}\bar{\sigma}^{\nu,\dot{\beta}_2\alpha_2}, \\
\langle p_1,\{I\}| \mathcal{O}^{(\mu\nu)} |p_3,\{A\}\rangle
&= \bigl\{\text{basis elements in Table~\ref{tab:spinhalf_irrep_1_1}}\bigr\}_{\{I\}}^{\{A\}}{}_{\alpha_1\alpha_2\dot{\beta}_1\dot{\beta}_2}
\bar{\sigma}^{\mu,\dot{\beta}_1\alpha_1}\bar{\sigma}^{\nu,\dot{\beta}_2\alpha_2}
\qquad (\text{traceless terms}) \\
&\quad + g^{\mu\nu}\bigl\{\text{basis elements in Table~\ref{tab:spinhalf_irrep_0_0}}\bigr\}_{\{I\}}^{\{A\}}
\qquad (\text{trace terms}), \\
\langle p_1,\{I\}| \mathcal{O}^{\mu\nu} |p_3,\{A\}\rangle
&= \bigl\{\text{basis elements in Tables~\ref{tab:spinhalf_irrep_1_0} and \ref{tab:spinhalf_irrep_0_1}}\bigr\}_{\{I\}}^{\{A\}}{}_{\alpha_1\alpha_2\dot{\beta}_1\dot{\beta}_2}
\bar{\sigma}^{\mu,\dot{\beta}_1\alpha_1}\bar{\sigma}^{\nu,\dot{\beta}_2\alpha_2} \\
&\quad + \bigl\{\text{basis elements in Table~\ref{tab:spinhalf_irrep_1_1}}\bigr\}_{\{I\}}^{\{A\}}{}_{\alpha_1\alpha_2\dot{\beta}_1\dot{\beta}_2}
\bar{\sigma}^{\mu,\dot{\beta}_1\alpha_1}\bar{\sigma}^{\nu,\dot{\beta}_2\alpha_2}
\qquad (\text{traceless terms}) \\
&\quad + g^{\mu\nu}\bigl\{\text{basis elements in Table~\ref{tab:spinhalf_irrep_0_0}}\bigr\}_{\{I\}}^{\{A\}}
\qquad (\text{trace terms}), \\
\langle p_1,\{I\}| \mathcal{O}^{\{\mu\nu\rho\}} |p_3,\{A\}\rangle
&= \bigl\{\text{basis elements in Table~\ref{tab:spinhalf_irrep_3half_3half}}\bigr\}_{\{I\}}^{\{A\}}{}_{\alpha_1\alpha_2\alpha_3\dot{\beta}_1\dot{\beta}_2\dot{\beta}_3}
\bar{\sigma}^{\mu,\dot{\beta}_1\alpha_1}\bar{\sigma}^{\nu,\dot{\beta}_2\alpha_2}\bar{\sigma}^{\rho,\dot{\beta}_3\alpha_3} .
\end{align*}

For spin-$1$ particles,
\begin{align*}
\langle p_1,\{I\}| \mathcal{O} |p_3,\{A\}\rangle
&= \bigl\{\text{basis elements in Table~\ref{tab:spin1_irrep_0_0}}\bigr\}_{\{I\}}^{\{A\}}, \\
\langle p_1,\{I\}| \mathcal{J}^\mu |p_3,\{A\}\rangle
&= \bigl\{\text{basis elements in Table~\ref{tab:spin1_irrep_half_half}}\bigr\}_{\{I\}}^{\{A\}}{}_{\alpha\dot{\beta}}
\bar{\sigma}^{\mu,\dot{\beta}\alpha}, \\
\langle p_1,\{I\}| \mathcal{A}^{[\mu\nu]} |p_3,\{A\}\rangle
&= \bigl\{\text{basis elements in Tables~\ref{tab:spin1_irrep_1_0} and \ref{tab:spin1_irrep_0_1}}\bigr\}_{\{I\}}^{\{A\}}{}_{\alpha_1\alpha_2\dot{\beta}_1\dot{\beta}_2}
\bar{\sigma}^{\mu,\dot{\beta}_1\alpha_1}\bar{\sigma}^{\nu,\dot{\beta}_2\alpha_2}, \\
\langle p_1,\{I\}| \mathcal{O}^{(\mu\nu)} |p_3,\{A\}\rangle
&= \bigl\{\text{basis elements in Table~\ref{tab:spin1_irrep_1_1}}\bigr\}_{\{I\}}^{\{A\}}{}_{\alpha_1\alpha_2\dot{\beta}_1\dot{\beta}_2}
\bar{\sigma}^{\mu,\dot{\beta}_1\alpha_1}\bar{\sigma}^{\nu,\dot{\beta}_2\alpha_2}
\qquad (\text{traceless terms}) \\
&\quad + g^{\mu\nu}\bigl\{\text{basis elements in Table~\ref{tab:spin1_irrep_0_0}}\bigr\}_{\{I\}}^{\{A\}}
\qquad (\text{trace terms}), \\
\langle p_1,\{I\}| \mathcal{O}^{\mu\nu} |p_3,\{A\}\rangle
&= \bigl\{\text{basis elements in Tables~\ref{tab:spin1_irrep_1_0} and \ref{tab:spin1_irrep_0_1}}\bigr\}_{\{I\}}^{\{A\}}{}_{\alpha_1\alpha_2\dot{\beta}_1\dot{\beta}_2}
\bar{\sigma}^{\mu,\dot{\beta}_1\alpha_1}\bar{\sigma}^{\nu,\dot{\beta}_2\alpha_2} \\
&\quad + \bigl\{\text{basis elements in Table~\ref{tab:spin1_irrep_1_1}}\bigr\}_{\{I\}}^{\{A\}}{}_{\alpha_1\alpha_2\dot{\beta}_1\dot{\beta}_2}
\bar{\sigma}^{\mu,\dot{\beta}_1\alpha_1}\bar{\sigma}^{\nu,\dot{\beta}_2\alpha_2}
\qquad (\text{traceless terms}) \\
&\quad + g^{\mu\nu}\bigl\{\text{basis elements in Table~\ref{tab:spin1_irrep_0_0}}\bigr\}_{\{I\}}^{\{A\}}
\qquad (\text{trace terms}), \\
\langle p_1,\{I\}| \mathcal{O}^{\{\mu\nu\rho\}} |p_3,\{A\}\rangle
&= \bigl\{\text{basis elements in Table~\ref{tab:spin1_irrep_3half_3half}}\bigr\}_{\{I\}}^{\{A\}}{}_{\alpha_1\alpha_2\alpha_3\dot{\beta}_1\dot{\beta}_2\dot{\beta}_3}
\bar{\sigma}^{\mu,\dot{\beta}_1\alpha_1}\bar{\sigma}^{\nu,\dot{\beta}_2\alpha_2}\bar{\sigma}^{\rho,\dot{\beta}_3\alpha_3} .
\end{align*}

For spin-$\tfrac{3}{2}$ particles,
\begin{align*}
\langle p_1,\{I\}| \mathcal{O} |p_3,\{A\}\rangle
&= \bigl\{\text{basis elements in Table~\ref{tab:spin3half_irrep_0_0}}\bigr\}_{\{I\}}^{\{A\}}, \\
\langle p_1,\{I\}| \mathcal{J}^\mu |p_3,\{A\}\rangle
&= \bigl\{\text{basis elements in Table~\ref{tab:spin3half_irrep_half_half}}\bigr\}_{\{I\}}^{\{A\}}{}_{\alpha\dot{\beta}}
\bar{\sigma}^{\mu,\dot{\beta}\alpha}, \\
\langle p_1,\{I\}| \mathcal{A}^{[\mu\nu]} |p_3,\{A\}\rangle
&= \bigl\{\text{basis elements in Tables~\ref{tab:spin3half_irrep_1_0} and \ref{tab:spin3half_irrep_0_1}}\bigr\}_{\{I\}}^{\{A\}}{}_{\alpha_1\alpha_2\dot{\beta}_1\dot{\beta}_2}
\bar{\sigma}^{\mu,\dot{\beta}_1\alpha_1}\bar{\sigma}^{\nu,\dot{\beta}_2\alpha_2}, \\
\langle p_1,\{I\}| \mathcal{O}^{(\mu\nu)} |p_3,\{A\}\rangle
&= \bigl\{\text{basis elements in Table~\ref{tab:spin3half_irrep_1_1}}\bigr\}_{\{I\}}^{\{A\}}{}_{\alpha_1\alpha_2\dot{\beta}_1\dot{\beta}_2}
\bar{\sigma}^{\mu,\dot{\beta}_1\alpha_1}\bar{\sigma}^{\nu,\dot{\beta}_2\alpha_2}
\qquad (\text{traceless terms}) \\
&\quad + g^{\mu\nu}\bigl\{\text{basis elements in Table~\ref{tab:spin3half_irrep_0_0}}\bigr\}_{\{I\}}^{\{A\}}
\qquad (\text{trace terms}), \\
\langle p_1,\{I\}| \mathcal{O}^{\mu\nu} |p_3,\{A\}\rangle
&= \bigl\{\text{basis elements in Tables~\ref{tab:spin3half_irrep_1_0} and \ref{tab:spin3half_irrep_0_1}}\bigr\}_{\{I\}}^{\{A\}}{}_{\alpha_1\alpha_2\dot{\beta}_1\dot{\beta}_2}
\bar{\sigma}^{\mu,\dot{\beta}_1\alpha_1}\bar{\sigma}^{\nu,\dot{\beta}_2\alpha_2} \\
&\quad + \bigl\{\text{basis elements in Table~\ref{tab:spin3half_irrep_1_1}}\bigr\}_{\{I\}}^{\{A\}}{}_{\alpha_1\alpha_2\dot{\beta}_1\dot{\beta}_2}
\bar{\sigma}^{\mu,\dot{\beta}_1\alpha_1}\bar{\sigma}^{\nu,\dot{\beta}_2\alpha_2}
\qquad (\text{traceless terms}) \\
&\quad + g^{\mu\nu}\bigl\{\text{basis elements in Table~\ref{tab:spin3half_irrep_0_0}}\bigr\}_{\{I\}}^{\{A\}}
\qquad (\text{trace terms}), \\
\langle p_1,\{I\}| \mathcal{O}^{\{\mu\nu\rho\}} |p_3,\{A\}\rangle
&= \bigl\{\text{basis elements in Table~\ref{tab:spin3half_irrep_3half_3half}}\bigr\}_{\{I\}}^{\{A\}}{}_{\alpha_1\alpha_2\alpha_3\dot{\beta}_1\dot{\beta}_2\dot{\beta}_3}
\bar{\sigma}^{\mu,\dot{\beta}_1\alpha_1}\bar{\sigma}^{\nu,\dot{\beta}_2\alpha_2}\bar{\sigma}^{\rho,\dot{\beta}_3\alpha_3} .
\end{align*}
}

Here the outer braces $\{\cdots\}$ denote linear combinations of the basis elements collected in the cited tables, with the corresponding FFs as coefficients. The index sets $\{I\}$ and $\{A\}$ are the little group indices describing the spins of particle 1 and particle 3, respectively. In the tables, the total spin of particle 1 is denoted by $s_1$, while the total spin of particle 3, namely the incoming particle, is denoted by $s_{p_3}$. The notation $\{\mu\nu\rho\}$ means complete symmetrization together with trace subtraction on the Lorentz indices. For the vector, rank-2, and rank-3 operators, the Lorentz indices are obtained from the corresponding spinor structures by contraction with $\bar{\sigma}^{\mu,\dot{\beta}\alpha}$, $\bar{\sigma}^{\mu,\dot{\beta}_1\alpha_1}\bar{\sigma}^{\nu,\dot{\beta}_2\alpha_2}$, and $\bar{\sigma}^{\mu,\dot{\beta}_1\alpha_1}\bar{\sigma}^{\nu,\dot{\beta}_2\alpha_2}\bar{\sigma}^{\rho,\dot{\beta}_3\alpha_3}$, respectively. In $\mathcal{O}^{(\mu\nu)}$, the term proportional to $g^{\mu\nu}$ gives the trace part, while the remaining term gives the traceless part. In $\mathcal{O}^{\mu\nu}$, the full basis is the sum of the scalar trace sector, the antisymmetric sector, and the symmetric traceless sector.

We now turn to the detailed tables in the alternative coupling form.

For spin-$\tfrac{1}{2}$ external particles and operators in the Lorentz representation $(0,0)$, the basis in the alternative coupling form is shown in Table~\ref{tab:spinhalf_irrep_0_0}.
{\renewcommand{\arraystretch}{2.35}
\setlength{\tabcolsep}{2pt}
\scriptsize
\begin{longtable}{|c|c|c|c|c|>{\centering\arraybackslash}p{4.5cm}|>{\centering\arraybackslash}p{7.5cm}|}
\caption{spin-$\frac{1}{2}$ particle \quad irrep $(0,0)$} \label{tab:spinhalf_irrep_0_0} \\
\hline
$s_1$ & $L$ & $s_3$ & $J$ & $s_{p_3}$ & amplitudes & basis elements \\
\hline
\endfirsthead
\hline
$s_1$ & $L$ & $s_3$ & $J$ & $s_{p_3}$ & amplitudes & basis elements \\
\hline
\endhead
$\frac{1}{2}$ & $0$ & $\frac{1}{2}$ & $0$ & $\frac{1}{2}$ & 
$(\langle \mathbf{1} \mathbf{3} \rangle - [\mathbf{1} \mathbf{3}]){}_I{}^A$ & 
$(\langle \mathbf{1} \mathbf{3} \rangle - [\mathbf{1} \mathbf{3}]){}_I{}^A$ \\
\hline
$\frac{1}{2}$ & $1$ & $\frac{1}{2}$ & $0$ & $\frac{1}{2}$ & 
$\begin{aligned} & (\langle \mathbf{1} \mathbf{3} \rangle - [\mathbf{1} \mathbf{3}]){}_I{}^K \\ & (\langle \mathbf{3} | p_1 | \mathbf{3} ])_K{}^A \end{aligned}$ & 
$\begin{aligned} & (\langle \mathbf{1} \mathbf{3} \rangle - [\mathbf{1} \mathbf{3}]){}_I{}^K \\ & (\langle \mathbf{3} | p_1 | \mathbf{3} ])_K{}^A \end{aligned}$ \\
\hline
\end{longtable}}

For spin-$\tfrac{1}{2}$ external particles and operators in the Lorentz representation $(\tfrac{1}{2},\tfrac{1}{2})$, the basis in the alternative coupling form is shown in Table~\ref{tab:spinhalf_irrep_half_half}.
{\renewcommand{\arraystretch}{2.35}
\setlength{\tabcolsep}{2pt}
\scriptsize
\begin{longtable}{|c|c|c|c|c|>{\centering\arraybackslash}p{4.5cm}|>{\centering\arraybackslash}p{7.5cm}|}
\caption{irrep: $(\frac{1}{2}, \frac{1}{2})$} \label{tab:spinhalf_irrep_half_half} \\
\hline
$s_1$ & $L$ & $s_3$ & $J$ & $s_{p_3}$ & amplitudes & basis elements \\
\hline
\endfirsthead
\hline
$s_1$ & $L$ & $s_3$ & $J$ & $s_{p_3}$ & amplitudes & basis elements \\
\hline
\endhead
$\frac{1}{2}$ & $0$ & $\frac{1}{2}$ & $0$ & $\frac{1}{2}$ & 
$(\langle \mathbf{1} \mathbf{3} \rangle - [\mathbf{1} \mathbf{3}]){}_I{}^A$ & 
$(\langle \mathbf{1} \mathbf{3} \rangle - [\mathbf{1} \mathbf{3}]){}_I{}^A (p_3)_{\alpha \dot{\beta}}$ \\
\hline
$\frac{1}{2}$ & $1$ & $\frac{1}{2}$ & $0$ & $\frac{1}{2}$ & 
$\begin{aligned} & (\langle \mathbf{1} \mathbf{3} \rangle - [\mathbf{1} \mathbf{3}]){}_I{}^K \\ & (\langle \mathbf{3} | p_1 | \mathbf{3} ])_K{}^A \end{aligned}$ & 
$\begin{aligned} & (\langle \mathbf{1} \mathbf{3} \rangle - [\mathbf{1} \mathbf{3}]){}_I{}^K \\ & (\langle \mathbf{3} | p_1 | \mathbf{3} ])_K{}^A (p_3)_{\alpha \dot{\beta}} \end{aligned}$ \\
\hline
$\frac{1}{2}$ & $0$ & $\frac{1}{2}$ & $1$ & $\frac{1}{2}$ & 
$(\langle \mathbf{1} \mathbf{3} \rangle - [\mathbf{1} \mathbf{3}]){}_I{}^A$ & 
$\begin{aligned} & (\langle \mathbf{1} \mathbf{3} \rangle - [\mathbf{1} \mathbf{3}]){}_I{}^{(J_1} \epsilon^{J_2) A} \\ & |\mathbf{3}_{J_1}\rangle_\alpha [\mathbf{3}_{J_2}|_{\dot{\beta}} \end{aligned}$ \\
\hline
$\frac{1}{2}$ & $1$ & $\frac{1}{2}$ & $1$ & $\frac{1}{2}$ & 
$\begin{aligned} & (\langle \mathbf{1} \mathbf{3} \rangle - [\mathbf{1} \mathbf{3}]){}_I{}^K \\ & (\langle \mathbf{3} | p_1 | \mathbf{3} ])_K{}^{J_1} \end{aligned}$ & 
$\begin{aligned} & (\langle \mathbf{1} \mathbf{3} \rangle - [\mathbf{1} \mathbf{3}]){}_I{}^K (\langle \mathbf{3} | p_1 | \mathbf{3} ])_K{}^{(J_1} \epsilon^{J_2) A} \\ & |\mathbf{3}_{J_1}\rangle_\alpha [\mathbf{3}_{J_2}|_{\dot{\beta}} \end{aligned}$ \\
\hline
$\frac{1}{2}$ & $1$ & $\frac{1}{2}$ & $1$ & $\frac{3}{2}$ & 
$\begin{aligned} & (\langle \mathbf{1} \mathbf{3} \rangle - [\mathbf{1} \mathbf{3}]){}_I{}^{(J_1} \\ & (\langle \mathbf{3} | p_1 | \mathbf{3} ])^{J_2 A)} \end{aligned}$ & 
$\begin{aligned} & (\langle \mathbf{1} \mathbf{3} \rangle - [\mathbf{1} \mathbf{3}]){}_I{}^{(J_1} (\langle \mathbf{3} | p_1 | \mathbf{3} ])^{J_2 A)} \\ & |\mathbf{3}_{J_1}\rangle_\alpha [\mathbf{3}_{J_2}|_{\dot{\beta}} \end{aligned}$ \\
\hline
\end{longtable}}

For spin-$\tfrac{1}{2}$ external particles and operators in the Lorentz representation $(1,1)$, the basis in the alternative coupling form is shown in Table~\ref{tab:spinhalf_irrep_1_1}.
{\renewcommand{\arraystretch}{2.35}
\setlength{\tabcolsep}{2pt}
\scriptsize
\begin{longtable}{|c|c|c|c|c|>{\centering\arraybackslash}p{4.5cm}|>{\centering\arraybackslash}p{7.5cm}|}
\caption{irrep: $(1, 1)$} \label{tab:spinhalf_irrep_1_1} \\
\hline
$s_1$ & $L$ & $s_3$ & $J$ & $s_{p_3}$ & amplitudes & basis elements \\
\hline
\endfirsthead
\hline
$s_1$ & $L$ & $s_3$ & $J$ & $s_{p_3}$ & amplitudes & basis elements \\
\hline
\endhead
$\frac{1}{2}$ & $0$ & $\frac{1}{2}$ & $0$ & $\frac{1}{2}$ & 
$(\langle \mathbf{1} \mathbf{3} \rangle - [\mathbf{1} \mathbf{3}]){}_I{}^A$ & 
$\begin{aligned} & (\langle \mathbf{1} \mathbf{3} \rangle - [\mathbf{1} \mathbf{3}]){}_I{}^A \\ & (p_3)^2_{(\alpha_1 \alpha_2), (\dot{\beta}_1 \dot{\beta}_2)} \end{aligned}$ \\
\hline
$\frac{1}{2}$ & $1$ & $\frac{1}{2}$ & $0$ & $\frac{1}{2}$ & 
$\begin{aligned} & (\langle \mathbf{1} \mathbf{3} \rangle - [\mathbf{1} \mathbf{3}]){}_I{}^K \\ & (\langle \mathbf{3} | p_1 | \mathbf{3} ])_K{}^A \end{aligned}$ & 
$\begin{aligned} & (\langle \mathbf{1} \mathbf{3} \rangle - [\mathbf{1} \mathbf{3}]){}_I{}^K (\langle \mathbf{3} | p_1 | \mathbf{3} ])_K{}^A \\ & (p_3)^2_{(\alpha_1 \alpha_2), (\dot{\beta}_1 \dot{\beta}_2)} \end{aligned}$ \\
\hline
$\frac{1}{2}$ & $0$ & $\frac{1}{2}$ & $1$ & $\frac{1}{2}$ & 
$(\langle \mathbf{1} \mathbf{3} \rangle - [\mathbf{1} \mathbf{3}]){}_I{}^A$ & 
$\begin{aligned} & (\langle \mathbf{1} \mathbf{3} \rangle - [\mathbf{1} \mathbf{3}]){}_I{}^{(J_1} \epsilon^{J_2) A} \\ & |\mathbf{3}_{J_1}\rangle_{(\alpha_1} (p_3)_{\alpha_2), (\dot{\beta}_1} [\mathbf{3}_{J_2}|_{\dot{\beta}_2)} \end{aligned}$ \\
\hline
$\frac{1}{2}$ & $1$ & $\frac{1}{2}$ & $1$ & $\frac{1}{2}$ & 
$\begin{aligned} & (\langle \mathbf{1} \mathbf{3} \rangle - [\mathbf{1} \mathbf{3}]){}_I{}^K \\ & (\langle \mathbf{3} | p_1 | \mathbf{3} ])_K{}^{J_1} \end{aligned}$ & 
$\begin{aligned} & (\langle \mathbf{1} \mathbf{3} \rangle - [\mathbf{1} \mathbf{3}]){}_I{}^K (\langle \mathbf{3} | p_1 | \mathbf{3} ])_K{}^{(J_1} \epsilon^{J_2) A} \\ & |\mathbf{3}_{J_1}\rangle_{(\alpha_1} (p_3)_{\alpha_2), (\dot{\beta}_1} [\mathbf{3}_{J_2}|_{\dot{\beta}_2)} \end{aligned}$ \\
\hline
\end{longtable}}

For spin-$\tfrac{1}{2}$ external particles and operators in the Lorentz representation $(1,0)$, the basis in the alternative coupling form is shown in Table~\ref{tab:spinhalf_irrep_1_0}.
{\renewcommand{\arraystretch}{2.35}
\setlength{\tabcolsep}{2pt}
\scriptsize
\begin{longtable}{|c|c|c|c|c|>{\centering\arraybackslash}p{4.5cm}|>{\centering\arraybackslash}p{7.5cm}|}
\caption{irrep: $(1, 0)$} \label{tab:spinhalf_irrep_1_0} \\
\hline
$s_1$ & $L$ & $s_3$ & $J$ & $s_{p_3}$ & amplitudes & basis elements \\
\hline
\endfirsthead
\hline
$s_1$ & $L$ & $s_3$ & $J$ & $s_{p_3}$ & amplitudes & basis elements \\
\hline
\endhead
$\frac{1}{2}$ & $0$ & $\frac{1}{2}$ & $1$ & $\frac{1}{2}$ & 
$(\langle \mathbf{1} \mathbf{3} \rangle - [\mathbf{1} \mathbf{3}]){}_I{}^A$ & 
$\begin{aligned} & (\langle \mathbf{1} \mathbf{3} \rangle - [\mathbf{1} \mathbf{3}]){}_I{}^{(J_1} \epsilon^{J_2) A} \\ & |\mathbf{3}_{J_1}\rangle_{\alpha_1} |\mathbf{3}_{J_2}\rangle_{\alpha_2} \epsilon_{\dot{\beta}_1 \dot{\beta}_2} \end{aligned}$ \\
\hline
$\frac{1}{2}$ & $1$ & $\frac{1}{2}$ & $1$ & $\frac{1}{2}$ & 
$\begin{aligned} & (\langle \mathbf{1} \mathbf{3} \rangle - [\mathbf{1} \mathbf{3}]){}_I{}^K \\ & (\langle \mathbf{3} | p_1 | \mathbf{3} ])_K{}^{J_1} \end{aligned}$ & 
$\begin{aligned} & (\langle \mathbf{1} \mathbf{3} \rangle - [\mathbf{1} \mathbf{3}]){}_I{}^K (\langle \mathbf{3} | p_1 | \mathbf{3} ])_K{}^{(J_1} \epsilon^{J_2) A} \\ & |\mathbf{3}_{J_1}\rangle_{\alpha_1} |\mathbf{3}_{J_2}\rangle_{\alpha_2} \epsilon_{\dot{\beta}_1 \dot{\beta}_2} \end{aligned}$ \\
\hline
$\frac{1}{2}$ & $1$ & $\frac{1}{2}$ & $1$ & $\frac{3}{2}$ & 
$\begin{aligned} & (\langle \mathbf{1} \mathbf{3} \rangle - [\mathbf{1} \mathbf{3}]){}_I{}^{(J_1} \\ & (\langle \mathbf{3} | p_1 | \mathbf{3} ])^{J_2 A)} \end{aligned}$ & 
$\begin{aligned} & (\langle \mathbf{1} \mathbf{3} \rangle - [\mathbf{1} \mathbf{3}]){}_I{}^{(J_1} (\langle \mathbf{3} | p_1 | \mathbf{3} ])^{J_2 A)} \\ & |\mathbf{3}_{J_1}\rangle_{\alpha_1} |\mathbf{3}_{J_2}\rangle_{\alpha_2} \epsilon_{\dot{\beta}_1 \dot{\beta}_2} \end{aligned}$ \\
\hline
$\frac{1}{2}$ & $2$ & $\frac{1}{2}$ & $1$ & $\frac{3}{2}$ & 
$\begin{aligned} & (\langle \mathbf{1} \mathbf{3} \rangle - [\mathbf{1} \mathbf{3}]){}_I{}^K \\ & (\langle \mathbf{3} | p_1 | \mathbf{3} ]^2)_K{}^{A J_1 J_2} \end{aligned}$ & 
$\begin{aligned} & (\langle \mathbf{1} \mathbf{3} \rangle - [\mathbf{1} \mathbf{3}]){}_I{}^K (\langle \mathbf{3} | p_1 | \mathbf{3} ]^2)_K{}^{(A J_1 J_2)} \\ & |\mathbf{3}_{J_1}\rangle_{\alpha_1} |\mathbf{3}_{J_2}\rangle_{\alpha_2} \epsilon_{\dot{\beta}_1 \dot{\beta}_2} \end{aligned}$ \\
\hline
\end{longtable}}

For spin-$\tfrac{1}{2}$ external particles and operators in the Lorentz representation $(0,1)$, the basis in the alternative coupling form is shown in Table~\ref{tab:spinhalf_irrep_0_1}.
{\renewcommand{\arraystretch}{2.35}
\setlength{\tabcolsep}{2pt}
\scriptsize
\begin{longtable}{|c|c|c|c|c|>{\centering\arraybackslash}p{4.5cm}|>{\centering\arraybackslash}p{7.5cm}|}
\caption{irrep: $(0, 1)$} \label{tab:spinhalf_irrep_0_1} \\
\hline
$s_1$ & $L$ & $s_3$ & $J$ & $s_{p_3}$ & amplitudes & basis elements \\
\hline
\endfirsthead
\hline
$s_1$ & $L$ & $s_3$ & $J$ & $s_{p_3}$ & amplitudes & basis elements \\
\hline
\endhead
$\frac{1}{2}$ & $0$ & $\frac{1}{2}$ & $1$ & $\frac{1}{2}$ & 
$(\langle \mathbf{1} \mathbf{3} \rangle - [\mathbf{1} \mathbf{3}]){}_I{}^A$ & 
$\begin{aligned} & (\langle \mathbf{1} \mathbf{3} \rangle - [\mathbf{1} \mathbf{3}]){}_I{}^{(J_1} \epsilon^{J_2) A} \\ & [\mathbf{3}_{J_1}|_{\dot{\beta}_1} [\mathbf{3}_{J_2}|_{\dot{\beta}_2} \epsilon_{\alpha_1 \alpha_2} \end{aligned}$ \\
\hline
$\frac{1}{2}$ & $1$ & $\frac{1}{2}$ & $1$ & $\frac{1}{2}$ & 
$\begin{aligned} & (\langle \mathbf{1} \mathbf{3} \rangle - [\mathbf{1} \mathbf{3}]){}_I{}^K \\ & (\langle \mathbf{3} | p_1 | \mathbf{3} ])_K{}^{J_1} \end{aligned}$ & 
$\begin{aligned} & (\langle \mathbf{1} \mathbf{3} \rangle - [\mathbf{1} \mathbf{3}]){}_I{}^K (\langle \mathbf{3} | p_1 | \mathbf{3} ])_K{}^{(J_1} \epsilon^{J_2) A} \\ & [\mathbf{3}_{J_1}|_{\dot{\beta}_1} [\mathbf{3}_{J_2}|_{\dot{\beta}_2} \epsilon_{\alpha_1 \alpha_2} \end{aligned}$ \\
\hline
$\frac{1}{2}$ & $1$ & $\frac{1}{2}$ & $1$ & $\frac{3}{2}$ & 
$\begin{aligned} & (\langle \mathbf{1} \mathbf{3} \rangle - [\mathbf{1} \mathbf{3}]){}_I{}^{(J_1} \\ & (\langle \mathbf{3} | p_1 | \mathbf{3} ])^{J_2 A)} \end{aligned}$ & 
$\begin{aligned} & (\langle \mathbf{1} \mathbf{3} \rangle - [\mathbf{1} \mathbf{3}]){}_I{}^{(J_1} (\langle \mathbf{3} | p_1 | \mathbf{3} ])^{J_2 A)} \\ & [\mathbf{3}_{J_1}|_{\dot{\beta}_1} [\mathbf{3}_{J_2}|_{\dot{\beta}_2} \epsilon_{\alpha_1 \alpha_2} \end{aligned}$ \\
\hline
$\frac{1}{2}$ & $2$ & $\frac{1}{2}$ & $1$ & $\frac{3}{2}$ & 
$\begin{aligned} & (\langle \mathbf{1} \mathbf{3} \rangle - [\mathbf{1} \mathbf{3}]){}_I{}^K \\ & (\langle \mathbf{3} | p_1 | \mathbf{3} ]^2)_K{}^{A J_1 J_2} \end{aligned}$ & 
$\begin{aligned} & (\langle \mathbf{1} \mathbf{3} \rangle - [\mathbf{1} \mathbf{3}]){}_I{}^K (\langle \mathbf{3} | p_1 | \mathbf{3} ]^2)_K{}^{(A J_1 J_2)} \\ & [\mathbf{3}_{J_1}|_{\dot{\beta}_1} [\mathbf{3}_{J_2}|_{\dot{\beta}_2} \epsilon_{\alpha_1 \alpha_2} \end{aligned}$ \\
\hline
\end{longtable}}

For spin-$\tfrac{1}{2}$ external particles and operators in the Lorentz representation $(\tfrac{3}{2},\tfrac{3}{2})$, the basis in the alternative coupling form is shown in Table~\ref{tab:spinhalf_irrep_3half_3half}.
{\renewcommand{\arraystretch}{2.35}
\setlength{\tabcolsep}{2pt}
\scriptsize
\begin{longtable}{|c|c|c|c|c|>{\centering\arraybackslash}p{4.5cm}|>{\centering\arraybackslash}p{7.5cm}|}
\caption{irrep: $(\frac{3}{2}, \frac{3}{2})$} \label{tab:spinhalf_irrep_3half_3half} \\
\hline
$s_1$ & $L$ & $s_3$ & $J$ & $s_{p_3}$ & amplitudes & basis elements \\
\hline
\endfirsthead
\hline
$s_1$ & $L$ & $s_3$ & $J$ & $s_{p_3}$ & amplitudes & basis elements \\
\hline
\endhead
$\frac{1}{2}$ & $0$ & $\frac{1}{2}$ & $0$ & $\frac{1}{2}$ & 
$(\langle \mathbf{1} \mathbf{3} \rangle - [\mathbf{1} \mathbf{3}]){}_I{}^A$ & 
$\begin{aligned} & (\langle \mathbf{1} \mathbf{3} \rangle - [\mathbf{1} \mathbf{3}]){}_I{}^A \\ & (p_3)^3_{(\alpha_1 \alpha_2 \alpha_3), (\dot{\beta}_1 \dot{\beta}_2 \dot{\beta}_3)} \end{aligned}$ \\
\hline
$\frac{1}{2}$ & $1$ & $\frac{1}{2}$ & $0$ & $\frac{1}{2}$ & 
$\begin{aligned} & (\langle \mathbf{1} \mathbf{3} \rangle - [\mathbf{1} \mathbf{3}]){}_I{}^K \\ & (\langle \mathbf{3} | p_1 | \mathbf{3} ])_K{}^A \end{aligned}$ & 
$\begin{aligned} & (\langle \mathbf{1} \mathbf{3} \rangle - [\mathbf{1} \mathbf{3}]){}_I{}^K (\langle \mathbf{3} | p_1 | \mathbf{3} ])_K{}^A \\ & (p_3)^3_{(\alpha_1 \alpha_2 \alpha_3), (\dot{\beta}_1 \dot{\beta}_2 \dot{\beta}_3)} \end{aligned}$ \\
\hline
$\frac{1}{2}$ & $0$ & $\frac{1}{2}$ & $1$ & $\frac{1}{2}$ & 
$(\langle \mathbf{1} \mathbf{3} \rangle - [\mathbf{1} \mathbf{3}]){}_I{}^A$ & 
$\begin{aligned} & (\langle \mathbf{1} \mathbf{3} \rangle - [\mathbf{1} \mathbf{3}]){}_I{}^{(J_1} \epsilon^{J_2) A} \\ & |\mathbf{3}_{J_1}\rangle_{(\alpha_1} (p_3)^2_{\alpha_2 \alpha_3), (\dot{\beta}_1 \dot{\beta}_2} [\mathbf{3}_{J_2}|_{\dot{\beta}_3)} \end{aligned}$ \\
\hline
$\frac{1}{2}$ & $1$ & $\frac{1}{2}$ & $1$ & $\frac{1}{2}$ & 
$\begin{aligned} & (\langle \mathbf{1} \mathbf{3} \rangle - [\mathbf{1} \mathbf{3}]){}_I{}^K \\ & (\langle \mathbf{3} | p_1 | \mathbf{3} ])_K{}^{J_1} \end{aligned}$ & 
$\begin{aligned} & (\langle \mathbf{1} \mathbf{3} \rangle - [\mathbf{1} \mathbf{3}]){}_I{}^K (\langle \mathbf{3} | p_1 | \mathbf{3} ])_K{}^{(J_1} \epsilon^{J_2) A} \\ & |\mathbf{3}_{J_1}\rangle_{(\alpha_1} (p_3)^2_{\alpha_2 \alpha_3), (\dot{\beta}_1 \dot{\beta}_2} [\mathbf{3}_{J_2}|_{\dot{\beta}_3)} \end{aligned}$ \\
\hline
\end{longtable}}

For spin-$1$ external particles and operators in the Lorentz representation $(0,0)$, the Method II basis is shown in Table~\ref{tab:spin1_irrep_0_0}.
{\renewcommand{\arraystretch}{2.35}
\setlength{\tabcolsep}{2pt}
\scriptsize
\begin{longtable}{|c|c|c|c|c|>{\centering\arraybackslash}p{4.5cm}|>{\centering\arraybackslash}p{7.5cm}|}
\caption{spin-$1$ particle \quad irrep $(0,0)$} \label{tab:spin1_irrep_0_0} \\
\hline
$s_1$ & $L$ & $s_3$ & $J$ & $s_{p_3}$ & amplitudes & basis elements \\
\hline
\endfirsthead
\hline
$s_1$ & $L$ & $s_3$ & $J$ & $s_{p_3}$ & amplitudes & basis elements \\
\hline
\endhead
$1$ & $0$ & $1$ & $0$ & $1$ & 
$\begin{aligned} & (\langle \mathbf{1} \mathbf{3} \rangle - [\mathbf{1} \mathbf{3}])^2{}_{I_1 I_2}{}^{A_1 A_2} \end{aligned}$ & 
$\begin{aligned} & (\langle \mathbf{1} \mathbf{3} \rangle - [\mathbf{1} \mathbf{3}])^2{}_{I_1 I_2}{}^{A_1 A_2} \end{aligned}$ \\
\hline
$1$ & $1$ & $1$ & $0$ & $1$ & 
$\begin{aligned} & (\langle \mathbf{1} \mathbf{3} \rangle - [\mathbf{1} \mathbf{3}])^2{}_{I_1 I_2}{}^{K(A_1} \\ & (\langle \mathbf{3} | p_1 | \mathbf{3} ])_K{}^{A_2)} \end{aligned}$ & 
$\begin{aligned} & (\langle \mathbf{1} \mathbf{3} \rangle - [\mathbf{1} \mathbf{3}])^2{}_{I_1 I_2}{}^{K(A_1} \\ & (\langle \mathbf{3} | p_1 | \mathbf{3} ])_K{}^{A_2)} \end{aligned}$ \\
\hline
$1$ & $2$ & $1$ & $0$ & $1$ & 
$\begin{aligned} & (\langle \mathbf{1} \mathbf{3} \rangle - [\mathbf{1} \mathbf{3}])^2{}_{I_1 I_2}{}^{K_1 K_2} \\ & (\langle \mathbf{3} | p_1 | \mathbf{3} ]^2)_{K_1 K_2}{}^{(A_1 A_2)} \end{aligned}$ & 
$\begin{aligned} & (\langle \mathbf{1} \mathbf{3} \rangle - [\mathbf{1} \mathbf{3}])^2{}_{I_1 I_2}{}^{K_1 K_2} \\ & (\langle \mathbf{3} | p_1 | \mathbf{3} ]^2)_{K_1 K_2}{}^{(A_1 A_2)} \end{aligned}$ \\
\hline
\end{longtable}}

For spin-$1$ external particles and operators in the Lorentz representation $(\tfrac{1}{2},\tfrac{1}{2})$, the Method II basis is shown in Table~\ref{tab:spin1_irrep_half_half}.
{\renewcommand{\arraystretch}{2.35}
\setlength{\tabcolsep}{2pt}
\scriptsize
\begin{longtable}{|c|c|c|c|c|>{\centering\arraybackslash}p{4.5cm}|>{\centering\arraybackslash}p{7.5cm}|}
\caption{irrep: $(\frac{1}{2}, \frac{1}{2})$} \label{tab:spin1_irrep_half_half} \\
\hline
$s_1$ & $L$ & $s_3$ & $J$ & $s_{p_3}$ & amplitudes & basis elements \\
\hline
\endfirsthead
\hline
$s_1$ & $L$ & $s_3$ & $J$ & $s_{p_3}$ & amplitudes & basis elements \\
\hline
\endhead
$1$ & $0$ & $1$ & $0$ & $1$ & 
$\begin{aligned} & (\langle \mathbf{1} \mathbf{3} \rangle - [\mathbf{1} \mathbf{3}])^2{}_{I_1 I_2}{}^{A_1 A_2} \end{aligned}$ & 
$\begin{aligned} & (\langle \mathbf{1} \mathbf{3} \rangle - [\mathbf{1} \mathbf{3}])^2{}_{I_1 I_2}{}^{(A_1 A_2)} \\ & (p_3)_{\alpha \dot{\beta}} \end{aligned}$ \\
\hline
$1$ & $1$ & $1$ & $0$ & $1$ & 
$\begin{aligned} & (\langle \mathbf{1} \mathbf{3} \rangle - [\mathbf{1} \mathbf{3}])^2{}_{I_1 I_2}{}^{K(A_1} \\ & (\langle \mathbf{3} | p_1 | \mathbf{3} ])_K{}^{A_2)} \end{aligned}$ & 
$\begin{aligned} & (\langle \mathbf{1} \mathbf{3} \rangle - [\mathbf{1} \mathbf{3}])^2{}_{I_1 I_2}{}^{K(A_1} \\ & (\langle \mathbf{3} | p_1 | \mathbf{3} ])_K{}^{A_2)} (p_3)_{\alpha \dot{\beta}} \end{aligned}$ \\
\hline
$1$ & $2$ & $1$ & $0$ & $1$ & 
$\begin{aligned} & (\langle \mathbf{1} \mathbf{3} \rangle - [\mathbf{1} \mathbf{3}])^2{}_{I_1 I_2}{}^{K_1 K_2} \\ & (\langle \mathbf{3} | p_1 | \mathbf{3} ]^2)_{K_1 K_2}{}^{(A_1 A_2)} \end{aligned}$ & 
$\begin{aligned} & (\langle \mathbf{1} \mathbf{3} \rangle - [\mathbf{1} \mathbf{3}])^2{}_{I_1 I_2}{}^{K_1 K_2} \\ & (\langle \mathbf{3} | p_1 | \mathbf{3} ]^2)_{K_1 K_2}{}^{(A_1 A_2)} (p_3)_{\alpha \dot{\beta}} \end{aligned}$ \\
\hline
$1$ & $1$ & $1$ & $1$ & $0$ & 
$\begin{aligned} & (\langle \mathbf{1} \mathbf{3} \rangle - [\mathbf{1} \mathbf{3}])^2{}_{I_1 I_2}{}^{K_1 K_2} \\ & (\langle \mathbf{3} | p_1 | \mathbf{3} ])_{K_1 K_2} \end{aligned}$ & 
$\begin{aligned} & (\langle \mathbf{1} \mathbf{3} \rangle - [\mathbf{1} \mathbf{3}])^2{}_{I_1 I_2}{}^{K_1 K_2} (\langle \mathbf{3} | p_1 | \mathbf{3} ])_{K_1 K_2} \\ & |\mathbf{3}^{(A_1}\rangle_\alpha [\mathbf{3}^{A_2)}|_{\dot{\beta}} \end{aligned}$ \\
\hline
\end{longtable}}

For spin-$1$ external particles and operators in the Lorentz representation $(1,0)$, the Method II basis is shown in Table~\ref{tab:spin1_irrep_1_0}.
{\renewcommand{\arraystretch}{2.35}
\setlength{\tabcolsep}{2pt}
\scriptsize
\begin{longtable}{|c|c|c|c|c|>{\centering\arraybackslash}p{4.5cm}|>{\centering\arraybackslash}p{7.5cm}|}
\caption{irrep: $(1, 0)$} \label{tab:spin1_irrep_1_0} \\
\hline
$s_1$ & $L$ & $s_3$ & $J$ & $s_{p_3}$ & amplitudes & basis elements \\
\hline
\endfirsthead
\hline
$s_1$ & $L$ & $s_3$ & $J$ & $s_{p_3}$ & amplitudes & basis elements \\
\hline
\endhead
$1$ & $1$ & $1$ & $1$ & $0$ & 
$\begin{aligned} & (\langle \mathbf{1} \mathbf{3} \rangle - [\mathbf{1} \mathbf{3}])^2{}_{I_1 I_2}{}^{K_1 K_2} \\ & (\langle \mathbf{3} | p_1 | \mathbf{3} ])_{K_1 K_2} \end{aligned}$ & 
$\begin{aligned} & (\langle \mathbf{1} \mathbf{3} \rangle - [\mathbf{1} \mathbf{3}])^2{}_{I_1 I_2}{}^{K_1 K_2} (\langle \mathbf{3} | p_1 | \mathbf{3} ])_{K_1 K_2} \\ & |\mathbf{3}^{(A_1}\rangle_{\alpha_1} |\mathbf{3}^{A_2)}\rangle_{\alpha_2} \epsilon_{\dot{\beta}_1 \dot{\beta}_2} \end{aligned}$ \\
\hline
$1$ & $0$ & $1$ & $1$ & $1$ & 
$\begin{aligned} & (\langle \mathbf{1} \mathbf{3} \rangle - [\mathbf{1} \mathbf{3}])^2{}_{I_1 I_2}{}^{(A_1 A_2)} \end{aligned}$ & 
$\begin{aligned} & (\langle \mathbf{1} \mathbf{3} \rangle - [\mathbf{1} \mathbf{3}])^2{}_{I_1 I_2}{}^{(A_1 C_1} \epsilon^{A_2) C_2} \\ & |\mathbf{3}_{(C_1}\rangle_{\alpha_1} |\mathbf{3}_{C_2)}\rangle_{\alpha_2} \epsilon_{\dot{\beta}_1 \dot{\beta}_2} \end{aligned}$ \\
\hline
$1$ & $1$ & $1$ & $1$ & $1$ & 
$\begin{aligned} & (\langle \mathbf{1} \mathbf{3} \rangle - [\mathbf{1} \mathbf{3}])^2{}_{I_1 I_2}{}^{K(A_1} \\ & (\langle \mathbf{3} | p_1 | \mathbf{3} ])_K{}^{A_2)} \end{aligned}$ & 
$\begin{aligned} & (\langle \mathbf{1} \mathbf{3} \rangle - [\mathbf{1} \mathbf{3}])^2{}_{I_1 I_2}{}^{K(A_1} (\langle \mathbf{3} | p_1 | \mathbf{3} ])_K{}^{C_1} \\ & \epsilon^{A_2) C_2} |\mathbf{3}_{(C_1}\rangle_{\alpha_1} |\mathbf{3}_{C_2)}\rangle_{\alpha_2} \epsilon_{\dot{\beta}_1 \dot{\beta}_2} \end{aligned}$ \\
\hline
\end{longtable}}

For spin-$1$ external particles and operators in the Lorentz representation $(0,1)$, the Method II basis is shown in Table~\ref{tab:spin1_irrep_0_1}.
{\renewcommand{\arraystretch}{2.35}
\setlength{\tabcolsep}{2pt}
\scriptsize
\begin{longtable}{|c|c|c|c|c|>{\centering\arraybackslash}p{4.5cm}|>{\centering\arraybackslash}p{7.5cm}|}
\caption{irrep: $(0, 1)$} \label{tab:spin1_irrep_0_1} \\
\hline
$s_1$ & $L$ & $s_3$ & $J$ & $s_{p_3}$ & amplitudes & basis elements \\
\hline
\endfirsthead
\hline
$s_1$ & $L$ & $s_3$ & $J$ & $s_{p_3}$ & amplitudes & basis elements \\
\hline
\endhead
$1$ & $1$ & $1$ & $1$ & $0$ & 
$\begin{aligned} & (\langle \mathbf{1} \mathbf{3} \rangle - [\mathbf{1} \mathbf{3}])^2{}_{I_1 I_2}{}^{K_1 K_2} \\ & (\langle \mathbf{3} | p_1 | \mathbf{3} ])_{K_1 K_2} \end{aligned}$ & 
$\begin{aligned} & (\langle \mathbf{1} \mathbf{3} \rangle - [\mathbf{1} \mathbf{3}])^2{}_{I_1 I_2}{}^{K_1 K_2} (\langle \mathbf{3} | p_1 | \mathbf{3} ])_{K_1 K_2} \\ & [\mathbf{3}^{(A_1}|_{\dot{\beta}_1} [\mathbf{3}^{A_2)}|_{\dot{\beta}_2} \epsilon_{\alpha_1 \alpha_2} \end{aligned}$ \\
\hline
$1$ & $0$ & $1$ & $1$ & $1$ & 
$\begin{aligned} & (\langle \mathbf{1} \mathbf{3} \rangle - [\mathbf{1} \mathbf{3}])^2{}_{I_1 I_2}{}^{(A_1 A_2)} \end{aligned}$ & 
$\begin{aligned} & (\langle \mathbf{1} \mathbf{3} \rangle - [\mathbf{1} \mathbf{3}])^2{}_{I_1 I_2}{}^{(A_1 C_1} \epsilon^{A_2) C_2} \\ & [\mathbf{3}_{(C_1}|_{\dot{\beta}_1} [\mathbf{3}_{C_2)}|_{\dot{\beta}_2} \epsilon_{\alpha_1 \alpha_2} \end{aligned}$ \\
\hline
$1$ & $1$ & $1$ & $1$ & $1$ & 
$\begin{aligned} & (\langle \mathbf{1} \mathbf{3} \rangle - [\mathbf{1} \mathbf{3}])^2{}_{I_1 I_2}{}^{K(A_1} \\ & (\langle \mathbf{3} | p_1 | \mathbf{3} ])_K{}^{A_2)} \end{aligned}$ & 
$\begin{aligned} & (\langle \mathbf{1} \mathbf{3} \rangle - [\mathbf{1} \mathbf{3}])^2{}_{I_1 I_2}{}^{K(A_1} (\langle \mathbf{3} | p_1 | \mathbf{3} ])_K{}^{C_1} \\ & \epsilon^{A_2) C_2} [\mathbf{3}_{(C_1}|_{\dot{\beta}_1} [\mathbf{3}_{C_2)}|_{\dot{\beta}_2} \epsilon_{\alpha_1 \alpha_2} \end{aligned}$ \\
\hline
\end{longtable}}

For spin-$1$ external particles and operators in the Lorentz representation $(1,1)$, the Method II basis is shown in Table~\ref{tab:spin1_irrep_1_1}.
{\renewcommand{\arraystretch}{2.35}
\setlength{\tabcolsep}{2pt}
\scriptsize
\begin{longtable}{|c|c|c|c|c|>{\centering\arraybackslash}p{4.5cm}|>{\centering\arraybackslash}p{7.5cm}|}
\caption{irrep: $(1, 1)$} \label{tab:spin1_irrep_1_1} \\
\hline
$s_1$ & $L$ & $s_3$ & $J$ & $s_{p_3}$ & amplitudes & basis elements \\
\hline
\endfirsthead
\hline
$s_1$ & $L$ & $s_3$ & $J$ & $s_{p_3}$ & amplitudes & basis elements \\
\hline
\endhead
$1$ & $0$ & $1$ & $0$ & $1$ & 
$\begin{aligned} & (\langle \mathbf{1} \mathbf{3} \rangle - [\mathbf{1} \mathbf{3}])^2{}_{I_1 I_2}{}^{(A_1 A_2)} \end{aligned}$ & 
$\begin{aligned} & (\langle \mathbf{1} \mathbf{3} \rangle - [\mathbf{1} \mathbf{3}])^2{}_{I_1 I_2}{}^{(A_1 A_2)} \\ & (p_3)^2_{(\alpha_1 \alpha_2), (\dot{\beta}_1 \dot{\beta}_2)} \end{aligned}$ \\
\hline
$1$ & $1$ & $1$ & $0$ & $1$ & 
$\begin{aligned} & (\langle \mathbf{1} \mathbf{3} \rangle - [\mathbf{1} \mathbf{3}])^2{}_{I_1 I_2}{}^{K(A_1} \\ & (\langle \mathbf{3} | p_1 | \mathbf{3} ])_K{}^{A_2)} \end{aligned}$ & 
$\begin{aligned} & (\langle \mathbf{1} \mathbf{3} \rangle - [\mathbf{1} \mathbf{3}])^2{}_{I_1 I_2}{}^{K(A_1} (\langle \mathbf{3} | p_1 | \mathbf{3} ])_K{}^{A_2)} \\ & (p_3)^2_{(\alpha_1 \alpha_2), (\dot{\beta}_1 \dot{\beta}_2)} \end{aligned}$ \\
\hline
$1$ & $2$ & $1$ & $0$ & $1$ & 
$\begin{aligned} & (\langle \mathbf{1} \mathbf{3} \rangle - [\mathbf{1} \mathbf{3}])^2{}_{I_1 I_2}{}^{K_1 K_2} \\ & (\langle \mathbf{3} | p_1 | \mathbf{3} ]^2)_{K_1 K_2}{}^{(A_1 A_2)} \end{aligned}$ & 
$\begin{aligned} & (\langle \mathbf{1} \mathbf{3} \rangle - [\mathbf{1} \mathbf{3}])^2{}_{I_1 I_2}{}^{K_1 K_2} (\langle \mathbf{3} | p_1 | \mathbf{3} ]^2)_{K_1 K_2}{}^{(A_1 A_2)} \\ & (p_3)^2_{(\alpha_1 \alpha_2), (\dot{\beta}_1 \dot{\beta}_2)} \end{aligned}$ \\
\hline
$1$ & $1$ & $1$ & $1$ & $0$ & 
$\begin{aligned} & (\langle \mathbf{1} \mathbf{3} \rangle - [\mathbf{1} \mathbf{3}])^2{}_{I_1 I_2}{}^{K_1 K_2} \\ & (\langle \mathbf{3} | p_1 | \mathbf{3} ])_{K_1 K_2} \end{aligned}$ & 
$\begin{aligned} & (\langle \mathbf{1} \mathbf{3} \rangle - [\mathbf{1} \mathbf{3}])^2{}_{I_1 I_2}{}^{K_1 K_2} (\langle \mathbf{3} | p_1 | \mathbf{3} ])_{K_1 K_2} \\ & |\mathbf{3}^{(A_1}\rangle_{(\alpha_1} (p_3)_{\alpha_2), (\dot{\beta}_1} [\mathbf{3}^{A_2)}|_{\dot{\beta}_2)} \end{aligned}$ \\
\hline
\end{longtable}}

For spin-$1$ external particles and operators in the Lorentz representation $(\tfrac{3}{2},\tfrac{3}{2})$, the Method II basis is shown in Table~\ref{tab:spin1_irrep_3half_3half}.
{\renewcommand{\arraystretch}{2.35}
\setlength{\tabcolsep}{2pt}
\scriptsize
\begin{longtable}{|c|c|c|c|c|>{\centering\arraybackslash}p{4.5cm}|>{\centering\arraybackslash}p{7.5cm}|}
\caption{spin-$1$ particle \quad irrep $(\frac{3}{2}, \frac{3}{2})$} \label{tab:spin1_irrep_3half_3half} \\
\hline
$s_1$ & $L$ & $s_3$ & $J$ & $s_{p_3}$ & amplitudes & basis elements \\
\hline
\endfirsthead
\hline
$s_1$ & $L$ & $s_3$ & $J$ & $s_{p_3}$ & amplitudes & basis elements \\
\hline
\endhead
$1$ & $0$ & $1$ & $0$ & $1$ & 
$\begin{aligned}&(\langle \mathbf{1} \mathbf{3} \rangle - [\mathbf{1} \mathbf{3}])^2{}_{I_1 I_2}{}^{(A_1 A_2)}\end{aligned}$ & 
$\begin{aligned}&(\langle \mathbf{1} \mathbf{3} \rangle - [\mathbf{1} \mathbf{3}])^2{}_{I_1 I_2}{}^{(A_1 A_2)}\\&(p_3)^3_{(\alpha_1 \alpha_2 \alpha_3), (\dot{\beta}_1 \dot{\beta}_2 \dot{\beta}_3)}\end{aligned}$ \\
\hline
$1$ & $1$ & $1$ & $0$ & $1$ & 
$\begin{aligned}&(\langle \mathbf{1} \mathbf{3} \rangle - [\mathbf{1} \mathbf{3}])^2{}_{I_1 I_2}{}^{K(A_1}\\&(\langle \mathbf{3} | p_1 | \mathbf{3} ])_K{}^{A_2)}\end{aligned}$ & 
$\begin{aligned}&(\langle \mathbf{1} \mathbf{3} \rangle - [\mathbf{1} \mathbf{3}])^2{}_{I_1 I_2}{}^{K(A_1}(\langle \mathbf{3} | p_1 | \mathbf{3} ])_K{}^{A_2)}\\&(p_3)^3_{(\alpha_1 \alpha_2 \alpha_3), (\dot{\beta}_1 \dot{\beta}_2 \dot{\beta}_3)}\end{aligned}$ \\
\hline
$1$ & $2$ & $1$ & $0$ & $1$ & 
$\begin{aligned}&(\langle \mathbf{1} \mathbf{3} \rangle - [\mathbf{1} \mathbf{3}])^2{}_{I_1 I_2}{}^{K_1 K_2}\\&(\langle \mathbf{3} | p_1 | \mathbf{3} ]^2)_{K_1 K_2}{}^{(A_1 A_2)}\end{aligned}$ & 
$\begin{aligned}&(\langle \mathbf{1} \mathbf{3} \rangle - [\mathbf{1} \mathbf{3}])^2{}_{I_1 I_2}{}^{K_1 K_2}(\langle \mathbf{3} | p_1 | \mathbf{3} ]^2)_{K_1 K_2}{}^{(A_1 A_2)}\\&(p_3)^3_{(\alpha_1 \alpha_2 \alpha_3), (\dot{\beta}_1 \dot{\beta}_2 \dot{\beta}_3)}\end{aligned}$ \\
\hline
$1$ & $1$ & $1$ & $1$ & $0$ & 
$\begin{aligned}&(\langle \mathbf{1} \mathbf{3} \rangle - [\mathbf{1} \mathbf{3}])^2{}_{I_1 I_2}{}^{K_1 K_2}\\&(\langle \mathbf{3} | p_1 | \mathbf{3} ])_{K_1 K_2}\end{aligned}$ & 
$\begin{aligned}&(\langle \mathbf{1} \mathbf{3} \rangle - [\mathbf{1} \mathbf{3}])^2{}_{I_1 I_2}{}^{K_1 K_2}(\langle \mathbf{3} | p_1 | \mathbf{3} ])_{K_1 K_2}\\&|\mathbf{3}^{(A_1}\rangle_{(\alpha_1}(p_3)^2_{\alpha_2 \alpha_3), (\dot{\beta}_1 \dot{\beta}_2}[\mathbf{3}^{A_2)}|_{\dot{\beta}_3)}\end{aligned}$ \\
\hline
\end{longtable}}

For spin-$\tfrac{3}{2}$ external particles and operators in the Lorentz representation $(0,0)$, the Method II basis is shown in Table~\ref{tab:spin3half_irrep_0_0}.
{\renewcommand{\arraystretch}{2.35}
\setlength{\tabcolsep}{2pt}
\scriptsize
\begin{longtable}{|c|c|c|c|c|>{\centering\arraybackslash}p{5.5cm}|>{\centering\arraybackslash}p{8.5cm}|}
\caption{spin-$3/2$ particle \quad irrep $(0,0)$} \label{tab:spin3half_irrep_0_0} \\
\hline
$s_1$ & $L$ & $s_3$ & $J$ & $s_{p_3}$ & amplitudes & basis elements \\
\hline
\endfirsthead
\hline
$s_1$ & $L$ & $s_3$ & $J$ & $s_{p_3}$ & amplitudes & basis elements \\
\hline
\endhead
$\frac{3}{2}$ & $0$ & $\frac{3}{2}$ & $0$ & $\frac{3}{2}$ & 
$\begin{aligned} & (\langle \mathbf{1} \mathbf{3} \rangle - [\mathbf{1} \mathbf{3}])^3{}_{I_1 I_2 I_3}{}^{A_1 A_2 A_3} \end{aligned}$ & 
$\begin{aligned} & (\langle \mathbf{1} \mathbf{3} \rangle - [\mathbf{1} \mathbf{3}])^3{}_{I_1 I_2 I_3}{}^{A_1 A_2 A_3} \end{aligned}$ \\
\hline
$\frac{3}{2}$ & $1$ & $\frac{3}{2}$ & $0$ & $\frac{3}{2}$ & 
$\begin{aligned} & (\langle \mathbf{1} \mathbf{3} \rangle - [\mathbf{1} \mathbf{3}])^3{}_{I_1 I_2 I_3}{}^{K(A_1 A_2} \\ & (\langle \mathbf{3} | p_1 | \mathbf{3} ])_K{}^{A_3)} \end{aligned}$ & 
$\begin{aligned} & (\langle \mathbf{1} \mathbf{3} \rangle - [\mathbf{1} \mathbf{3}])^3{}_{I_1 I_2 I_3}{}^{K(A_1 A_2} \\ & (\langle \mathbf{3} | p_1 | \mathbf{3} ])_K{}^{A_3)} \end{aligned}$ \\
\hline
$\frac{3}{2}$ & $2$ & $\frac{3}{2}$ & $0$ & $\frac{3}{2}$ & 
$\begin{aligned} & (\langle \mathbf{1} \mathbf{3} \rangle - [\mathbf{1} \mathbf{3}])^3{}_{I_1 I_2 I_3}{}^{K_1 K_2(A_1} \\ & (\langle \mathbf{3} | p_1 | \mathbf{3} ]^2)_{K_1 K_2}{}^{A_2 A_3)} \end{aligned}$ & 
$\begin{aligned} & (\langle \mathbf{1} \mathbf{3} \rangle - [\mathbf{1} \mathbf{3}])^3{}_{I_1 I_2 I_3}{}^{K_1 K_2(A_1} \\ & (\langle \mathbf{3} | p_1 | \mathbf{3} ]^2)_{K_1 K_2}{}^{A_2 A_3)} \end{aligned}$ \\
\hline
$\frac{3}{2}$ & $3$ & $\frac{3}{2}$ & $0$ & $\frac{3}{2}$ & 
$\begin{aligned} & (\langle \mathbf{1} \mathbf{3} \rangle - [\mathbf{1} \mathbf{3}])^3{}_{I_1 I_2 I_3}{}^{K_1 K_2 K_3} \\ & (\langle \mathbf{3} | p_1 | \mathbf{3} ]^3)_{K_1 K_2 K_3}{}^{(A_1 A_2 A_3)} \end{aligned}$ & 
$\begin{aligned} & (\langle \mathbf{1} \mathbf{3} \rangle - [\mathbf{1} \mathbf{3}])^3{}_{I_1 I_2 I_3}{}^{K_1 K_2 K_3} \\ & (\langle \mathbf{3} | p_1 | \mathbf{3} ]^3)_{K_1 K_2 K_3}{}^{(A_1 A_2 A_3)} \end{aligned}$ \\
\hline
\end{longtable}}

For spin-$\tfrac{3}{2}$ external particles and operators in the Lorentz representation $(\tfrac{1}{2},\tfrac{1}{2})$, the Method II basis is shown in Table~\ref{tab:spin3half_irrep_half_half}.
{\renewcommand{\arraystretch}{2.35}
\setlength{\tabcolsep}{2pt}
\scriptsize
\begin{longtable}{|c|c|c|c|c|>{\centering\arraybackslash}p{5.5cm}|>{\centering\arraybackslash}p{8.5cm}|}
\caption{irrep: $(\frac{1}{2}, \frac{1}{2})$} \label{tab:spin3half_irrep_half_half} \\
\hline
$s_1$ & $L$ & $s_3$ & $J$ & $s_{p_3}$ & amplitudes & basis elements \\
\hline
\endfirsthead
\hline
$s_1$ & $L$ & $s_3$ & $J$ & $s_{p_3}$ & amplitudes & basis elements \\
\hline
\endhead
$\frac{3}{2}$ & $0$ & $\frac{3}{2}$ & $0$ & $\frac{3}{2}$ & 
$\begin{aligned} & (\langle \mathbf{1} \mathbf{3} \rangle - [\mathbf{1} \mathbf{3}])^3{}_{I_1 I_2 I_3}{}^{A_1 A_2 A_3} \end{aligned}$ & 
$\begin{aligned} & (\langle \mathbf{1} \mathbf{3} \rangle - [\mathbf{1} \mathbf{3}])^3{}_{I_1 I_2 I_3}{}^{(A_1 A_2 A_3)} \\ & (p_3)_{\alpha \dot{\beta}} \end{aligned}$ \\
\hline
$\frac{3}{2}$ & $1$ & $\frac{3}{2}$ & $0$ & $\frac{3}{2}$ & 
$\begin{aligned} & (\langle \mathbf{1} \mathbf{3} \rangle - [\mathbf{1} \mathbf{3}])^3{}_{I_1 I_2 I_3}{}^{K(A_1 A_2} \\ & (\langle \mathbf{3} | p_1 | \mathbf{3} ])_K{}^{A_3)} \end{aligned}$ & 
$\begin{aligned} & (\langle \mathbf{1} \mathbf{3} \rangle - [\mathbf{1} \mathbf{3}])^3{}_{I_1 I_2 I_3}{}^{K(A_1 A_2} \\ & (\langle \mathbf{3} | p_1 | \mathbf{3} ])_K{}^{A_3)} (p_3)_{\alpha \dot{\beta}} \end{aligned}$ \\
\hline
$\frac{3}{2}$ & $2$ & $\frac{3}{2}$ & $0$ & $\frac{3}{2}$ & 
$\begin{aligned} & (\langle \mathbf{1} \mathbf{3} \rangle - [\mathbf{1} \mathbf{3}])^3{}_{I_1 I_2 I_3}{}^{K_1 K_2(A_1} \\ & (\langle \mathbf{3} | p_1 | \mathbf{3} ]^2)_{K_1 K_2}{}^{A_2 A_3)} \end{aligned}$ & 
$\begin{aligned} & (\langle \mathbf{1} \mathbf{3} \rangle - [\mathbf{1} \mathbf{3}])^3{}_{I_1 I_2 I_3}{}^{K_1 K_2(A_1} \\ & (\langle \mathbf{3} | p_1 | \mathbf{3} ]^2)_{K_1 K_2}{}^{A_2 A_3)} (p_3)_{\alpha \dot{\beta}} \end{aligned}$ \\
\hline
$\frac{3}{2}$ & $3$ & $\frac{3}{2}$ & $0$ & $\frac{3}{2}$ & 
$\begin{aligned} & (\langle \mathbf{1} \mathbf{3} \rangle - [\mathbf{1} \mathbf{3}])^3{}_{I_1 I_2 I_3}{}^{K_1 K_2 K_3} \\ & (\langle \mathbf{3} | p_1 | \mathbf{3} ]^3)_{K_1 K_2 K_3}{}^{(A_1 A_2 A_3)} \end{aligned}$ & 
$\begin{aligned} & (\langle \mathbf{1} \mathbf{3} \rangle - [\mathbf{1} \mathbf{3}])^3{}_{I_1 I_2 I_3}{}^{K_1 K_2 K_3} \\ & (\langle \mathbf{3} | p_1 | \mathbf{3} ]^3)_{K_1 K_2 K_3}{}^{(A_1 A_2 A_3)} (p_3)_{\alpha \dot{\beta}} \end{aligned}$ \\
\hline
\end{longtable}}

For spin-$\tfrac{3}{2}$ external particles and operators in the Lorentz representation $(1,0)$, the Method II basis is shown in Table~\ref{tab:spin3half_irrep_1_0}.
{\renewcommand{\arraystretch}{2.35}
\setlength{\tabcolsep}{2pt}
\scriptsize
\begin{longtable}{|c|c|c|c|c|>{\centering\arraybackslash}p{5.5cm}|>{\centering\arraybackslash}p{8.5cm}|}
\caption{irrep: $(1, 0)$} \label{tab:spin3half_irrep_1_0} \\
\hline
$s_1$ & $L$ & $s_3$ & $J$ & $s_{p_3}$ & amplitudes & basis elements \\
\hline
\endfirsthead
\hline
$s_1$ & $L$ & $s_3$ & $J$ & $s_{p_3}$ & amplitudes & basis elements \\
\hline
\endhead
$\frac{3}{2}$ & $1$ & $\frac{3}{2}$ & $1$ & $\frac{1}{2}$ & 
$\begin{aligned} & (\langle \mathbf{1} \mathbf{3} \rangle - [\mathbf{1} \mathbf{3}])^3{}_{I_1 I_2 I_3}{}^{K_1 K_2 A_1} \\ & (\langle \mathbf{3} | p_1 | \mathbf{3} ])_{K_1 K_2} \end{aligned}$ & 
$\begin{aligned} & (\langle \mathbf{1} \mathbf{3} \rangle - [\mathbf{1} \mathbf{3}])^3{}_{I_1 I_2 I_3}{}^{K_1 K_2 (A_1} (\langle \mathbf{3} | p_1 | \mathbf{3} ])_{K_1 K_2} \\ & |\mathbf{3}^{A_2}\rangle_{\alpha_1} |\mathbf{3}^{A_3)}\rangle_{\alpha_2} \epsilon_{\dot{\beta}_1 \dot{\beta}_2} \end{aligned}$ \\
\hline
$\frac{3}{2}$ & $2$ & $\frac{3}{2}$ & $1$ & $\frac{1}{2}$ & 
$\begin{aligned} & (\langle \mathbf{1} \mathbf{3} \rangle - [\mathbf{1} \mathbf{3}])^3{}_{I_1 I_2 I_3}{}^{K_1 K_2 K_3} \\ & (\langle \mathbf{3} | p_1 | \mathbf{3} ]^2)_{K_1 K_2 K_3}{}^{A_1} \end{aligned}$ & 
$\begin{aligned} & (\langle \mathbf{1} \mathbf{3} \rangle - [\mathbf{1} \mathbf{3}])^3{}_{I_1 I_2 I_3}{}^{K_1 K_2 K_3} (\langle \mathbf{3} | p_1 | \mathbf{3} ]^2)_{K_1 K_2 K_3}{}^{(A_1} \\ & |\mathbf{3}^{A_2}\rangle_{\alpha_1} |\mathbf{3}^{A_3)}\rangle_{\alpha_2} \epsilon_{\dot{\beta}_1 \dot{\beta}_2} \end{aligned}$ \\
\hline
$\frac{3}{2}$ & $0$ & $\frac{3}{2}$ & $1$ & $\frac{3}{2}$ & 
$\begin{aligned} & (\langle \mathbf{1} \mathbf{3} \rangle - [\mathbf{1} \mathbf{3}])^3{}_{I_1 I_2 I_3}{}^{(A_1 A_2 C_1)} \end{aligned}$ & 
$\begin{aligned} & (\langle \mathbf{1} \mathbf{3} \rangle - [\mathbf{1} \mathbf{3}])^3{}_{I_1 I_2 I_3}{}^{(A_1 A_2 C_1} \\ & \epsilon^{A_3) C_2} |\mathbf{3}_{(C_1}\rangle_{\alpha_1} |\mathbf{3}_{C_2)}\rangle_{\alpha_2} \epsilon_{\dot{\beta}_1 \dot{\beta}_2} \end{aligned}$ \\
\hline
\end{longtable}}

For spin-$\tfrac{3}{2}$ external particles and an operator in the Lorentz representation $(0,1)$, the Method II basis is shown in Table~\ref{tab:spin3half_irrep_0_1}.
{\renewcommand{\arraystretch}{2.35}
\setlength{\tabcolsep}{2pt}
\scriptsize
\begin{longtable}{|c|c|c|c|c|>{\centering\arraybackslash}p{5.5cm}|>{\centering\arraybackslash}p{8.5cm}|}
\caption{irrep: $(0, 1)$} \label{tab:spin3half_irrep_0_1} \\
\hline
$s_1$ & $L$ & $s_3$ & $J$ & $s_{p_3}$ & amplitudes & basis elements \\
\hline
\endfirsthead
\hline
$s_1$ & $L$ & $s_3$ & $J$ & $s_{p_3}$ & amplitudes & basis elements \\
\hline
\endhead
$\frac{3}{2}$ & $1$ & $\frac{3}{2}$ & $1$ & $\frac{1}{2}$ & 
$\begin{aligned} & (\langle \mathbf{1} \mathbf{3} \rangle - [\mathbf{1} \mathbf{3}])^3{}_{I_1 I_2 I_3}{}^{K_1 K_2 A_1} \\ & (\langle \mathbf{3} | p_1 | \mathbf{3} ])_{K_1 K_2} \end{aligned}$ & 
$\begin{aligned} & (\langle \mathbf{1} \mathbf{3} \rangle - [\mathbf{1} \mathbf{3}])^3{}_{I_1 I_2 I_3}{}^{K_1 K_2 (A_1} (\langle \mathbf{3} | p_1 | \mathbf{3} ])_{K_1 K_2} \\ & [\mathbf{3}^{A_2}|_{\dot{\beta}_1} [\mathbf{3}^{A_3)}|_{\dot{\beta}_2} \epsilon_{\alpha_1 \alpha_2} \end{aligned}$ \\
\hline
$\frac{3}{2}$ & $2$ & $\frac{3}{2}$ & $1$ & $\frac{1}{2}$ & 
$\begin{aligned} & (\langle \mathbf{1} \mathbf{3} \rangle - [\mathbf{1} \mathbf{3}])^3{}_{I_1 I_2 I_3}{}^{K_1 K_2 K_3} \\ & (\langle \mathbf{3} | p_1 | \mathbf{3} ]^2)_{K_1 K_2 K_3}{}^{A_1} \end{aligned}$ & 
$\begin{aligned} & (\langle \mathbf{1} \mathbf{3} \rangle - [\mathbf{1} \mathbf{3}])^3{}_{I_1 I_2 I_3}{}^{K_1 K_2 K_3} (\langle \mathbf{3} | p_1 | \mathbf{3} ]^2)_{K_1 K_2 K_3}{}^{(A_1} \\ & [\mathbf{3}^{A_2}|_{\dot{\beta}_1} [\mathbf{3}^{A_3)}|_{\dot{\beta}_2} \epsilon_{\alpha_1 \alpha_2} \end{aligned}$ \\
\hline
$\frac{3}{2}$ & $0$ & $\frac{3}{2}$ & $1$ & $\frac{3}{2}$ & 
$\begin{aligned} & (\langle \mathbf{1} \mathbf{3} \rangle - [\mathbf{1} \mathbf{3}])^3{}_{I_1 I_2 I_3}{}^{(A_1 A_2 C_1)} \end{aligned}$ & 
$\begin{aligned} & (\langle \mathbf{1} \mathbf{3} \rangle - [\mathbf{1} \mathbf{3}])^3{}_{I_1 I_2 I_3}{}^{(A_1 A_2 C_1} \\ & \epsilon^{A_3) C_2} [\mathbf{3}_{(C_1}|_{\dot{\beta}_1} [\mathbf{3}_{C_2)}|_{\dot{\beta}_2} \epsilon_{\alpha_1 \alpha_2} \end{aligned}$ \\
\hline
\end{longtable}}

For spin-$\tfrac{3}{2}$ external particles and operators in the Lorentz representation $(1,1)$, the Method II basis is shown in Table~\ref{tab:spin3half_irrep_1_1}.
{\renewcommand{\arraystretch}{2.35}
\setlength{\tabcolsep}{2pt}
\scriptsize
\begin{longtable}{|c|c|c|c|c|>{\centering\arraybackslash}p{5.5cm}|>{\centering\arraybackslash}p{8.5cm}|}
\caption{irrep: $(1, 1)$} \label{tab:spin3half_irrep_1_1} \\
\hline
$s_1$ & $L$ & $s_3$ & $J$ & $s_{p_3}$ & amplitudes & basis elements \\
\hline
\endfirsthead
\hline
$s_1$ & $L$ & $s_3$ & $J$ & $s_{p_3}$ & amplitudes & basis elements \\
\hline
\endhead
$\frac{3}{2}$ & $0$ & $\frac{3}{2}$ & $0$ & $\frac{3}{2}$ & 
$\begin{aligned} & (\langle \mathbf{1} \mathbf{3} \rangle - [\mathbf{1} \mathbf{3}])^3{}_{I_1 I_2 I_3}{}^{(A_1 A_2 A_3)} \end{aligned}$ & 
$\begin{aligned} & (\langle \mathbf{1} \mathbf{3} \rangle - [\mathbf{1} \mathbf{3}])^3{}_{I_1 I_2 I_3}{}^{(A_1 A_2 A_3)} \\ & (p_3)^2_{(\alpha_1 \alpha_2), (\dot{\beta}_1 \dot{\beta}_2)} \end{aligned}$ \\
\hline
$\frac{3}{2}$ & $1$ & $\frac{3}{2}$ & $0$ & $\frac{3}{2}$ & 
$\begin{aligned} & (\langle \mathbf{1} \mathbf{3} \rangle - [\mathbf{1} \mathbf{3}])^3{}_{I_1 I_2 I_3}{}^{K(A_1 A_2} \\ & (\langle \mathbf{3} | p_1 | \mathbf{3} ])_K{}^{A_3)} \end{aligned}$ & 
$\begin{aligned} & (\langle \mathbf{1} \mathbf{3} \rangle - [\mathbf{1} \mathbf{3}])^3{}_{I_1 I_2 I_3}{}^{K(A_1 A_2} (\langle \mathbf{3} | p_1 | \mathbf{3} ])_K{}^{A_3)} \\ & (p_3)^2_{(\alpha_1 \alpha_2), (\dot{\beta}_1 \dot{\beta}_2)} \end{aligned}$ \\
\hline
$\frac{3}{2}$ & $2$ & $\frac{3}{2}$ & $0$ & $\frac{3}{2}$ & 
$\begin{aligned} & (\langle \mathbf{1} \mathbf{3} \rangle - [\mathbf{1} \mathbf{3}])^3{}_{I_1 I_2 I_3}{}^{K_1 K_2(A_1} \\ & (\langle \mathbf{3} | p_1 | \mathbf{3} ]^2)_{K_1 K_2}{}^{A_2 A_3)} \end{aligned}$ & 
$\begin{aligned} & (\langle \mathbf{1} \mathbf{3} \rangle - [\mathbf{1} \mathbf{3}])^3{}_{I_1 I_2 I_3}{}^{K_1 K_2(A_1} (\langle \mathbf{3} | p_1 | \mathbf{3} ]^2)_{K_1 K_2}{}^{(A_2 A_3)} \\ & (p_3)^2_{(\alpha_1 \alpha_2), (\dot{\beta}_1 \dot{\beta}_2)} \end{aligned}$ \\
\hline
\end{longtable}}

For spin-$\tfrac{3}{2}$ external particles and operators in the Lorentz representation $(\tfrac{3}{2},\tfrac{3}{2})$, the Method II basis is shown in Table~\ref{tab:spin3half_irrep_3half_3half}.
{\renewcommand{\arraystretch}{2.35}
\setlength{\tabcolsep}{2pt}
\scriptsize
\begin{longtable}{|c|c|c|c|c|>{\centering\arraybackslash}p{5.5cm}|>{\centering\arraybackslash}p{8.5cm}|}
\caption{irrep: $(\frac{3}{2}, \frac{3}{2})$} \label{tab:spin3half_irrep_3half_3half} \\
\hline
$s_1$ & $L$ & $s_3$ & $J$ & $s_{p_3}$ & amplitudes & basis elements \\
\hline
\endfirsthead
\hline
$s_1$ & $L$ & $s_3$ & $J$ & $s_{p_3}$ & amplitudes & basis elements \\
\hline
\endhead
$\frac{3}{2}$ & $0$ & $\frac{3}{2}$ & $0$ & $\frac{3}{2}$ & 
$\begin{aligned} & (\langle \mathbf{1} \mathbf{3} \rangle - [\mathbf{1} \mathbf{3}])^3{}_{I_1 I_2 I_3}{}^{(A_1 A_2 A_3)} \end{aligned}$ & 
$\begin{aligned} & (\langle \mathbf{1} \mathbf{3} \rangle - [\mathbf{1} \mathbf{3}])^3{}_{I_1 I_2 I_3}{}^{(A_1 A_2 A_3)} \\ & (p_3)^3_{(\alpha_1 \alpha_2 \alpha_3), (\dot{\beta}_1 \dot{\beta}_2 \dot{\beta}_3)} \end{aligned}$ \\
\hline
$\frac{3}{2}$ & $1$ & $\frac{3}{2}$ & $0$ & $\frac{3}{2}$ & 
$\begin{aligned} & (\langle \mathbf{1} \mathbf{3} \rangle - [\mathbf{1} \mathbf{3}])^3{}_{I_1 I_2 I_3}{}^{K(A_1 A_2} \\ & (\langle \mathbf{3} | p_1 | \mathbf{3} ])_K{}^{A_3)} \end{aligned}$ & 
$\begin{aligned} & (\langle \mathbf{1} \mathbf{3} \rangle - [\mathbf{1} \mathbf{3}])^3{}_{I_1 I_2 I_3}{}^{K(A_1 A_2} (\langle \mathbf{3} | p_1 | \mathbf{3} ])_K{}^{A_3)} \\ & (p_3)^3_{(\alpha_1 \alpha_2 \alpha_3), (\dot{\beta}_1 \dot{\beta}_2 \dot{\beta}_3)} \end{aligned}$ \\
\hline
$\frac{3}{2}$ & $2$ & $\frac{3}{2}$ & $0$ & $\frac{3}{2}$ & 
$\begin{aligned} & (\langle \mathbf{1} \mathbf{3} \rangle - [\mathbf{1} \mathbf{3}])^3{}_{I_1 I_2 I_3}{}^{K_1 K_2(A_1} \\ & (\langle \mathbf{3} | p_1 | \mathbf{3} ]^2)_{K_1 K_2}{}^{A_2 A_3)} \end{aligned}$ & 
$\begin{aligned} & (\langle \mathbf{1} \mathbf{3} \rangle - [\mathbf{1} \mathbf{3}])^3{}_{I_1 I_2 I_3}{}^{K_1 K_2(A_1} (\langle \mathbf{3} | p_1 | \mathbf{3} ]^2)_{K_1 K_2}{}^{(A_2 A_3)} \\ & (p_3)^3_{(\alpha_1 \alpha_2 \alpha_3), (\dot{\beta}_1 \dot{\beta}_2 \dot{\beta}_3)} \end{aligned}$ \\
\hline
\end{longtable}}

\section{Conclusion}
\label{sec:conclusion}

In this work, we have developed a covariant construction of FF bases for matrix elements by starting from the complete basis of massive three point amplitudes organized by the $LS$ coupling. Instead of writing down a large Lorentz covariant ansatz and then eliminating redundant structures afterwards, we relate the matrix elements to the complete $LS$ three-point amplitude basis through two fixed covariant projections, one projecting the matrix elements onto the corresponding three-point amplitudes and the other projecting the amplitude basis back onto the matrix-element basis. Combined with the $LS$ counting, this provides a unified technique for both counting and explicit construction. From this perspective, the $LS$ coupling and multipole expansion of matrix elements are extended to a Lorentz covariant setting.

On the one hand, an important part of the present work is the physical interpretation of the multipole construction. In section~\ref{sec:Non-relativistic multipole expansion}, we first review the traditional electromagnetic multipole expansion, where the current is decomposed into charge, longitudinal, electric, and magnetic multipoles. 
We then reformulate the same matrix elements in the $LS$ basis. In the Breit frame, the momentum transfer $Q^\mu$ is purely spatial, so the angular dependence is described by $\hat q$. The orbital harmonic $Y^L_m(\hat q)$ is coupled with the crossed-channel spin tensor $S$ as $L\otimes S\to J $. Each allowed $(L,S)$ channel defines an independent tensor structure, with scalar coefficient $F^{(J)}_{LS}(q^2)$.
The Zemach tensor formulation gives the Cartesian realization of the same $LS$ construction. Thus the traditional electromagnetic multipoles, the $LS$ basis, and the Zemach tensor formulation are different bases for the same $\mathrm{SO}(3)$ tensor decomposition. The spin-$\frac12\to\frac12$ and spin-$1\to1$ examples make this relation explicit.

To keep both Lorentz covariance and a clear $LS$ interpretation, section~\ref{sec:Canonical-spinor formulation of the multipole expansion} introduces the canonical-spinor method. In this method, the $\mathrm{SL}(2,\mathbb C)$ spinor indices carry Lorentz covariance, while the little group indices carry the spin degrees of freedom. Since these two kinds of indices are separated, the spin coupling can be performed directly in little group space without losing covariance. The local operator matrix element can then be viewed as a structure induced by a three-point $LS$ amplitude. Each allowed $(L,S)$ channel gives one covariant structure. In this way, the construction of the matrix element basis becomes systematic, and its counting, completeness, and linear independence are inherited from the three-point $LS$ amplitude basis.

In application, this construction has several practical advantages. Once the complete three-point amplitude basis is known, the derivation of the matrix element basis becomes simple and algorithmic. The linear independence of the resulting structures is inherited directly from the amplitude basis, so no separate redundancy removal procedure is needed. The method is fully general in the Lorentz representation of the operator and does not assume $P$ or $T$ conservation. At the same time, the counting of independent structures follows directly from the allowed $LS$ couplings, and the basis itself admits a universal explicit expression. In section~\ref{sec:Covariant Bases from Method II}, we summarize the explicit construction results for spin-$\tfrac{1}{2}$, spin-$1$, and spin-$\tfrac{3}{2}$ external particles.

Overall, our results provide a systematic covariant multipole expansion for generalized FFs that is explicit, complete, and free of hidden redundancies. We expect this framework to be useful for the parameterization of local operator matrix elements for a broad range of spins and Lorentz representations, and to offer a practical starting point for future studies of higher spin FFs and related covariant observables.

\section*{Acknowledgments}
This work is supported by the National Science Foundation of China under Grants No. 12347105, No. 12375099 and No. 12047503, and the National Key Research and Development Program of China Grant No. 2020YFC2201501, No. 2021YFA0718304.

\appendix

\section{Spherical and tensor basis}
\label{app:Spherical and tensor basis}

The spherical basis used to convert between spherical components and Cartesian tensor components is specified. The vector spherical basis is
\begin{eqnarray}
    e^i_{+1}
    &=&
    -\frac{1}{\sqrt{2}}(1,i,0),
    \qquad
    e^i_0
    =
    (0,0,1),
    \qquad
    e^i_{-1}
    =
    \frac{1}{\sqrt{2}}(1,-i,0),
    \label{eq:spherical_basis_vector}
\end{eqnarray}
where $i$ denotes the Cartesian components $(x,y,z)$. The dual basis is defined by
\begin{eqnarray}
    \bar e^i_m
    &\equiv&
    (e^i_m)^*
    =
    (-1)^m e^i_{-m}.
    \label{eq:dual_spherical_basis_vector}
\end{eqnarray}
For any Cartesian vector $V^i$, the spherical and Cartesian components are related by
\begin{eqnarray}
    V_m
    &=&
    e^i_m V^i,
    \qquad
    V^i
    =
    \sum_{m=-1}^{1}\bar e^i_m V_m .
    \label{eq:vector_spherical_cartesian_conversion}
\end{eqnarray}
Higher-rank spherical basis tensors are obtained by coupling several vector spherical bases with CGCs,
\begin{eqnarray}
    e^{i_1\cdots i_L}_{LM}
    &=&
    \sum_{m_1,\ldots,m_L}
    C^{LM}_{1m_1;\cdots;1m_L}
    e^{i_1}_{m_1}\cdots e^{i_L}_{m_L},
    \label{eq:spherical_tensor_basis_general}
\end{eqnarray}
where $C^{LM}_{1m_1;\cdots;1m_L}$ denotes the successive CGCs projection of $L$ spin-one indices onto the irreducible rank-$L$ representation. The first few cases are
\begin{eqnarray}
    e_{00}
    &=&
    1,
    \qquad
    e^i_{1m}
    =
    e^i_m,\qquad e^{ij}_{2M}
    =
    \sum_{m_1,m_2}
    C^{2,M}_{1,m_1;1,m_2}
    e^i_{m_1}e^j_{m_2}.
    \label{eq:rank2_spherical_tensor_basis}
\end{eqnarray}
The corresponding dual tensors used for Cartesian reconstruction are
\begin{eqnarray}
    \bar e^{i_1\cdots i_L}_{LM}
    &\equiv&
    \left(e^{i_1\cdots i_L}_{LM}\right)^*,
    \qquad
    \bar e^{ij}_{2M}
    =
    \left(e^{ij}_{2M}\right)^* .
    \label{eq:dual_spherical_tensor_basis}
\end{eqnarray}
Both $e^{ij}_{2M}$ and $\bar e^{ij}_{2M}$ are symmetric and traceless in their Cartesian indices. The tensor $e^{ij}_{2M}$ projects a Cartesian tensor onto its spherical component, while $\bar e^{ij}_{2M}$ reconstructs the Cartesian tensor from its spherical components.

\section{Clebsch-Gordan coefficients}

\subsection{CGCs calculation}
Consider two canonical states $\left|s_1 \sigma_1\right\rangle$ and $\left|s_2 \sigma_2\right\rangle$. Our goal is to couple these two states into a total spin-$S$, such that,
\begin{equation}
  |s_1-s_2|\leq S \leq s_1 + s_2.
\end{equation}
The states are related by a linear relation given by
\begin{equation}
  \left|S \sigma\right\rangle
  = \sum_{\sigma_1 \sigma_2} C^{s_1,\sigma_1;s_2,\sigma_2}_{S,\sigma}\left|s_1 \sigma_1\right\rangle \left|s_2 \sigma_2\right\rangle,
\end{equation}
where the coefficients $C^{s_1,\sigma_1;s_2,\sigma_2}_{S,\sigma}$ are called CGCs.
We have that $\sigma = \sigma_1 + \sigma_2$.
An expression for the CG coefficients was also derived by Wigner~\cite{Rose1995AngularMomentum}
\begin{equation}
\begin{aligned}
      C^{s_1,\sigma_1;s_2,\sigma_2}_{S,\sigma}
  =& \delta_{\sigma}^{\sigma_1+\sigma_2}\\
  &\times
  \left[
    (2S+1)
    \frac{
      (S+s_1-s_2)!(S-s_1+s_2)!(s_1+s_2-S)!(S+\sigma)!(S-\sigma)!
    }{
      (S+s_1+s_2+1)!(s_1-\sigma_1)!(s_1+\sigma_1)!(s_2-\sigma_2)!(s_2+\sigma_2)!
    }
  \right]^{\frac12}\\
  &\times
  \sum_{t}
  \frac{(-1)^{t+S+\sigma}}{t!}
  \frac{
    (S+\sigma_1+s_2-t)!(s_1-\sigma_1+t)!
  }{
    S-(s_1+s_2-t)!(S+\sigma-t)!(t+s_1-s_2-\sigma)!
  }.
\end{aligned}
\end{equation}
The sum is taken over all values of $t$ which lead to non negative factorials.
Using the orthogonality relations, we can get the inverse relations
\begin{equation}
  \left|s_1 \sigma_1\right\rangle \left|s_2 \sigma_2\right\rangle
  = \sum_{S} C_{s_1,\sigma_1;s_2,\sigma_2}^{S,\sigma} \left|S \sigma\right\rangle.
\end{equation}
The CGCs have the following symmetry properties:
\begin{align}\label{eq:CGCtran}
  C^{s_1,\sigma_1;s_2,\sigma_2}_{S,\sigma}
  &= (-1)^{s_1+s_2-S}C^{s_1,-\sigma_1;s_2,-\sigma_2}_{S,-\sigma},\\
  C^{s_1,\sigma_1;s_2,\sigma_2}_{S,\sigma}
  &= (-1)^{s_1+s_2-S}C^{s_2,\sigma_2;s_1,\sigma_1}_{S,\sigma},\\
  C^{s_1,\sigma_1;s_2,\sigma_2}_{S,\sigma}
  &= (-1)^{s_1-\sigma_1}
  \left[\frac{2S+1}{2s_2+1}\right]^{\frac12}C^{s_1,\sigma_1;S,-\sigma}_{s_2,-\sigma_2}.
\end{align}

\subsection{Examples of $\mathrm{SU}(2)$ CGCs calculation}
\label{subsec:Examples of CGC calculation}
We first present some examples of $\mathrm{SU}(2)$ CGC calculations. The $\mathrm{SU}(2)$ CGCs are
\begin{equation}
	C_{s,\{J\}}^{s_{1},\{I\};s_{2},\{K\}}\equiv\sqrt{\dfrac{(a+b)!(b+c)!(a+c+1)!}{b!(a+b+c+1)!a!c!}}\delta_{(J_{1}\cdots J_{a}}^{(I_{1}\cdots I_{a}}\varepsilon^{I_{a+1}\cdots I_{a+b}),(K_{1}\cdots K_{b}}\delta_{J_{a+1}\cdots J_{a+c})}^{K_{b+1}\cdots K_{b+c})},
\end{equation}
with
\begin{equation}
	\begin{aligned}
		a=&s_{1}+s-s_{2},\\
		b=&s_{1}+s_{2}-s,\\
		c=&s_{2}+s-s_{1},\\
		\varepsilon^{I_{1}\cdots I_{b},K_{1}\cdots K_{b}}=&\varepsilon^{I_{1}K_{1}}\cdots \varepsilon^{I_{b}K_{b}}.
	\end{aligned}
\end{equation}

\begin{enumerate}
\item $s=\frac{1}{2}, s_1=\frac{1}{2}, s_2=0$ with $a=1,b=0,c=0$. The CGC is
\begin{equation}
	C_{\frac{1}{2},\{J_{1}\}}^{\frac{1}{2},\{I_{1}\};0}=\delta_{J_{1}}^{I_{1}}.
\end{equation}
\item $s=1,s_1=1,s_2=0$ with $a=2,b=0,c=0$. The CGC is
\begin{equation}
\begin{aligned}
    C_{1,\{J_{1}J_2\}}^{1,\{I_{1}I_2\};0}=&\delta_{(J_{1}J_2)}^{(I_{1}I_2)}\\
    =&\frac{1}{2}(\delta_{J_{1}}^{I_{1}}\delta_{J_2}^{I_2}+\delta_{J_{1}}^{I_{2}}\delta_{J_{2}}^{I_{1}}).
\end{aligned}
\end{equation}
\item $s=\frac{3}{2},s_1=\frac{3}{2},s_2=0$ with $a=3,b=0,c=0$. The CGC is
\begin{equation}
    \begin{aligned}
        C_{\frac{3}{2},\{J_{1}J_2J_3\}}^{\frac{3}{2},\{I_1I_2I_3\};0}=&\delta_{(J_1J_2J_3)}^{(I_1I_2I_3)}\\
        =&\frac{1}{6}(\delta_{J_1}^{I_1}\delta_{J_2}^{I_2}\delta_{J_3}^{I_3}+\delta_{J_1}^{I_1}\delta_{J_2}^{I_3}\delta_{J_3}^{I_2}+\delta_{J_1}^{I_2}\delta_{J_2}^{I_1}\delta_{J_3}^{I_3}\\
        &+\delta_{J_1}^{I_2}\delta_{J_2}^{I_3}\delta_{J_3}^{I_1}+\delta_{J_1}^{I_3}\delta_{J_2}^{I_1}\delta_{J_3}^{I_2}+\delta_{J_1}^{I_3}\delta_{J_2}^{I_2}\delta_{J_3}^{I_1}).
    \end{aligned}
\end{equation}
\item $s=\frac{1}{2},s_1=\frac{1}{2},s_2=1$ with $a=0,b=1,c=1$. The CGC is
\begin{equation}
\begin{aligned}
    	C_{\frac{1}{2},\{J_{1}\}}^{\frac{1}{2},\{I_{1}\};1,\{K_{1}K_{2}\}}=&\sqrt{\frac{2}{3}}\varepsilon^{I_{1}(K_{1}}\delta_{J_{1}}^{K_{2})}\\
        =&\frac{1}{2}\sqrt{\frac{2}{3}}\left(\varepsilon^{I_{1}K_{1}}\delta_{J_{1}}^{K_{2}}+\varepsilon^{I_{1}K_{2}}\delta_{J_{1}}^{K_{1}}\right).
\end{aligned}
\end{equation}
\item $s=\frac{3}{2},s_1=\frac{1}{2},s_2=1$ with $a=1,b=0,c=2$. The CGC is
\begin{equation}
\begin{aligned}
    	C_{\frac{3}{2},\{J_{1}J_2J_3\}}^{\frac{1}{2},\{I_{1}\};1,\{K_{1}K_{2}\}}=&\delta_{(J_1J_2J_3)}^{I_1(K_1K_2)}\\
        =&\frac{1}{6}(\delta_{J_1}^{I_1}\delta_{J_2}^{K_1}\delta_{J_3}^{K_2}+\delta_{J_1}^{I_1}\delta_{J_3}^{K_1}\delta_{J_2}^{K_2}+\delta_{J_2}^{I_1}\delta_{J_1}^{K_1}\delta_{J_3}^{K_2}\\
        &+\delta_{J_2}^{I_1}\delta_{J_3}^{K_1}\delta_{J_1}^{K_2}+\delta_{J_3}^{I_1}\delta_{J_1}^{K_1}\delta_{J_2}^{K_2}+\delta_{J_3}^{I_1}\delta_{J_2}^{K_1}\delta_{J_1}^{K_2}).
\end{aligned}
\end{equation}
\item $s=\frac{1}{2},s_1=\frac{3}{2},s_2=1$ with $a=1,b=2,c=0$. The CGC is
\begin{equation}
\begin{aligned}
    	C_{\frac{1}{2},\{J_{1}\}}^{\frac{3}{2},\{I_{1}I_2I_3\};1,\{K_{1}K_{2}\}}=&\sqrt{\frac{1}{2}}\delta_{J_1}^{(I_{1}}\varepsilon^{I_{2}I_{3}),(K_{1}K_{2})}\\
        =&\frac{1}{6}\sqrt{\frac{1}{2}}(\delta_{J_1}^{I_{1}}\varepsilon^{I_{2}K_{1}}\varepsilon^{I_3K_2} +\delta_{J_1}^{I_{1}}\varepsilon^{I_{3}K_{1}}\varepsilon^{I_2K_2} +\delta_{J_1}^{I_{2}}\varepsilon^{I_{1}K_{1}}\varepsilon^{I_{3}K_{2}}\\
        &+\delta_{J_1}^{I_{2}}\varepsilon^{I_{3}K_{1}}\varepsilon^{I_{1}K_{2}} +\delta_{J_1}^{I_{3}}\varepsilon^{I_{1}K_{1}}\varepsilon^{I_{2}K_{2}} +\delta_{J_1}^{I_{3}}\varepsilon^{I_{2}K_{1}}\varepsilon^{I_{1}K_{2}}.\\
\end{aligned}
\end{equation}
\item $s=1,s_1=1,s_2=1$ with $a=1,b=1,c=1$. The CGC is
\begin{equation}
\begin{aligned}
    	C_{1,\{J_{1}J_2\}}^{1,\{I_{1}I_2\};1,\{K_{1}K_{2}\}}=&\delta_{(J_1}^{(I_{1}}\varepsilon^{I_{2})(K_{1}}\delta_{J_{2})}^{K_{2})}\\
        =&\frac{1}{8}(\delta_{J_1}^{I_{1}}\varepsilon^{I_{2}K_{1}}\delta_{J_{2}}^{K_{2}} + \delta_{J_1}^{I_{1}}\varepsilon^{I_{2}K_{2}}\delta_{J_{2}}^{K_{1}} +\delta_{J_1}^{I_{2}}\varepsilon^{I_{1}K_{1}}\delta_{J_{2}}^{K_{2}} +\delta_{J_1}^{I_{2}}\varepsilon^{I_{1}K_{2}}\delta_{J_{2}}^{K_{1}}\\
        &+\delta_{J_2}^{I_{1}}\varepsilon^{I_{2}K_{1}}\delta_{J_{1}}^{K_{2}} +\delta_{J_2}^{I_{1}}\varepsilon^{I_{2}K_{2}}\delta_{J_{1}}^{K_{1}} +\delta_{J_2}^{I_{2}}\varepsilon^{I_{1}K_{1}}\delta_{J_{1}}^{K_{2}} + \delta_{J_2}^{I_{2}}\varepsilon^{I_{1}K_{2}}\delta_{J_{1}}^{K_{1}}).
\end{aligned}
\end{equation}
\end{enumerate}

\section{Construction result for arbitrary spin}

For completeness, the following table collects the universal pattern of the $LS$ basis elements for arbitrary external spin and arbitrary operator representation $(j_L,j_R)$ with each row representing one class of allowed couplings.

{\renewcommand{\arraystretch}{2.8}
\setlength{\tabcolsep}{1pt}
\tiny
\setlength{\LTleft}{0pt}
\setlength{\LTright}{0pt}

\begin{longtable}{|c|c|c|c|c|>{\raggedright\arraybackslash}p{8.5cm}|}
\caption{Method II basis for $\mathcal{M}$ with general spin and general Lorentz irrep} \label{tab:structure_full_CG_narrow} \\
\hline\hline
$s_1$ & $L$ & $s_{p_3}$ & $J$ & $s_3$ & \multicolumn{1}{|c|}{basis elements} \\
\hline\hline
\endfirsthead

\multicolumn{6}{c}{{\bfseries  \thetable\ Irrep $\mathcal{M}$}} \\
\hline\hline
$s_1$ & $L$ & $s_{p_3}$ & $J$ & $s_3$ & \multicolumn{1}{|c|}{basis elements} \\
\hline\hline
\endhead

\hline \multicolumn{6}{|r|}{{next page...}} \\ \hline
\endfoot
\hline\hline
\endlastfoot

$s_1$ & $||s_{p_3} - J| - s_1|$ & $s_{p_3}$ & $J$ & $|s_{p_3} - J|$ & 
$\begin{aligned}
& \left[ \left( [\mathbf{1}\mathbf{3}] - \langle\mathbf{1}\mathbf{3}\rangle \right)^{2s_1; (I_1 \dots I_{2s_1}), (A_1 \dots A_{2s_1})} \right] \\
& \left[ \left( [\mathbf{3}| p_1 |\mathbf{3}\rangle \right)^{||s_{p_3} - J| - s_1|; (N_1 \dots N_{2||s_{p_3} - J| - s_1|})} \right] \\
& C^{|s_{p_3} - J|, \, (C_1 \dots C_{s_{p_3} - J + |s_{p_3} - J|} \,\, D_1 \dots D_{-s_{p_3} + J + |s_{p_3} - J|}}_{s_1, (A_1 \dots A_{2s_1}); \,\, ||s_{p_3} - J| - s_1|, (N_1 \dots N_{2||s_{p_3} - J| - s_1|})} \\
& \epsilon^{s_{p_3}+J-|s_{p_3} - J|;\, C_{s_{p_3}-J+|s_{p_3} - J|+1} \dots C_{2s_{p_3}}), D_{-s_{p_3}+J+|s_{p_3} - J|+1} \dots D_{2J}} \\
& | \mathbf{3}_{(D_1 \dots D_{j_L-j_R+J}} \rangle^{(j_L-j_R+J)}_{(\alpha_1 \dots \alpha_{j_L-j_R+J}} \\
& (p_3)^{(j_L+j_R-J)}_{\alpha_{j_L-j_R+J+1} \dots \alpha_{2j_L}), (\dot{\beta}_1 \dots \dot{\beta}_{j_L+j_R-J}} \\
& [\mathbf{3}_{D_{j_L-j_R+J+1} \dots D_{2J})} |^{(-j_L+j_R+J)}_{\dot{\beta}_{j_L+j_R-J+1} \dots \dot{\beta}_{2j_R})} \\[3ex]
\end{aligned}$ \\
\hline\hline

\end{longtable}}

\bibliographystyle{JHEP}
\bibliography{ref}
\end{document}